\documentclass[a4paper,twocolumn,11pt,accepted=2022-09-15]{quantumarticle}

\pdfoutput=1
\usepackage[utf8]{inputenc}
\usepackage[english]{babel}
\usepackage[T1]{fontenc}
\usepackage{amsmath}
\usepackage{hyperref}

\usepackage{tikz}
\usepackage{lipsum}

\usepackage{braket}
\usepackage{amsmath}
\usepackage{amssymb}
\usepackage{graphicx}
\usepackage{esint}
\usepackage{epstopdf}
\usepackage{color}
\usepackage{hyperref}
\usepackage{mathtools}
\usepackage{stackengine}

\usepackage[numbers,sort&compress]{natbib}

\begin{document}

\title{Quantum information with top quarks in QCD}

\author{Yoav Afik}
\email{yoavafik@gmail.com}
\affiliation{Experimental Physics Department, CERN, 1211 Geneva, Switzerland}
\orcid{0000-0001-8102-356X}

\author{Juan Ram\'on Mu\~noz de Nova}
\email{jrmnova@fis.ucm.es}
\affiliation{Departamento de F\'isica de Materiales, Universidad Complutense de Madrid, E-28040 Madrid, Spain}
\orcid{0000-0001-6229-9640}

\begin{abstract}
Top quarks represent unique high-energy systems since their spin correlations can be measured, thus allowing to study fundamental aspects of quantum mechanics with qubits at high-energy colliders. We present here the general framework of the quantum state of a top-antitop ($t\bar{t}$) quark pair produced through quantum chromodynamics (QCD) in a high-energy collider. We argue that, in general, the total quantum state that can be probed in a collider is given in terms of the production spin density matrix, which necessarily gives rise to a mixed state. We compute the quantum state of a $t\bar{t}$ pair produced from the most elementary QCD processes, finding the presence of entanglement and CHSH violation in different regions of phase space. We show that any realistic hadronic production of a $t\bar{t}$ pair is a statistical mixture of these elementary QCD processes. We focus on the experimentally relevant cases of proton-proton and proton-antiproton collisions, performed at the LHC and the Tevatron, analyzing the dependence of the quantum state with the energy of the collisions. We provide experimental observables for entanglement and CHSH-violation signatures. At the LHC, these signatures are given by the measurement of a single observable, which in the case of entanglement represents the violation of a Cauchy-Schwarz inequality. We extend the validity of the quantum tomography protocol for the $t\bar{t}$ pair proposed in the literature to more general quantum states, and for any production mechanism. Finally, we argue that a CHSH violation measured in a collider is only a weak form of violation of Bell's theorem, necessarily containing a number of loopholes.
\end{abstract}


\maketitle

\section{Introduction}\label{sec:intro}

The Standard Model of particle physics is a relativistic quantum field theory, based on special relativity and quantum mechanics. Therefore, it allows to study fundamental properties of quantum mechanics in a genuinely relativistic environment, at the frontier of the known Physics. However, even though the Standard Model is inherently a quantum theory, observing basic quantum phenomena in a high-energy collider can become quite challenging due to the nature of the measurement process.

Some light on this problem can be shed by the field of quantum information, where most of the foundations of quantum mechanics find a direct application. There, a characteristic signature of quantumness is provided by the existence of correlations that cannot be accounted by a classical probability theory, arising due to the intrinsic wave nature of quantum mechanics. Here emerges the concept of entanglement~\cite{Einstein1935,Schrodinger1935,Bell1964}, perhaps the most genuine feature of quantum mechanics. Entanglement plays a key role in quantum technologies like quantum computation, cryptography, metrology and teleportation~\cite{Bennett1993,Bouwmeester_1997,Gottesman_1999,Bennett2000,Raussendorf2001,Gisin2002,Giovannetti2004}. In particular, the study of entanglement in high-energy setups is of fundamental interest since entanglement is expected to be critically affected by relativistic effects~\cite{Gingrich2002,Peres2004,Friis2010,Friis2013,Giacomini2019,Kurashvili2022}. A number of works have already addressed the role of entanglement in the context of high-energy physics~\cite{Bramon2002,Shi2004,Kayser2010,Alba2017,Tu2020,Feal2021}.

The simplest system which can exhibit entanglement is that formed by two qubits. In high-energy physics, an interesting realization of a two-qubit system is a pair of decaying spin-1/2 particles, since their spin correlations can be measured from the kinematical distribution of their decay products. The best candidate from the Standard Model to carry out such a measurement is the top quark, the most massive fundamental particle known to exist ($m_tc^2 \approx 173~\textrm{GeV}$), first discovered by the D0 and CDF collaborations at the Tevatron in the Fermilab in 1995~\cite{Abachi1995,Abe1995}. Top quarks are typically produced in top-antitop ($t\bar{t}$) pairs which, due to their high mass, quickly decay well before any other process can affect their spin correlations. As a result, the spins of the $t\bar{t}$ pair are correlated with the kinematical distribution of the decay products, from where the original top spin quantum state can be reconstructed.

This feature has rendered top spin correlations a rich subject of study within high-energy physics~\cite{Kane1992,Bernreuther1994,Parke:1996pr,Bernreuther1998,Bernreuther2004,Uwer2005,Baumgart2013,Bernreuther2015}. Indeed, spin correlations between $t\bar{t}$ pairs have already been measured by the D0 and CDF collaborations at the Tevatron with proton-antiproton ($p\bar{p}$) collisions~\cite{Aaltonen2010,Abazov2011ka,Abazov2015psg}, and by the ATLAS and CMS collaborations at the Large Hadron Collider (LHC) with proton-proton ($pp$) collisions~\cite{Aad2012,Chatrchyan2013,Aad2014mfk,Sirunyan2019,Aaboud2019hwz}.

However, despite the extensive literature on the topic, only a very recent work has studied the entanglement between $t\bar{t}$ pairs~\cite{Afik2021}. Specifically, it was proposed that the entanglement of a $t\bar{t}$ pair can be detected at the LHC with high statistical significance using the current data recorded during Run~2. This would represent the first measurement ever of entanglement in a quark pair, and the highest-energy observation of entanglement so far. Entanglement has been also recently postulated as potentially sensitive to effects of New Physics beyond the Standard Model~\cite{Aoude2022,Fabbrichesi2022}. An experimental protocol for the quantum tomography of a $t\bar{t}$ pair was also developed in Ref.~\cite{Afik2021}, a canonical technique in quantum information but novel in the high-energy context. In parallel, some recent works have also addressed the possibility of measuring the violation of Bell inequalities with $t\bar{t}$ pairs~\cite{Fabbrichesi2021,Severi2022,Aguilar2022}, and also with $W^+W^-$ pairs~\cite{Barr:2021zcp}, which are massive spin-1 bosons. Even the measurement of such a basic quantum phenomenon as quantum interference can become non-trivial in a high-energy collider~\cite{Larkoski2022}. Apart from the intrinsic interest of studying foundational aspects of quantum mechanics at the frontier of the known Physics, all these works are paving the way to also use high-energy colliders for the study of quantum information theory. Due to their genuine relativistic behavior, the exotic character of the symmetries and interactions involved, as well as their fundamental nature, high-energy colliders are extremely attractive systems for these purposes.

Here, we provide the general formalism of the quantum state of a $t\bar{t}$ pair created through quantum chromodynamics (QCD) within a genuine quantum information approach. We discuss that, in general, the total quantum state that can be probed in a scattering experiment in a collider is given in terms of the so-called production spin density matrix~\cite{Bernreuther1994}, which is necessarily a mixed state. 

For the specific case of a $t\bar{t}$ pair, we analyze in detail its quantum state for the most elementary QCD production processes: light quark-antiquark ($q\bar{q}$) or gluon-gluon ($gg$) interactions. We extend the entanglement analysis of Ref.~\cite{Afik2021} to also include the violation of Bell inequalities, finding that both quantum phenomena emerge in certain regions of phase space for both $q\bar{q}$ and $gg$ processes. Interestingly, the entanglement structure can be qualitatively understood in terms of basic conservation laws, without the need of knowledge of the particular details of QCD interactions.

We show that any realistic QCD mechanism of $t\bar{t}$ production can be seen as a statistical mixture of the previous building blocks. In particular, we focus on the case of $pp$ and $p\bar{p}$ collisions, corresponding to the LHC and the Tevatron, respectively. We study the dependence of the quantum state on the energy of the collisions, extending the current phenomenological literature restricted to Run~2 of LHC~\cite{Afik2021,Fabbrichesi2021,Severi2022,Aguilar2022}. We find that, for sufficiently high energies, both types of collisions converge to the same state due to the dominance of $gg$ processes. 

In a high-energy collider, spin correlations are measured from the fit of the differential cross-section describing the angular distribution of the decay products~\cite{Bernreuther2004,Bernreuther:2010ny}. We propose realistic experimental observables for the characterization of the $t\bar{t}$ quantum state by integrating the signal in certain regions of phase space. At the LHC, it was shown that an entanglement signature can be obtained from the measurement of one single magnitude, proportional to the trace of the spin correlation matrix~\cite{Afik2021}. We prove in this work that the violation of Bell inequality can be also signaled from the measurement of just one parameter, the transverse spin correlations. Moreover, we carry out a study of the energy dependence of both signatures. Regarding the Tevatron, we find that an entanglement signature is provided by direct integration over the whole phase space. 

Finally, we analyze in detail the conceptual significance of the experimental implementation of these techniques. We show explicitly that an entanglement measurement at the LHC represents the violation of a Cauchy-Schwarz inequality. In contrast, we do not expect a statistically significant observation of entanglement at the Tevatron. We argue that the quantum tomography protocol developed in Ref.~\cite{Afik2021} can be extended to more general quantum states, and to any $t\bar{t}$ production process. We discuss that, due to the nature of the detection process, only weak violations of Bell inequalities can be measured in a high-energy collider, since some loopholes, like those related to the free-will or to the detection efficiency, cannot be closed. 

Remarkably, while the probability and the spin density matrix of each $t\bar{t}$ production process are computed by the high-energy theory, once both are given, we are simply left with a typical problem in quantum information involving the convex sum of two-qubit quantum states, where the usual techniques of the field can be applied. This important observation is translated into the article presentation, fully developed within a quantum information language, aimed at making it easily understandable by the physics community outside the high-energy field.

The paper is arranged as follows. Section~\ref{sec:general} discusses in detail the general framework upon we build the results of this work, presenting the basic tools of two-qubit systems used throughout this work along with a general discussion about quantum states in colliders and relativistic particle-antiparticle production. Section~\ref{sec:entanglementfundamental} studies the $t\bar{t}$ quantum state in elementary QCD processes. Section~\ref{sec:EntanglementRealistic} extends the previous results to more realistic processes occurring in actual colliders, analyzing in detail the energy dependence. Section~\ref{sec:totalquantumstate} translates these ideas into relevant experimental observables. Section~\ref{sec:experimental} provides some technical remarks about the experimental implementation of the discussed quantum information techniques. Finally, Section~\ref{sec:conclusions} summarizes the main conclusions, and discusses future perspectives. Technical details are given in the Appendices.

\section{General formalism}\label{sec:general}

\subsection{Two-qubit systems}\label{subsec:Hilbert}

Quantum states are represented in general by a density matrix $\rho$, a Hermitian non-negative operator with unit trace in a certain Hilbert space $\mathcal{H}$, $\rm{tr}(\rho)=1$. These conditions imply that the number of real parameters characterizing $\rho$ for $\dim \mathcal{H}=N$ is $N^2-1$. Expectation values for observables $O$ are computed by taking the product trace, $\braket{O}=\rm{tr}(O\rho)$. 

The most simple example of density matrix is provided by a qubit, that is, a two-level quantum system. The density matrix of a qubit takes the simple form
\begin{equation}\label{eq:GeneralQubitQuantumState}
    \rho=\frac{I_2+\sum_{i}B_{i}\sigma^i}{2}
\end{equation}
with $I_n$ the $n\times n$ identity matrix and $\sigma^i$, $i=1,2,3$, the usual Pauli matrices. A physical (i.e., non-negative) density matrix $\rho$
is described by Bloch vectors $|\mathbf{B}|\leq 1$, with pure states given by unit vectors saturating the inequality. The $3$ coefficients $B_{i}$ completely determine the quantum state of the system. For a spin-1/2 particle, they represent the spin polarization, $B_i=\braket{\sigma^i}$. An alternative description of $\rho$ can be provided in terms of angular momentum coherent states $\ket{\mathbf{\hat{n}}}$ through the $P$-representation (see Appendix~\ref{app:Coherent} for a summary of its main properties):
\begin{equation}\label{eq:PRepresentationQubit}
    \rho=\int \mathrm{d}\Omega~P(\mathbf{\hat{n}})\ket{\mathbf{\hat{n}}}\bra{\mathbf{\hat{n}}},~\int \mathrm{d}\Omega~P(\mathbf{\hat{n}})=1
\end{equation}
where the angular momentum coherent states $\ket{\mathbf{\hat{n}}}$ satisfy $\mathbf{\hat{n}}\cdot\mathbf{\sigma}\ket{\mathbf{\hat{n}}}=\ket{\mathbf{\hat{n}}}$, and $\Omega$ is the solid angle associated to the unit vector $\mathbf{\hat{n}}$. The function $P(\mathbf{\hat{n}})$ is the qubit analogue of the celebrated Glauber-Sudarshan $P$-function in quantum optics~\cite{Walls2008}. It is easily seen that $P(\mathbf{n})$ can be always chosen as non-negative for any density matrix $\rho$ describing a qubit. 

More intriguing quantum states arise in bipartite Hilbert spaces $\mathcal{H}=\mathcal{H}_A\otimes\mathcal{H}_B$ composed of subsystems $A,B$. A quantum state is called separable \textit{iff} it can be written as a convex sum of product states
\begin{equation}\label{eq:Separability}
\rho=\sum_n p_n \rho^{A}_{n}\otimes\rho^{B}_{n},~\sum_n p_n=1,
\end{equation}
with $p_n\geq 0$. An \textit{entangled} state is defined as a non-separable state. 

To illustrate this concept, we consider the case of a pair of qubits, whose density matrix $\rho$ can be decomposed as
\begin{widetext}
\begin{equation}\label{eq:GeneralBipartiteStateRotations}
\rho=\frac{I_4+\sum_{i}\left(B^{+}_{i}\sigma^i\otimes I_2+B^{-}_{i}I_2\otimes\sigma^i \right)+\sum_{i,j}C_{ij}\sigma^{i}\otimes\sigma^{j}}{4}
\end{equation}
\end{widetext}
Now, the quantum state of the system is determined by $15$ parameters $B^{\pm}_{i},C_{ij}$, which for spin-1/2 particles are their spin polarizations $B^{+}_{i}=\braket{\sigma^i\otimes I_2},~B^{-}_{i}=\braket{I_2\otimes\sigma^i}$, and their spin correlations $C_{ij}=\braket{\sigma^{i}\otimes\sigma^{j}}$. 

An insightful way to understand entanglement for $2$-qubit systems arises by considering the $P$-representation 
\begin{equation}\label{eq:PRepresentation2Qubit}
    \rho=\int\mathrm{d}\Omega_A\mathrm{d}\Omega_B~P(\mathbf{n}_A,\mathbf{n}_B)\ket{\mathbf{n}_A\mathbf{n}_B}\bra{\mathbf{n}_A\mathbf{n}_B},
\end{equation}
where $\ket{\mathbf{n}_A\mathbf{n}_B}=\ket{\mathbf{n}_A}\otimes\ket{\mathbf{n}_B}$. From Eq.~(\ref{eq:Separability}), it is immediately seen that a state is separable \textit{iff}  it admits a non-negative $P$-representation. This implies that the spin correlations for a separable state
\begin{equation}
    C_{ij}=\braket{\sigma^{i} \otimes\sigma^{j}}=\int\mathrm{d}\Omega_A\mathrm{d}\Omega_B~P(\mathbf{n}_A,\mathbf{n}_B)n^i_An^j_B
\end{equation}
are purely classical, described by a classical probability distribution. The same applies to the spin polarizations. Thus, in the language of the $P$-function, entanglement is equivalent to a $P$-function that necessarily presents negative values, a genuine distinctive signature of non-classicality and quantum behavior as well known from quantum optics~\cite{Walls2008}.

In order to signal the presence of entanglement, several theoretical criteria can be used. Perhaps the most well-known one is the Peres-Horodecki criterion~\cite{Peres1996,Horodecki1997}, which simply states that if $\rho$ is separable, taking partial transpose with respect to the second subsystem
\begin{equation}\label{eq:SeparabilityPeresHorodecki}
\rho^{\rm{T_2}}=\sum_n p_n \rho^{a}_{n}\otimes\left(\rho^{b}_{n}\right)^{\rm{T}}
\end{equation}
also yields a non-negative operator. Hence, a not positive semi-definite $\rho^{\rm{T_2}}$ implies that $\rho$ is entangled. In two-qubit systems, the Peres-Horodecki criterion is also a necessary condition for entanglement.

A quantitative measurement of the entanglement of two qubits is provided by the concurrence~\cite{Wooter1998}:
\begin{equation}\label{eq:Concurrence}
\mathcal{C}[\rho]\equiv \max(0,\lambda_1-\lambda_2-\lambda_3-\lambda_4)
\end{equation}
with $\lambda_i$ the eigenvalues of the matrix $\sqrt{\sqrt{\rho}\tilde{\rho}\sqrt{\rho}}$ ordered in decreasing magnitude, $\tilde{\rho}=(\sigma_2\otimes\sigma_2)~\rho^*~(\sigma_2\otimes\sigma_2)$, and $\rho^*$ the complex conjugate of $\rho$ in the spin basis of $\sigma_3$. The concurrence is related to the entanglement of formation and satisfies $0\leq \mathcal{C}[\rho]\leq 1$, where a quantum state is entangled \textit{iff}  $\mathcal{C}[\rho]>0$. Hence, $\mathcal{C}[\rho]=1$ implies that $\rho$ is maximally entangled.

The computation of these magnitudes requires full knowledge of the quantum state. This can be achieved by means of quantum tomography, a technique able to reconstruct a quantum state from the measurement of a selected set of observables. For the case of a single qubit, characterized by $3$ parameters, it is enough to measure the Bloch vector $\mathbf{B}$, which is the spin polarization for a spin-1/2 particle. For the case of two qubits, the quantum tomography is performed by measuring the $15$ parameters determining $\mathbf{B}^{\pm},\mathbf{C}$, which for spin-1/2 particles represent the spin polarizations and spin correlations, respectively. In actual experiments, an additional measurement is typically required to ensure the proper normalization of the density matrix~\cite{James2001}. Nevertheless, in order to detect entanglement, simpler criteria which require the measurement of just a few parameters can be formulated. Appendix~\ref{app:criteria} contains some general entanglement criteria specifically developed for $t\bar{t}$ quantum states and used throughout this work.

An even stronger requirement than entanglement is the violation of Bell-type inequalities~\cite{Bell1964}. No local hidden-variable model can give rise to it, implying a violation of local realism. The Clauser-Horner-Shimoni-Holt (CHSH) inequality~\cite{Clauser1969} provides a particularly useful form of Bell inequality for $2\times 2$ systems:
\begin{equation}\label{eq:CHSHOriginal}
|C(a_1,b_1)-C(a_1,b_2)+C(a_2,b_1)+C(a_2,b_2)|\leq 2
\end{equation}
where $a_i,b_i$, $i=1,2$ are measurement settings in the Alice, Bob subsystems $A,B$, and $C(a_i,b_j)$ are their correlations. In the case of spin-1/2 particles, they can be interpreted as measurements of spin polarizations along certain directions defined by unit vectors $\mathbf{a}_i,\mathbf{b}_j$, $C(a_i,b_j)=\braket{(\mathbf{a}_i\cdot \mathbf{\sigma}) \otimes (\mathbf{b}_j\cdot \mathbf{\sigma})}$. Thus, Eq.~(\ref{eq:CHSHOriginal}) can be rewritten in vector notation as simply
\begin{equation}\label{eq:CHSHOriginalSpin}
    |\mathbf{a}^{\textrm{T}}_1\mathbf{C}\left(\mathbf{b}_1-\mathbf{b}_2\right)+\mathbf{a}^{\textrm{T}}_2\mathbf{C}\left(\mathbf{b}_1+\mathbf{b}_2\right)|\leq 2
\end{equation}
with $\mathbf{C}$ the same correlation matrix of Eq.~(\ref{eq:GeneralBipartiteStateRotations}). Any separable state of the form (\ref{eq:Separability}) satisfies Eq.~(\ref{eq:CHSHOriginalSpin}). Hence, only entangled states can violate a CHSH inequality.

In practice, the maximization of the l.h.s. of Eq.~(\ref{eq:CHSHOriginalSpin}) is computed from
\begin{align}\label{eq:CHSH}
\nonumber \mathcal{B}[\rho]&\equiv \max_{\mathbf{a}_i,\mathbf{b}_i}|\mathbf{a}^{\textrm{T}}_1\mathbf{C}\left(\mathbf{b}_1-\mathbf{b}_2\right)+\mathbf{a}^{\textrm{T}}_2\mathbf{C}\left(\mathbf{b}_1+\mathbf{b}_2\right)|\\
&=2\sqrt{\mu_1+\mu_2}
\end{align}
where $0\leq \mu_i\leq 1$ are the eigenvalues of $\mathbf{C}^{\textrm{T}}\mathbf{C}$, ordered in decreasing magnitude~\cite{Horodecki1995}. Therefore, the CHSH inequality can be violated \textit{iff}  $\mu_1+\mu_2>1$, where the maximum possible violation of the CHSH inequality is the Cirel'son bound $\mathcal{B}[\rho]=2\sqrt{2}$~\cite{Cirelson1980}.

\subsection{Quantum states in colliders}\label{subsec:QuantumColliders}

We now study how the physics in real colliders can be described in terms of quantum states. For illustrative purposes, we first address the simple non-relativistic scattering of an incident spinless particle with mass $m$ from a single fixed target, characterized by a potential $V$ (the interested reader is referred to Ref.~\cite{Taylor2006} for the basics of scattering theory). The process is determined by the scattering matrix $S$, whose elements in momentum representation are
\begin{equation}\label{eq:ScatteringMatrix}
\braket{\mathbf{p}'|S|\mathbf{p}}=\delta(\mathbf{p}-\mathbf{p}')-2\pi i\delta(E_{\mathbf{p}'}-E_{\mathbf{p}})
\braket{\mathbf{p}'|T|\mathbf{p}}
\end{equation}
where the $T$-matrix satisfies the Lippmann-Schwinger equation $T=V+VG_0T$, with $G_0$ the usual retarded free Green's function, and $E_{\mathbf{p}}=\mathbf{p}^2/2m$ is the kinetic energy. In the Dirac picture, the wave function of the scattered state resulting from an incident particle with well-defined momentum $\ket{\mathbf{p}}$ is
\begin{equation}
    \ket{\Psi}=S\ket{\mathbf{p}}
\end{equation}
The scattering amplitude $f(\mathbf{p}\rightarrow \mathbf{p}')\equiv -(2\pi)^2m\hbar\braket{\mathbf{p}'|T|\mathbf{p}}$ determines the differential cross-section characterizing the scattering to some momentum $\mathbf{p}'$ with the same energy, 
\begin{equation}
\frac{\mathrm{d}\sigma}{\mathrm{d}\Omega}=|f(\mathbf{p}\rightarrow \mathbf{p}')|^2
\end{equation}
$\Omega$ being the solid angle associated to $\mathbf{p}'$. Therefore, the differential cross-section is proportional to the probability of the process, given by the squared \textit{on-shell} $T$-matrix element connecting the initial to the final state. 

The scattered state $\ket{\Psi}$ can be also rewritten as a density matrix:
\begin{align}
\rho_S&=\ket{\Psi}\bra{\Psi}\\
\nonumber &=
\iint\mathrm{d}\mathbf{p}' \mathrm{d}\mathbf{p}''\ket{\mathbf{p}'}\bra{\mathbf{p}'}S\ket{\mathbf{p}}\bra{\mathbf{p}}S^{\dagger}\ket{\mathbf{p}''}\bra{\mathbf{p}''}
\end{align}


In a real collider, only momentum measurements of the scattered particle can be performed and thus, not all the information is needed to describe the quantum state in experiments. Moreover, scattering events along the beam direction are not typically measured. Therefore, the relevant quantum state that encodes all the information that can be probed in a collider results from projecting $\rho_S$ onto momentum states $\mathbf{p}'\neq \mathbf{p}$ with the operators $\Pi_{\mathbf{p}'}=\ket{\mathbf{p}'}\bra{\mathbf{p}'}$ as
\begin{align}\label{eq:QuantumStateColliderNR}
    \rho&=\frac{\int \mathrm{d}\mathbf{p}'~\Pi_{\mathbf{p}'}\ket{\Psi}\bra{\Psi}\Pi_{\mathbf{p}'}}{\int \mathrm{d}\mathbf{p}'~|\braket{\Psi|\mathbf{p}'}|^2\braket{\mathbf{p}'|\mathbf{p}'}}\\
    \nonumber &=\frac{\int \mathrm{d}\Omega~|\bra{\mathbf{p}'}T\ket{\mathbf{p}}|^2\ket{\mathbf{p}'}\bra{\mathbf{p}'}}{\int \mathrm{d}\Omega~|\bra{\mathbf{p}'}T\ket{\mathbf{p}}|^2\braket{\mathbf{p}'|\mathbf{p}'}}\\ \nonumber &=\frac{1}{\sigma}
    \int \mathrm{d}\Omega~\frac{\mathrm{d}\sigma}{\mathrm{d}\Omega}\frac{\ket{\mathbf{p}'}\bra{\mathbf{p}'}}{\braket{\mathbf{p}'|\mathbf{p}'}}
\end{align}
where the factor $\braket{\mathbf{p}'|\mathbf{p}'}$ in the denominator ensures proper normalization, $\textrm{tr}~\rho=1$.
The quantum state described by the density matrix $\rho$ is thus mixed, resulting from the incoherent sum of momentum states with a probability proportional to the differential cross-section, where the total cross-section $\sigma$ here plays an analogue role to the partition function in statistical mechanics. As a result, $\rho$ is written only in terms of differential cross-sections, the observables measured in actual colliders, computed in terms of $T$-matrix elements. In this way, the expectation value of any momentum observable $O(\mathbf{P})$ can be obtained from $\rho$ as
\begin{equation}
    \braket{O}=\textrm{tr}[O\rho]=\frac{1}{\sigma}
    \int \mathrm{d}\Omega~\frac{\mathrm{d}\sigma}{\mathrm{d}\Omega}O(\mathbf{p}')
\end{equation}

For a spin-half particle, things go along the same lines. By noticing that the spin of the scattered particle is not typically detected in a collider, we now project using $\Pi_{\mathbf{p}'}=\sum_\alpha \ket{\mathbf{p}'\alpha}\bra{\mathbf{p}'\alpha}$ with $\alpha$ labeling spin indices, finding that the quantum state characterizing the scattering of an incident particle with momentum $\mathbf{p}$ and spin $\lambda$ is
\begin{align}\label{eq:QuantumStateNRSpin}
    \nonumber  &\rho^\lambda=\frac{1}{Z}\sum_{\alpha\beta}\int \mathrm{d}\Omega~\bra{\mathbf{p}'\alpha}T\ket{\mathbf{p}\lambda}\bra{\mathbf{p}\lambda}T^{\dagger}\ket{\mathbf{p}'\beta}\\
    &\times\frac{\ket{\mathbf{p}'\alpha}\bra{\mathbf{p}'\beta}}{\braket{\mathbf{p}'|\mathbf{p}'}} =\frac{1}{Z}\sum_{\alpha\beta}\int \mathrm{d}\Omega~R^{\lambda}_{\alpha\beta}(\mathbf{p}')\frac{\ket{\mathbf{p}'\alpha}\bra{\mathbf{p}'\beta}}{\braket{\mathbf{p}'|\mathbf{p}'}}
\end{align}
where we have used $\braket{\mathbf{p}'\alpha|\mathbf{p}'\beta}=\braket{\mathbf{p}'|\mathbf{p}'}\delta_{\alpha\beta}$. In the equation above, the partition function $Z$ is defined to ensure normalization, and the production spin density matrix, in the following simply denoted as the $R$-matrix~\cite{Aoude2022}, is defined as
\begin{equation}\label{eq:ProductionSpinDensityMatrix}
    R^{\lambda}_{\alpha\beta}(\mathbf{p}')\equiv \bra{\mathbf{p}'\alpha}T\ket{\mathbf{p}\lambda}\bra{\mathbf{p}\lambda}T^{\dagger}\ket{\mathbf{p}'\beta}
\end{equation}
The $R$-matrix is not properly normalized, since its trace is proportional to the differential cross-section of the process,
\begin{align}
\textrm{tr}R^{\lambda}(\mathbf{p}')&=\sum_\alpha R^{\lambda}_{\alpha\alpha}(\mathbf{p}')\\
    \nonumber & =\sum_\alpha |\bra{\mathbf{p}'\alpha}T\ket{\mathbf{p}\lambda}|^2\propto\frac{\mathrm{d}\sigma^{\lambda}}{\mathrm{d}\Omega}
\end{align}
The proportionality factor can be chosen arbitrarily; for the present moment, we take $R^{\lambda}$ directly proportional to the \textit{on-shell} $T$-matrix elements. Thus, the partition function becomes once more proportional to the total cross-section
\begin{equation}
    Z=\int \mathrm{d}\Omega~\textrm{tr}\,R^{\lambda}(\mathbf{p}')\propto\int \mathrm{d}\Omega~\frac{\mathrm{d}\sigma^{\lambda}}{\mathrm{d}\Omega}=\sigma^\lambda
\end{equation}

As a $2\times 2$ Hermitian matrix, the most general form of the $R$-matrix is similar to that of Eq.~(\ref{eq:GeneralQubitQuantumState}),
\begin{equation}
    R^{\lambda}=\tilde{A}^{\lambda}+\sum_i \tilde{B}^{\lambda}_i\sigma^i,
\end{equation}
but with an extra parameter $\tilde{A}^{\lambda}$ that determines the probability of the process, $\textrm{tr}R^{\lambda}(\mathbf{p}')=2\tilde{A}^{\lambda}(\mathbf{p}')$.

The proper spin density matrix with unit trace describing the quantum state for a scattering process along a fixed direction is obtained by normalization of $R$, 
\begin{equation}
    \rho^{\lambda}_{\alpha\beta}(\mathbf{p}')=\frac{R^{\lambda}_{\alpha\beta}(\mathbf{p}')}{\textrm{tr}R^{\lambda}(\mathbf{p}')}=\frac{R^{\lambda}_{\alpha\beta}(\mathbf{p}')}{2\tilde{A}^{\lambda}(\mathbf{p}')},
\end{equation}
and whose spin polarization $\mathbf{B}^\lambda$ is given by $B^\lambda_i=\tilde{B}^\lambda_i/\tilde{A}^{\lambda}$.
In terms of these quantum substates, the total quantum state of Eq.~(\ref{eq:QuantumStateNRSpin}) can be written as simply
\begin{equation}\label{eq:QuantumStateNRSpinClear}
    \rho^{\lambda}=
    \frac{1}{\sigma^\lambda}\sum_{\alpha\beta}\int \mathrm{d}\Omega~\frac{\mathrm{d}\sigma^{\lambda}}{\mathrm{d}\Omega}\rho^{\lambda}_{\alpha\beta}(\mathbf{p}')\frac{\ket{\mathbf{p}'\alpha}\bra{\mathbf{p}'\beta}}{\braket{\mathbf{p}'|\mathbf{p}'}}
\end{equation}
Hence, we can intuitively understand the total spin-momentum quantum state $\rho^{\lambda}$ as the incoherent average in momentum of the spin quantum states $\rho^{\lambda}(\mathbf{p}')$ describing the scattering along a fixed direction, weighted with the probability of that scattering process (proportional to the differential cross-section). The total cross-section $\sigma^\lambda$ plays once more the role of the partition function.

Finally, we note that in a collider the initial state is typically unpolarized since spin degrees of freedom cannot be controlled. Thus, the real quantum state describing the experiment after many events results from the averaged $R$-matrix
\begin{equation}
   R(\mathbf{p}')=\frac{1}{2}\sum_\lambda R^{\lambda}(\mathbf{p}')
\end{equation}
In summary, the quantum state describing the scattered particles in a collider is a mixed state because of two fundamentally different reasons, related to the control over the degrees of freedom. (i) Regarding orbital variables: Momentum distributions of the scattered particles are the only measurable observables in colliders, given in terms of differential cross-sections. Thus, even if the scattered state is pure, one can only access to the diagonal part in momentum, which motivates the use of a reduced mixed density matrix. (ii) Regarding discrete variables: Internal degrees of freedom of the initial state cannot be typically controlled and one has to average uniformly over them. This average over all possible initial states after many scattering events results in an incoherent mixture that is described by a density matrix.

\subsection{Relativistic particle-antiparticle pair production}

We now switch to the actual case of high-energy colliders, whose physics is described by relativistic quantum field theories within the framework of the Standard Model. The Standard Model is composed of fundamental spin-1/2 particles (and their corresponding antiparticles) which interact through the mediation of gauge bosons of spin-1. These interactions can be either strong (mediated by massless gluons $g$), weak (mediated by the massive $Z^0,W^{\pm}$ bosons), or electromagnetic (mediated by the photon $\gamma$), corresponding to a Yang-Mills theory $\text{SU(3)} \otimes \text{SU(2)} \otimes \text{U(1)}$, respectively. Electromagnetic and weak interactions are both unified within the framework of electroweak theory, while strong interactions are described by quantum chromodynamics. The remaining particle of the Standard Model, the Higgs boson, has spin-0, and is the responsible for the occurrence of mass in massive particles. 

A natural Standard Model candidate for an entangled two-qubit system is a particle-antiparticle (denoted generically as $P\bar{P}$) pair produced from some initial state $I$:
\begin{equation}
    I\rightarrow P+\bar{P}
\end{equation}
In the following, we restrict to the case where $P$ is a Standard Model fermion of mass $m$, although the formalism can be easily generalized to the case where $P$ is some massless spin-1 gauge boson like the photon. Among the Standard Model fermions, we can distinguish between quarks and leptons, where the latter do not interact through QCD since they do not have color degrees of freedom. Typically, the components of the initial state $I$ annihilate themselves through gauge interactions, giving rise to the $P\bar{P}$ pair. In the process, total energy and momentum are conserved.

The kinematics of a $P\bar{P}$ pair is determined in the center-of-mass (c.m.) frame by the invariant mass $M$ and the direction of flight $\hat{k}$ of the particle $P$. In this frame, the particle/antiparticle four-momenta read $k^{\mu}=(k^0,\mathbf{k}), \bar{k}^{\mu}=(k^0,-\mathbf{k})$, with $\hat{k}=\mathbf{k}/|\mathbf{k}|$, and satisfy the Lorentz-invariant dispersion relation 
\begin{equation}
\bar{k}^2=k^2\equiv k^{\mu} k_{\mu}=\left(k^0\right)^2-\mathbf{k}^2=m^2,
\end{equation}
where in the following we work in natural units $\hbar=c=1$. 

The invariant mass $M$ is the c.m. energy of the pair, defined from the usual invariant Mandelstam variables as
\begin{equation}
    M^2\equiv s\equiv(k+\bar{k})^2,
\end{equation}
In the c.m. frame, $M^2=4\left(k^0\right)^2=4(m^2 +\mathbf{k}^2)$, where the momentum is related to the particle velocity $\beta$ by $|\mathbf{k}|=m\beta/\sqrt{1-\beta^2}$, so
\begin{equation}
    \beta=\sqrt{1-\frac{4m^2}{M^2}}
\end{equation}
Hence, threshold production ($\beta=0$) is at the minimum energy possible for a $P\bar{P}$ pair, $M=2m$, as naturally expected.

With respect to the quantum state of the $P\bar{P}$ pair, \textit{mutatis mutandis}, we define again an $R$-matrix in terms of the \textit{on-shell} relativistic $T$-matrix elements as
\begin{equation}\label{eq:ProductionSpinDensityMatrixT}
R^{I\lambda}_{\alpha\beta,\alpha'\beta'}(M,\hat{k})\equiv \bra{M\hat{k}\alpha\beta}T\ket{I\lambda}\bra{I\lambda}T^{\dagger}\ket{M\hat{k}\alpha'\beta'}
\end{equation}
with $\ket{I\lambda}$ labeling the initial state of the system and
\begin{equation}
    \ket{M\hat{k}\alpha\beta}\equiv \ket{k\alpha}\otimes\ket{\bar{k}\beta}
\end{equation}
where the first/second subspace correspond to the particle/antiparticle, respectively. The spins of the particles are computed in their respective rest frames, where they are well defined. Once more, if internal degrees of freedom $\lambda$ of the initial state cannot be controlled (like spin or color), one uses the averaged $R$-matrix
\begin{equation}\label{eq:ProductionSpinDensityMatrixTAveraged}
    R^{I}_{\alpha\beta,\alpha'\beta'}(M,\hat{k})=\frac{1}{N_\lambda}\sum_\lambda R^{I\lambda}_{\alpha\beta,\alpha'\beta'}(M,\hat{k})
\end{equation}
with $N_{\lambda}$ the number of internal degrees of freedom of $I$. In analogy to Eq.~(\ref{eq:QuantumStateNRSpin}), the resulting quantum state is 
\begin{align}
\rho^I&=\frac{1}{Z}\sum_{\alpha\beta,\alpha'\beta'}\int \mathrm{d}\Omega~R^{I}_{\alpha\beta,\alpha'\beta'}(M,\hat{k}) \\ \nonumber
&\times\frac{\ket{M\hat{k}\alpha\beta}\bra{M\hat{k}\alpha'\beta'}}{\braket{M\hat{k}|M\hat{k}}}
\end{align}
with the partition function $Z$ proportional once more to the total cross-section
\begin{equation}
    Z=\int \mathrm{d}\Omega~\textrm{tr}~R^I(M,\hat{k})\propto \sigma^I
\end{equation}
The $R$-matrix characterizes the quantum state of any $P\bar{P}$ pair produced in a relativistic process, which is a mixed state by the very same reasons as in the non-relativistic case. It takes the general form
\begin{align}\label{eq:GeneralDensityMatrix}
R&=\tilde{A}I_4+\sum_{i}\left(\tilde{B}^{+}_{i}\sigma^i\otimes I_2+\tilde{B}^{-}_{i}I_2\otimes\sigma^i \right)\\ \nonumber
&+\sum_{i,j}\tilde{C}_{ij}\sigma^{i}\otimes\sigma^{j}
\end{align}
in a similar fashion to Eq.~(\ref{eq:GeneralBipartiteStateRotations}), but once more with an extra coefficient $\tilde{A}$ proportional to the c.m. differential cross-section.

The actual spin density matrix is obtained from the normalization of $R$:
\begin{equation}\label{eq:PhysicalQuantu}
\rho=\frac{R}{\mathrm{tr}(R)}=\frac{R}{4\tilde{A}}
\end{equation}
As a result, the spin polarizations $B^{\pm}_{i}$ and spin correlations $C_{ij}$ of the $P\bar{P}$ pair are
\begin{equation}\label{eq:PhysicalPolCor}
B^{\pm}_{i}=\frac{\tilde{B}^{\pm}_{i}}{\tilde{A}},~C_{ij}=\frac{\tilde{C}_{ij}}{\tilde{A}} 
\end{equation}

In terms of these quantum states, one can retrieve the relativistic version of Eq.~(\ref{eq:QuantumStateNRSpinClear}),
\begin{align}\label{eq:QuantumStateRSpin}
\rho^I&=\frac{1}{\sigma^I}\sum_{\alpha\beta,\alpha'\beta'}\int \mathrm{d}\Omega~\frac{\mathrm{d}\sigma^I}{\mathrm{d}\Omega~}\rho^{I}_{\alpha\beta,\alpha'\beta'}(M,\hat{p})\\ \nonumber
&\times \frac{\ket{M\hat{k}\alpha\beta}\bra{M\hat{k}\alpha'\beta'}}{\braket{M\hat{k}|M\hat{k}}}
\end{align}

\section{Production of $t\bar{t}$ in elementary QCD processes} \label{sec:entanglementfundamental}

A particularly interesting example of particle-antiparticle pair is provided by a top-antitop quark pair, which will be the subject of study throughout the rest of the work. Specifically, we restrict to the production of $t\bar{t}$ pairs through QCD processes, although the concepts and techniques developed here can be straightforwardly extended to other types of interaction and/or particle-antiparticle pairs. For the theoretical computations, we employ leading-order (LO) QCD perturbation theory, since it provides analytical results and a clear picture of the underlying physics. Higher-order corrections are known to be small and do not change the main results~\cite{Bernreuther2004,Afik2021,Fabbrichesi2021,Severi2022,Aoude2022}.

\begin{figure}[t]
  \centering
     \includegraphics[width=\columnwidth]{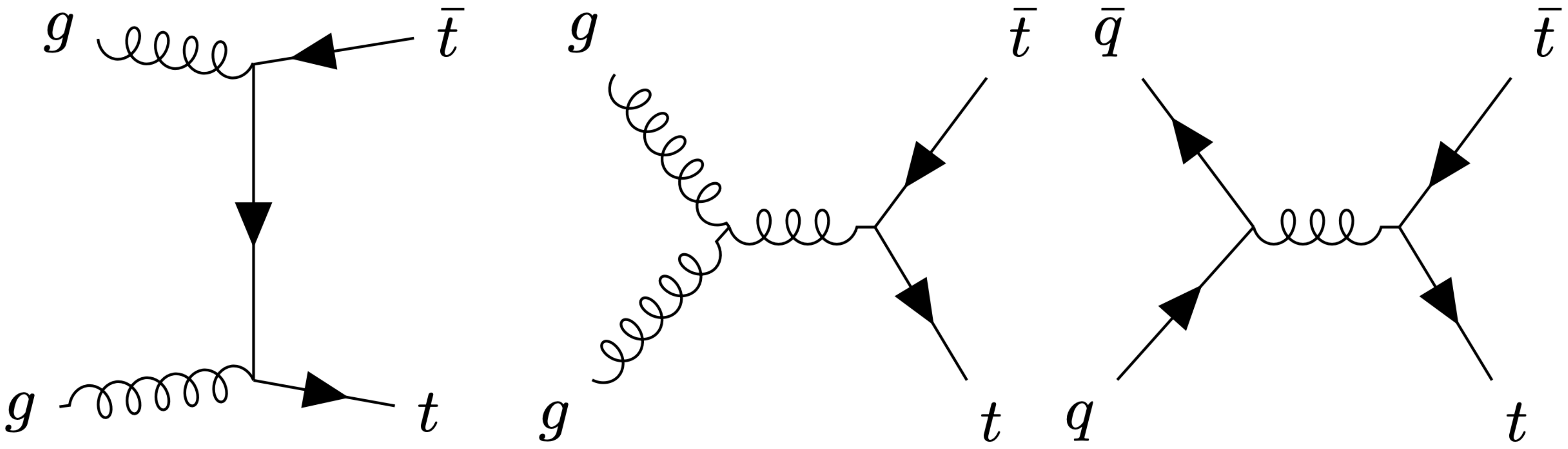} 
     \caption{Representative Feynman diagrams for $t\bar{t}$ production within the Standard Model. Spring lines represent gluons and straight lines represent quarks, all either real or virtual.}
     \label{fig:Feynman}
\end{figure}

At LO QCD, only two initial states can produce a $t\bar{t}$ pair: a light quark-antiquark ($q \bar{q}$) or a gluon ($gg$) pair,
\begin{align}\label{eq:partonreactions}
    q+\bar{q}&\rightarrow t+\bar{t} ,\\
\nonumber g+g&\rightarrow t+\bar{t} .
\end{align}
Representative Feynman diagrams for these processes in the Standard Model are depicted in Fig.~\ref{fig:Feynman}. 

Each initial state $I=q\bar{q},gg$ creates a $t\bar{t}$ pair in a spin quantum state described by the production spin density matrix $R^{I}(M_{t\bar{t}},\hat{k})$, where the spin and color degrees of freedom of the initial state have already been averaged, $M_{t\bar{t}}$ is the invariant mass of the $t\bar{t}$ pair, and $\hat{k}$ is the flight direction of the top quark. For the characterization of the $R$-matrix, an orthonormal basis needs to be fixed in order to compute the corresponding spin polarizations and correlations. The most common choice is the helicity basis~\cite{Baumgart2013}, defined in the c.m. frame as $\{\hat{k},\hat{n},\hat{r}\}$, with $\hat{r}=(\hat{p}-\cos\Theta\hat{k})/\sin\Theta$, $\hat{n}=\hat{r}\times\hat{k}$, where $\hat{p}$ is a unitary vector along the direction of the initial state and $\Theta$ is the production angle of the top quark with respect to the beam, $\cos\Theta=\hat{k}\cdot \hat{p}$. We note here that, due to momentum conservation, the c.m. frame of the initial state is the same as that of the $t\bar{t}$ pair. 

\begin{figure}[t]\includegraphics[width=0.8\columnwidth]{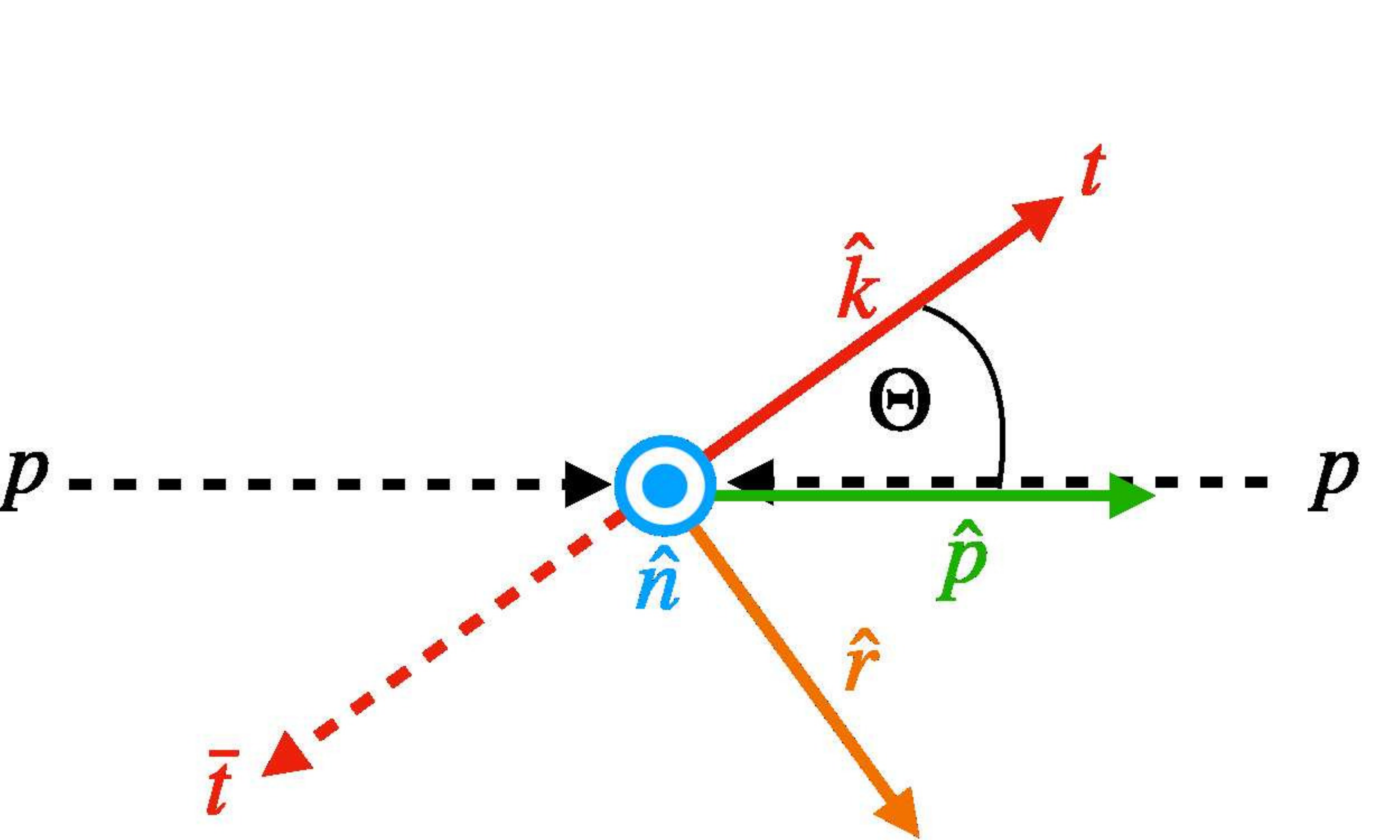}
\centering
     \caption{Orthonormal helicity basis, defined in the c.m. frame. $\hat{k}$ is the direction of the top and $\hat{p}$ is the direction of the initial beam. The vector $\hat{n}$ is perpendicular to the $\{\hat{k},\hat{p}\}$ plane and $\hat{r}=\hat{k}\times\hat{n}$ is the vector orthogonal to $\hat{k}$ within the $\{\hat{k},\hat{p}\}$ plane.} 
     \label{fig:HelicityBasis}
\end{figure}

The advantage of the helicity basis is that, although the $t\bar{t}$ relativistic spins are only well-defined in their respective rest frames, those are equivalent to the c.m. frame via a Lorentz transformation along the top direction, which does not change the orientation of the helicity basis. A schematic representation of this basis is provided in Fig.~\ref{fig:HelicityBasis}. 

The production spin density matrix $R^{I}(M_{t\bar{t}},\hat{k})$ in the helicity basis is only a function of $\beta$ (or, equivalently, of $M_{t\bar{t}}$) and $\cos\Theta$. Specifically, in the Standard Model, the correlation matrix $\tilde{C}^I_{ij}$ is symmetric and $\tilde{B}_i^{I,+}=\tilde{B}_i^{I,-}$. Furthermore, at LO, the $t\bar{t}$ pair is unpolarized, $\tilde{B}_i^{I,\pm}=0$, and, in QCD production, their spins along the $n$-axis are uncorrelated with respect to the remaining directions, $\tilde{C}^I_{nr}=\tilde{C}^I_{nk}=0$~\cite{Sirunyan2019}. Therefore, $R^{I}(M_{t\bar{t}},\hat{k})$ is characterized at LO QCD by only $5$ parameters in the helicity basis: $\tilde{A}^I,\tilde{C}^I_{kk},\tilde{C}^I_{nn},\tilde{C}^I_{rr},\tilde{C}^I_{kr}$. The values of those coefficients can be computed analytically and are well known~\cite{Bernreuther1998,Uwer2005,Baumgart2013}. Here, we fix the normalization of $R^{I}(M_{t\bar{t}},\hat{k})$ such that the c.m. differential cross-section for $t\bar{t}$ production from an initial state $I$ is~\cite{Bernreuther1998}
\begin{equation}\label{eq:CrossSectionDifferential}
\frac{\mathrm{d}\sigma^I}{\mathrm{d}\Omega}=\frac{\alpha^2_s\beta}{M^2_{t\bar{t}}}\tilde{A}^I(M_{t\bar{t}},\hat{k})
\end{equation}
with $\alpha_s\simeq 0.118$ the strong coupling constant characterizing the strength of QCD interactions. 

\begin{figure*}[tb!]
\begin{tabular}{@{}cc@{}}
    \includegraphics[width=1.1\columnwidth]{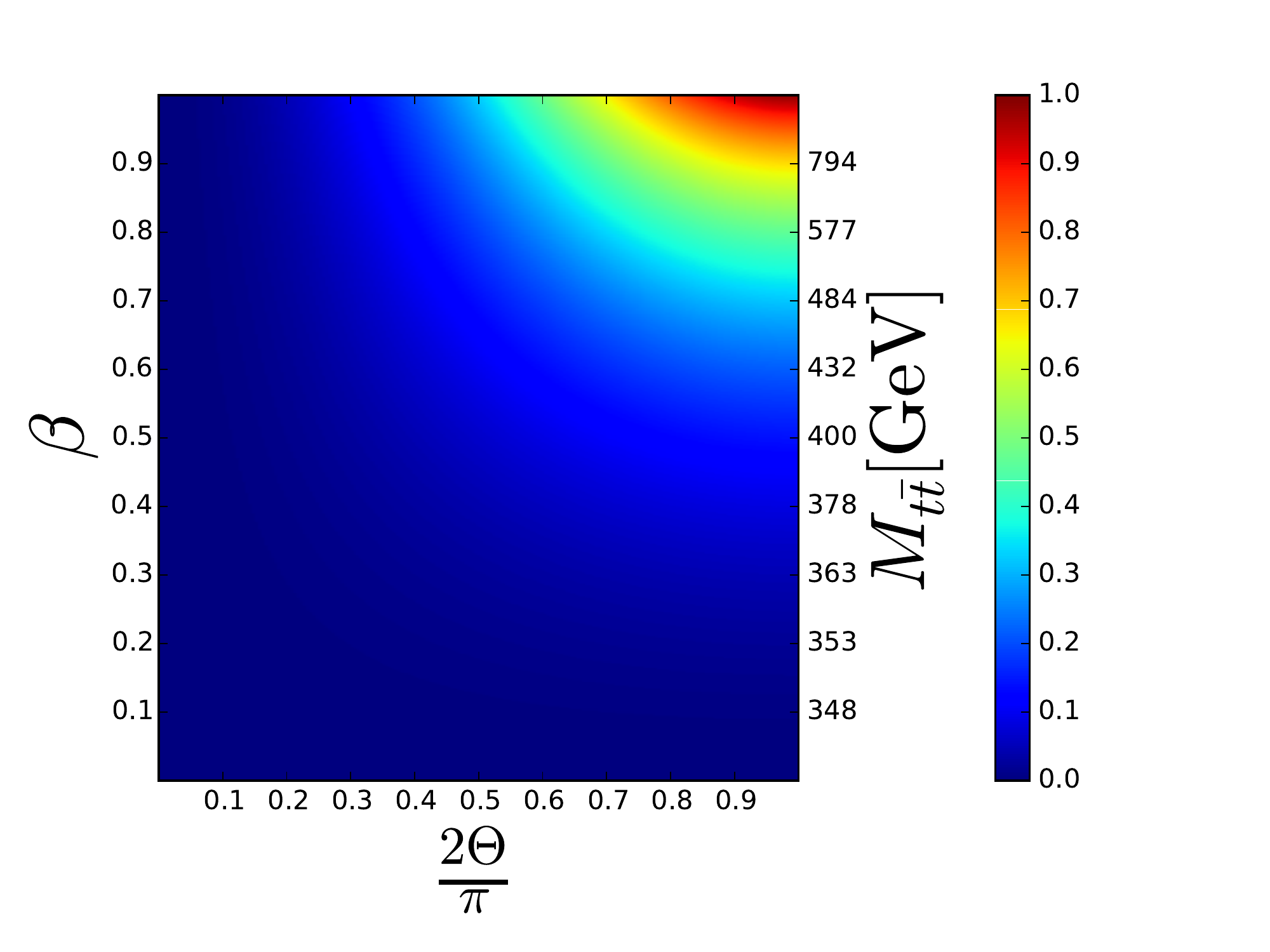} &
    \includegraphics[width=1.1\columnwidth]{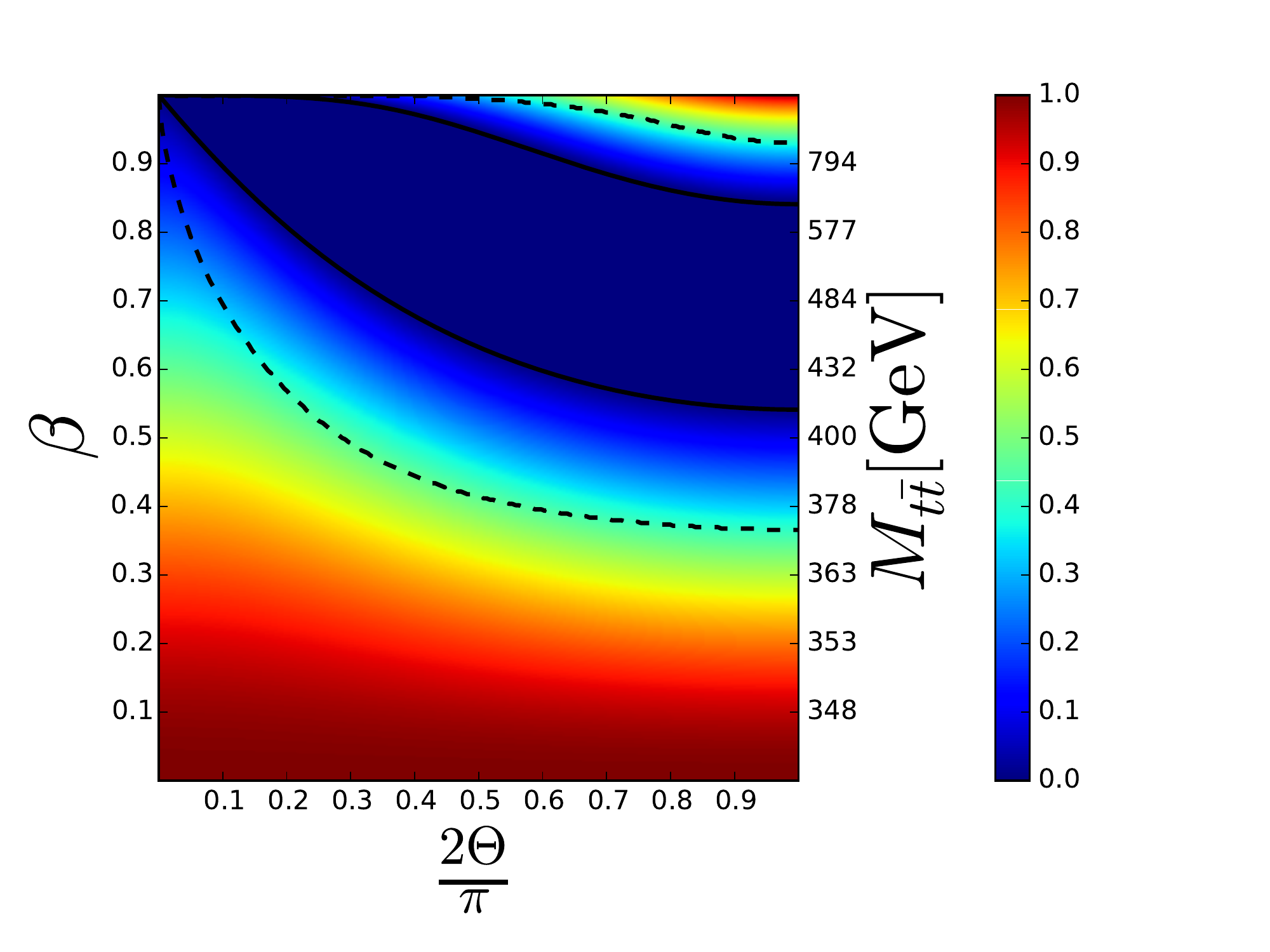} 
\end{tabular}
\caption{Concurrence of the spin density matrix $\rho^I(\beta,\hat{k})$  resulting from an initial state $I=q\bar{q},gg$ as a function of the top velocity $\beta$ and the production angle $\Theta$ in the $t\bar{t}$ c.m. frame. All plots are symmetric under the transformation $\Theta\rightarrow \pi-\Theta$. Left: $q\bar{q}\rightarrow t\bar{t}$. Right: $gg\rightarrow t\bar{t}$. Solid black lines represent the critical boundaries between separability and entanglement $\beta^\textrm{PH}_{c1,c2}(\Theta)$, while dashed black lines represent the critical boundaries for the violation of the CHSH inequality, $\beta^\textrm{CH}_{c1,c2}(\Theta)$.}
\label{fig:ConcurrenceFundamental}
\end{figure*}

The actual spin density matrices $\rho^I(M_{t\bar{t}},\hat{k})$ are computed from the normalization of $R^I(M_{t\bar{t}},\hat{k})$. As a result of the above considerations, at LO $\rho^I(M_{t\bar{t}},\hat{k})$ is unpolarized, its correlation matrix is symmetric, and it is already diagonal in the $\hat{n}$ direction. These last two properties imply that the correlation matrix can be diagonalized by an appropriated rotation in the $\{\hat{k},\hat{r}\}$ plane, with eigenvalues $\{C^I_{+},C^I_{nn},C^I_{-}\}$,   
\begin{equation}
    C^I_{\pm}=\frac{C^I_{kk}+C^I_{rr}}{2}\pm\sqrt{\left(\frac{C^I_{kk}-C^I_{rr}}{2}\right)^2+C^{I2}_{kr}}
\end{equation}
and orthonormal eigenvectors $\{\hat{u}_{+},\hat{n},\hat{u}_{-}\}$. We will refer to this orthonormal basis as the \textit{diagonal} basis.

Regarding entanglement, since $C^I_{nn}<0$ in all phase space and for both initial states [see Eqs. (\ref{eq:LOSpinCorrelationsqq}), (\ref{eq:LOSpinCorrelationsgg})], we find that the Peres-Horodecki criterion is equivalent to $\Delta^I>0$ (see Appendix~\ref{app:criteria}), where  
\begin{align}\label{eq:Deltatt}
\Delta^I&\equiv\frac{-C^I_{nn}+|C^I_{+}+C^I_{-}|-1}{2}\\ \nonumber
&=\frac{-C^I_{nn}+|C^I_{kk}+C^I_{rr}|-1}{2}
\end{align}
We note that this definition differs from that of Ref.~\cite{Afik2021} by a factor $2$, so the concurrence is directly equal to $\Delta^{I}$ for entangled states:
\begin{equation}\label{eq:ConcurrenceHelicity}
    \mathcal{C}[\rho^I]=\max(\Delta^I,0) 
\end{equation}
Previous approaches in high-energy physics based on the entanglement entropy~\cite{Kharzeev2017,Martens:2017cvj,Tu2020} are not useful here since they are only valid for pure states. We recall that the entanglement of the quantum state $\rho^I(M_{t\bar{t}},\hat{k})$ is Lorentz invariant because it has well-defined momentum~\cite{Gingrich2002,Peres2004}. 

Regarding the violation of the CHSH inequality, from  Eq.~(\ref{eq:CHSH}) it is immediately seen in the diagonal basis to be equivalent to 
\begin{align}
    \mu^I&\equiv \textrm{tr}\left[\mathbf{C}^{\textrm{T}}\mathbf{C}\right]-\left(C_{\textrm{min}}^{I}\right)^2-1>0\\
    \nonumber \left(C_{\textrm{min}}^{I}\right)^2&\equiv \min \left\{\left(C_{+}^{I}\right)^2,\left(C_{nn}^{I}\right)^2,\left(C_{-}^{I}\right)^2\right\}
\end{align}

\subsection{$q\bar{q}$ processes}

For $q\bar{q}$ processes, the coefficients of the $R$-matrix in the helicity basis are
\begin{align}\label{eq:LOSpinCorrelationsqq}
\tilde{A}^{q\bar{q}}&=F_q(2-\beta^2\sin^2\Theta)\\
\nonumber \tilde{C}^{q\bar{q}}_{rr}&=F_q(2-\beta^2)\sin^2\Theta\\
\nonumber \tilde{C}^{q\bar{q}}_{nn}&=-F_q\beta^2\sin^2\Theta\\
\nonumber \tilde{C}^{q\bar{q}}_{kk}&=F_q\left[2-(2-\beta^2)\sin^2\Theta\right]\\
\nonumber \tilde{C}^{q\bar{q}}_{rk}&=\tilde{C}^{q\bar{q}}_{kr}=F_q\sqrt{1-\beta^2}\sin2\Theta\\
\nonumber F_q&=\frac{1}{18}
\end{align}

The resulting $t\bar{t}$ quantum state $\rho^{q\bar{q}}(\beta,\hat{k})$ is entangled in the bulk of phase space as \footnote{We note that the result for $\Delta^{q\bar{q}}$ in Ref.~\cite{Afik2021} contained a typo, and a factor $2$ was missing.}
\begin{align}\label{}
\Delta^{q\bar{q}}=-C_{nn}^{q\bar{q}}=\frac{\beta^2\sin^2\Theta}{2-\beta^2\sin^2\Theta}\geq 0
\end{align}

The above inequality is saturated only at threshold ($\beta=0$) or for forward production ($\Theta=0$). In both limits, the $t\bar{t}$ spins are aligned along the beam axis in a maximally correlated but separable mixed state:
\begin{align}\label{eq:qqseparable}
C_{ij}^{q\bar{q}}(0,\hat{k})&=C_{ij}^{q\bar{q}}(\beta,\hat{p})=\hat{p}_i\hat{p}_j\\
\nonumber
\rho^{q\bar{q}}(0,\hat{k})&=\rho^{q\bar{q}}(\beta,\hat{p})\\
\nonumber &=\frac{\ket{\uparrow_{\hat{p}}\uparrow_{\hat{p}}}\bra{\uparrow_{\hat{p}}\uparrow_{\hat{p}}}+\ket{\downarrow_{\hat{p}}\downarrow_{\hat{p}}}\bra{\downarrow_{\hat{p}}\downarrow_{\hat{p}}}}{2}
\end{align}
where $\ket{\uparrow_{\hat{p}}},\ket{\downarrow_{\hat{p}}}$ are the spin eigenstates along the direction $\hat{p}$. This spin alignment at threshold is a consequence of spin conservation, since the $t\bar{t}$ pair is produced from the initial $q\bar{q}$ state via gluon exchange (see Fig.~\ref{fig:Feynman}), which is a massless spin-1 boson. 

Hence, for low production angles or close to threshold, the degree of entanglement is expected to be small, as can seen in left Fig.~\ref{fig:ConcurrenceFundamental}, where the concurrence of $\rho^{q\bar{q}}(\beta,\hat{k})$ is represented.

In the opposite limit of high transverse momentum $p_T$ (i.e., high energies and production angles close to $\Theta=\pi/2$), the $t\bar{t}$ pair is in a spin-triplet pure state,
\begin{align}\label{eq:triplet}
C_{ij}^{q\bar{q}}(1,\hat{n}\times\hat{p})&=\delta_{ij}-2\hat{n}_i\hat{n}_j\\
\nonumber \rho^{q\bar{q}}(1,\hat{n}\times\hat{p})&=\ket{\Psi_{\infty}}\bra{\Psi_{\infty}}\\
\nonumber \ket{\Psi_{\infty}}&=\frac{\ket{\uparrow_{\hat{n}}\downarrow_{\hat{n}}}+\ket{\downarrow_{\hat{n}}\uparrow_{\hat{n}}}}{\sqrt{2}}
\end{align}
This state is maximally entangled, $\mathcal{C}[\rho]=1$, as seen in upper right corner of left Fig.~\ref{fig:ConcurrenceFundamental}.

With respect to the CHSH inequality, it is also violated by $q\bar{q}$ production processes in the bulk of phase space. This is seen in the diagonal basis, where
\begin{equation}
    C^{q\bar{q}}_{+}=1,~C^{q\bar{q}}_{-}=-C^{q\bar{q}}_{nn}=\Delta^{q\bar{q}}
\end{equation}
This basis is often called the \textit{off-diagonal} basis~\cite{Mahlon1996} in the literature of $q\bar{q}$ processes, and is commonly used because the $t\bar{t}$ spins along the $\hat{u}_{+}$ direction are perfectly correlated. Because of this,
\begin{equation}
    \mu^{q\bar{q}}(\beta,\Theta)=\left[\Delta^{q\bar{q}}(\beta,\Theta)\right]^2\geq 0
\end{equation}
Therefore, remarkably, entanglement and CHSH violation are equivalent conditions for $q\bar{q}$ processes.

\subsection{$gg$ processes}

For $gg$ processes, the coefficients of the $R$-matrix in the helicity basis are
\begin{align}\label{eq:LOSpinCorrelationsgg}
\nonumber \tilde{A}^{gg}&=F_g\left[1+2\beta^2\sin^2\Theta-\beta^4(1+\sin^4\Theta)\right]\\
\nonumber \tilde{C}^{gg}_{rr}&=-F_g\left[1-\beta^2(2-\beta^2)(1+\sin^4\Theta)\right]\\
\nonumber \tilde{C}^{gg}_{nn}&=-F_g\left[1-2\beta^2+\beta^4(1+\sin^4\Theta)\right]\\
\nonumber \tilde{C}^{gg}_{kk}&=-F_g\left[1-\beta^2\frac{\sin^2 2\Theta}{2}-\beta^4(1+\sin^4\Theta)\right]\\
\nonumber \tilde{C}^{gg}_{rk}&=\tilde{C}^{gg}_{kr}=F_g\sqrt{1-\beta^2}\beta^2\sin2\Theta\sin^2\Theta\\
F_g&=\frac{7+9\beta^2\cos^2\Theta}{192(1-\beta^2\cos^2\Theta)^2}
\end{align}
The resulting $t\bar{t}$ quantum state $\rho^{gg}(\beta,\hat{k})$ is entangled \textit{iff}  $\Delta^{gg}(\beta,\Theta)>0$, where
\begin{align}
    \nonumber \Delta^{gg}&=\frac{\beta^4(1+\sin^4\Theta)-\beta^2(1+\sin^2\Theta)}{1+2\beta^2\sin^2\Theta-\beta^4(1+\sin^4\Theta)}\\
    &+\frac{|1-\beta^2(1+\sin^2\Theta)|}{1+2\beta^2\sin^2\Theta-\beta^4(1+\sin^4\Theta)}
\end{align}
The second line implies that $\rho^{gg}(\beta,\hat{k})$ is not entangled for $\beta^2(1+\sin^2\Theta)=1$, so there are finite regions of separability in phase space, in contrast to the $q\bar{q}$ case. In particular, two regions of entanglement can be distinguished: one close to threshold, and another one for high $p_T$, delimited by the lower and upper critical boundaries $\beta^\textrm{PH}_{c1}(\Theta),\beta^\textrm{PH}_{c2}(\Theta)$: 
\begin{align}\label{eq:BetaCriticals}
\nonumber \beta^\textrm{PH}_{c1}(\Theta)&=\sqrt{\frac{1+\sin^2\Theta-\sqrt{2}\sin\Theta}{1+\sin^4\Theta}}\\
\beta^\textrm{PH}_{c2}(\Theta)&=\frac{1}{(1+\sin^4\Theta)^{\frac{1}{4}}}
\end{align}
These features can be observed in right Fig.~\ref{fig:ConcurrenceFundamental}, where the concurrence of $\rho^{gg}(\beta,\hat{k})$ is represented, with the black solid lines signaling the critical boundaries $\beta^\textrm{PH}_{c1,c2}(\Theta)$ between entanglement and separability.

The strong entanglement signature close to threshold results from the fact that the spin polarizations of the $gg$ initial state are allowed to align in different directions. Due to angular momentum conservation, these features produce at threshold a $t\bar{t}$ pair in a spin singlet
\begin{align}\label{eq:singletgg}
C_{ij}^{gg}(0,\hat{k})&=-\delta_{ij}\\
\nonumber \rho^{gg}(0,\hat{k})&=\ket{\Psi_0}\bra{\Psi_0}\\
\nonumber \ket{\Psi_0}&=\frac{\ket{\uparrow_{\hat{n}}\downarrow_{\hat{n}}}-\ket{\downarrow_{\hat{n}}\uparrow_{\hat{n}}}}{\sqrt{2}}
\end{align}
which is maximally entangled. 

In the opposite limit of high $p_T$, gluon fusion produces the same maximally entangled triplet state as $q\bar{q}$ processes, Eq.~(\ref{eq:triplet}). The reason behind this similarity is that for high energies the orbital angular momentum contribution dominates over the spin contribution. 

With respect to the violation of the CHSH inequality, the eigenvalues of the correlation matrix are 
\begin{widetext}
\begin{equation}
\resizebox{\hsize}{!}{$C^{gg}_{\pm}=
    \frac{-1+\beta^2(1+\sin^2\Theta)\pm \sqrt{\beta^4(1-2\sin^2\Theta+5\sin^4\Theta)-2\beta^6(1-\sin^2\Theta+3\sin^4\Theta+\sin^6\Theta)+\beta^8\left[1+\sin^4\Theta\right]^2}}{1+2\beta^2\sin^2\Theta-\beta^4(1+\sin^4\Theta)}$}
\end{equation}
\end{widetext}
As a result, an explicit analytical computation of the critical boundaries $\beta^\textrm{CH}_{c1,c2}(\Theta)$ of the regions where the CHSH inequality is violated becomes much more cumbersome; even in the relatively simple case of $\Theta=\pi/2$, the critical values $\beta^\textrm{CH}_{c1,c2}(\pi/2)$ are obtained from the zeros of fourth order polynomials in $\beta^2$, finding $\beta^\textrm{CH}_{c1}(\pi/2)\approx 0.367<\beta^\textrm{PH}_{c1}(\pi/2)=2^{-1/4}\approx 0.541$, and $\beta^\textrm{CH}_{c2}(\pi/2)\approx 0.931>\beta^\textrm{PH}_{c2}(\pi/2)=\sqrt{1-1/\sqrt{2}}\approx 0.841$. The numerically computed boundaries $\beta^\textrm{CH}_{c1,c2}(\Theta)$ are represented by black dashed lines in right Fig.~\ref{fig:ConcurrenceFundamental}. As expected, the regions where CHSH inequality is violated are within entangled regions of phase space, $\beta^\textrm{CH}_{c1}(\Theta)\leq\beta^\textrm{PH}_{c1}(\Theta)\leq\beta^\textrm{PH}_{c2}(\Theta)\leq\beta^\textrm{CH}_{c2}(\Theta)$.

An important conclusion is that the entanglement structure of $t\bar{t}$ production can be mostly understood from basic laws of angular momentum conservation between the initial and final state, without the need to invoke technical details of the specific form of QCD interactions.

\section{Production of $t\bar{t}$ in realistic QCD processes}\label{sec:EntanglementRealistic}

Because of color confinement, quarks and gluons are not free particles and cannot be found isolated in nature. Instead, they are forming hadrons, which are bound states of quarks through QCD interactions. As a result, we can understand hadrons as a sea of quarks and gluons, which within this context are indistinctively denoted as partons~\cite{Feynman:1969wa,Bjorken1969}. Quantitatively, the composition of a hadron is modeled by its so-called parton distribution function (PDF), which determines the participation of each parton in a particular QCD process at a certain energy scale. Due to its non-perturbative character, PDF distributions are typically computed by fitting experimental data (see Appendix~\ref{app:PDF} for more technical details about PDF). Here, we focus on $t\bar{t}$ production from two main types of hadron processes which are of high relevance for experiments: proton-proton collisions, as in the LHC, and proton-antiproton collisions, as in the Tevatron.

The above considerations modify the results of the previous section as following. For given energy and top direction in the c.m. frame, the $R$-matrix describing the $t\bar{t}$ pair resulting from a hadronic process is computed in terms of each partonic matrix $R^{I}(M_{t\bar{t}},\hat{k})$ as
\begin{equation}\label{eq:Rtotal}
R(M_{t\bar{t}},\hat{k},\sqrt{s})=\sum_{I=q\bar{q},gg} L_{I}(M_{t\bar{t}},\sqrt{s})R^{I}(M_{t\bar{t}},\hat{k})
\end{equation}
The function $L_{I}(M_{t\bar{t}},\sqrt{s})$ is the so-called luminosity function~\cite{Bernreuther1998}, and is computed in terms of the PDF describing the colliding hadrons. The luminosity function can be regarded as the probability distribution of occurrence of each initial state $I$ in the total hadronic process for a given $M_{t\bar{t}}$, depending also on the hadron c.m. energy $\sqrt{s}$ [see Eq.~(\ref{eq:LuminosityPDF})]. We note that the c.m. frame of the colliding hadrons is not the same as the parton c.m. frame. However, at LO the direction of the parton initial state $I$ in its c.m. frame is to a very good approximation that of the initial hadron beam, so one can take safely $\hat{p}$ along the hadron beam in Fig.~\ref{fig:HelicityBasis}, which is the one whose direction is controlled in a collider~\cite{Bernreuther2004}.

With the help of the luminosities and the $R$-matrix, the differential cross-section characterizing $t\bar{t}$ production from a given hadronic process is computed as
\begin{align}\label{eq:CrossSectionDifferentialReal}
\frac{\mathrm{d}\sigma}{\mathrm{d}\Omega\mathrm{d}M_{t\bar{t}}}&=\frac{\alpha^2_s\beta}{M^2_{t\bar{t}}}\tilde{A}(M_{t\bar{t}},\hat{k},\sqrt{s})\\
\nonumber &=\sum_{I=q\bar{q},gg} L_{I}(M_{t\bar{t}},\sqrt{s})\frac{\mathrm{d}\sigma^I}{\mathrm{d}\Omega}(M_{t\bar{t}},\hat{k})
\end{align}
where the partonic differential cross-sections are those of Eq.~(\ref{eq:CrossSectionDifferential}). Thus, the differential cross-section per unit solid angle and per unit c.m. energy is just the sum of the partonic differential cross-sections from an initial state $I$, multiplied by the probability distribution $L_{I}$ of producing each initial state $I$ with energy $M_{t\bar{t}}$.

\begin{figure*}[t]
\begin{tabular}{@{}cccc@{}}
    \stackinset{l}{18pt}{t}{12pt}{\textcolor{white}{(a)}}{\includegraphics[width=0.5\columnwidth]{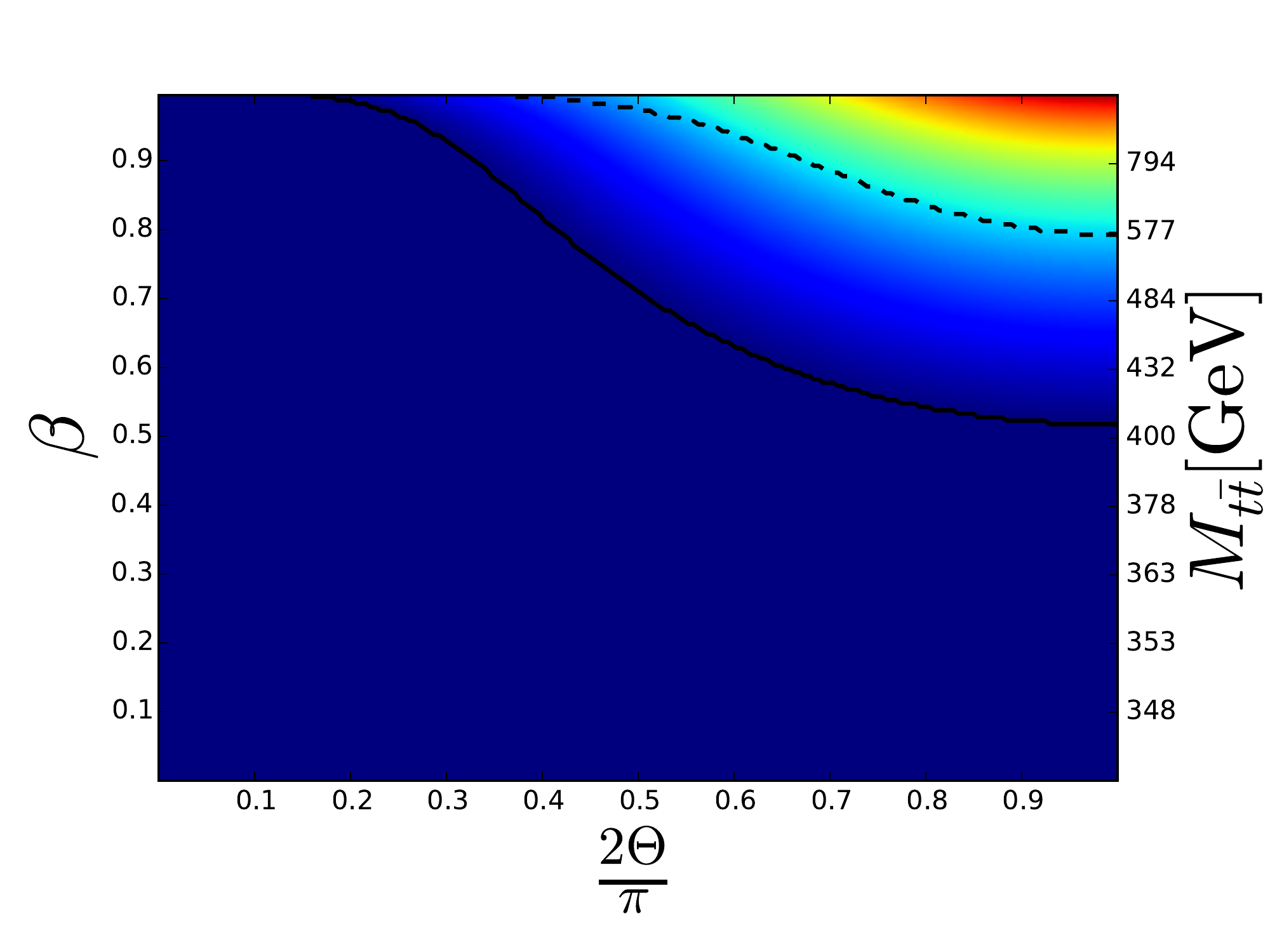}} &
    \stackinset{l}{18pt}{t}{12pt}{\textcolor{white}{(b)}}
    {\includegraphics[width=0.5\columnwidth]{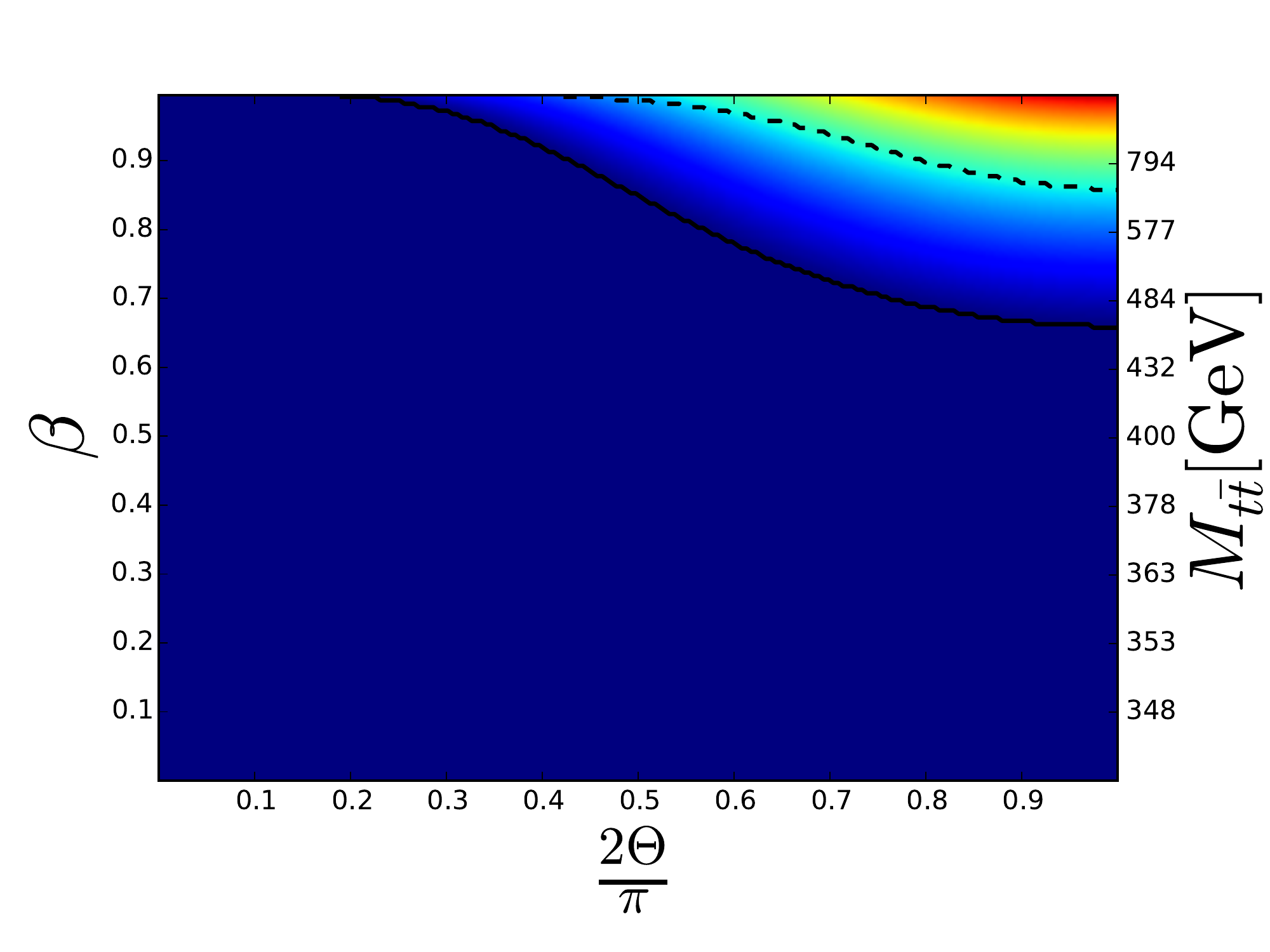}} & \stackinset{l}{18pt}{t}{12pt}{\textcolor{white}{(c)}}{\includegraphics[width=0.5\columnwidth]{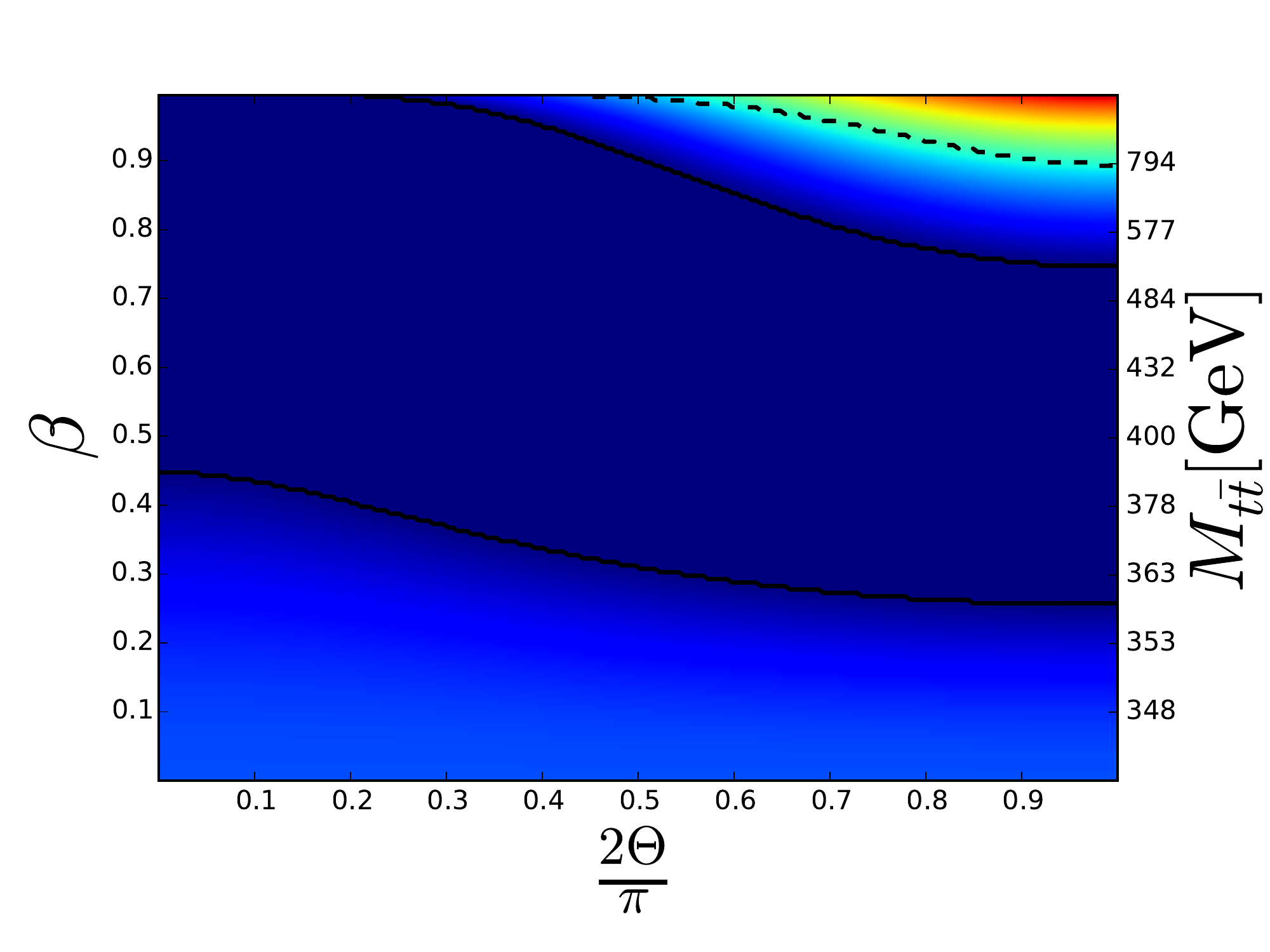}} &  
    \stackinset{l}{18pt}{t}{12pt}{\textcolor{white}{(d)}}
    {\includegraphics[width=0.5\columnwidth]{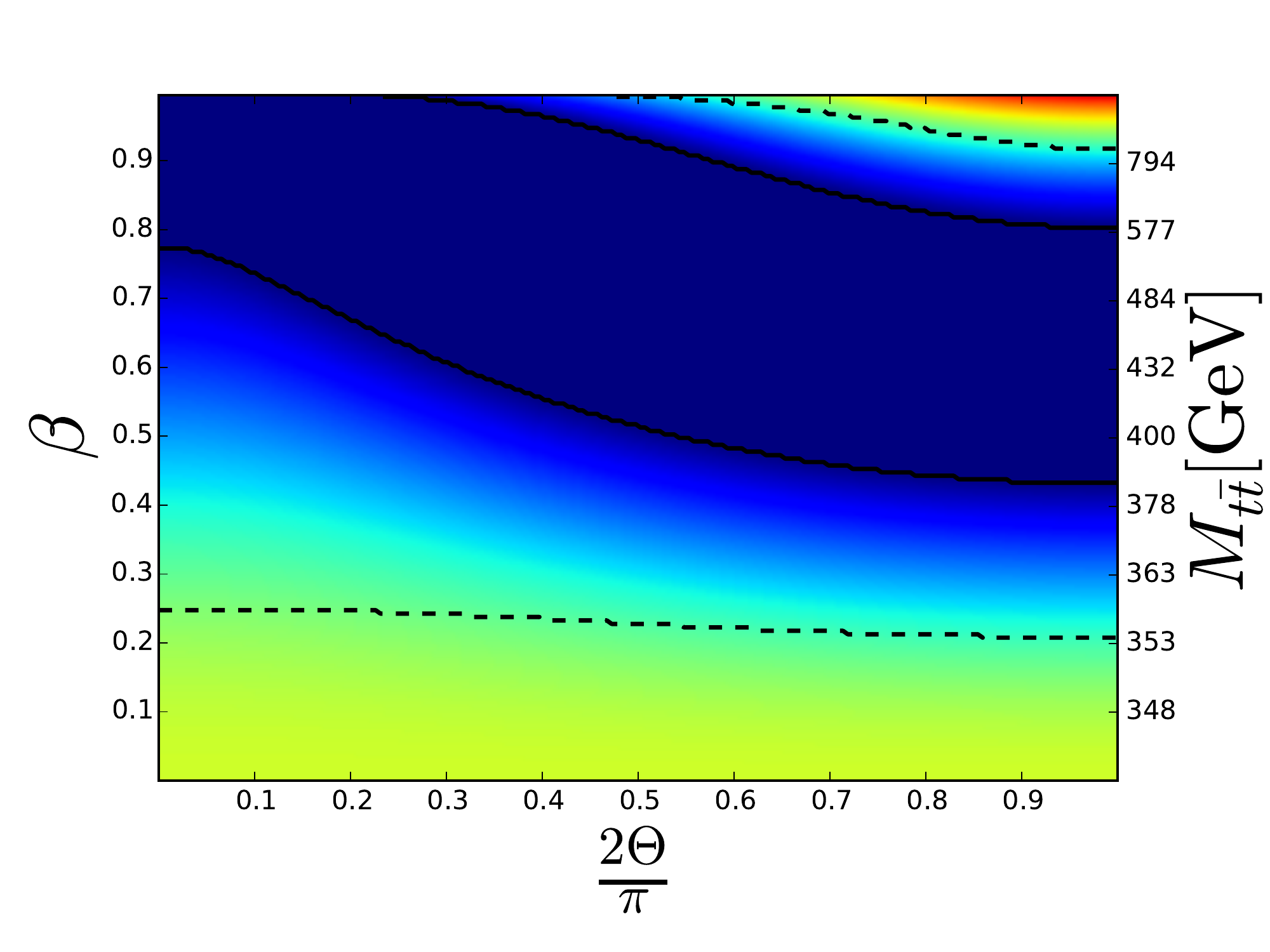}}\\
    \stackinset{l}{18pt}{t}{12pt}{\textcolor{white}{(e)}}{\includegraphics[width=0.5\columnwidth]{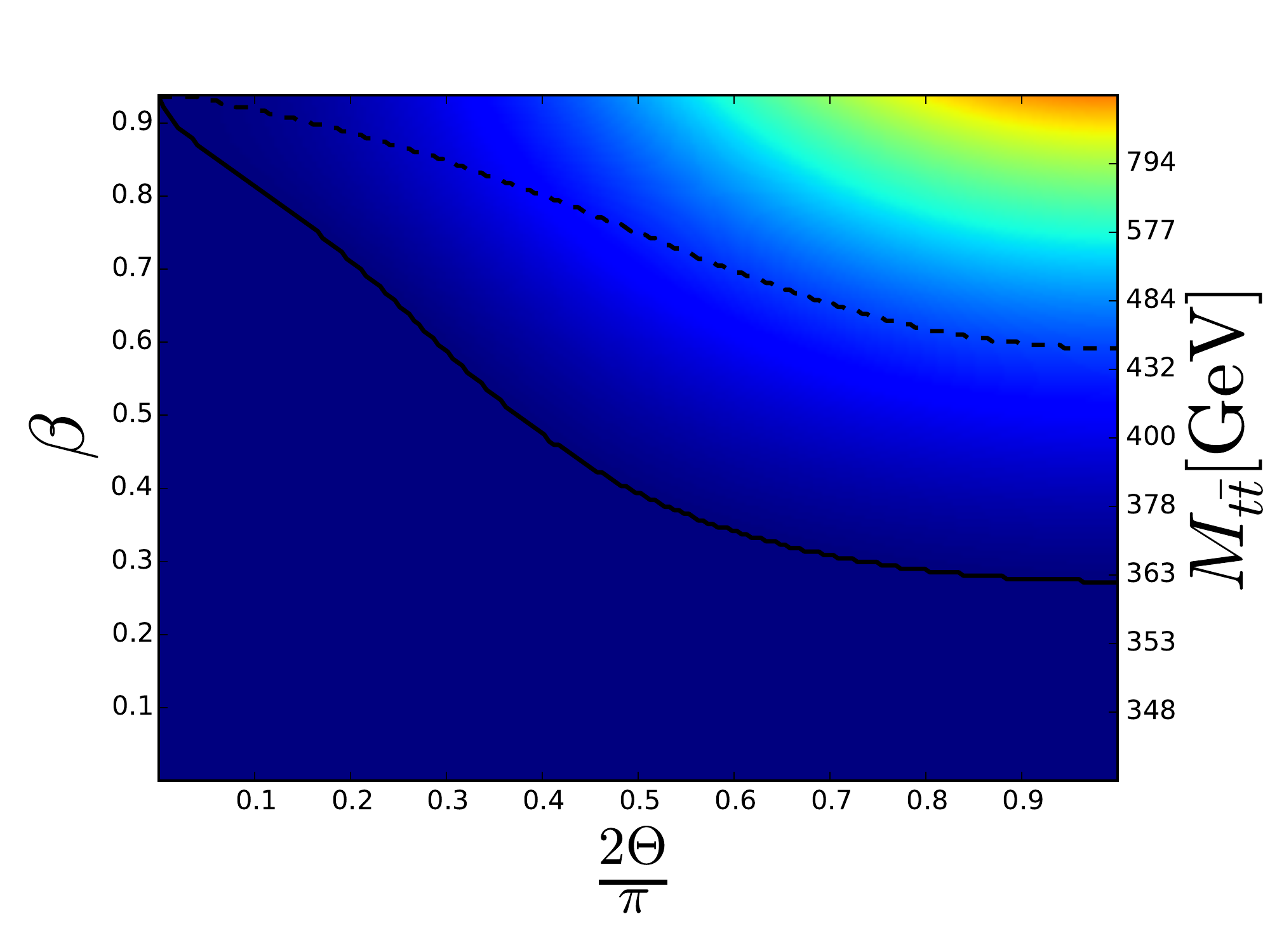}} &
    \stackinset{l}{18pt}{t}{12pt}{\textcolor{white}{(f)}}
    {\includegraphics[width=0.5\columnwidth]{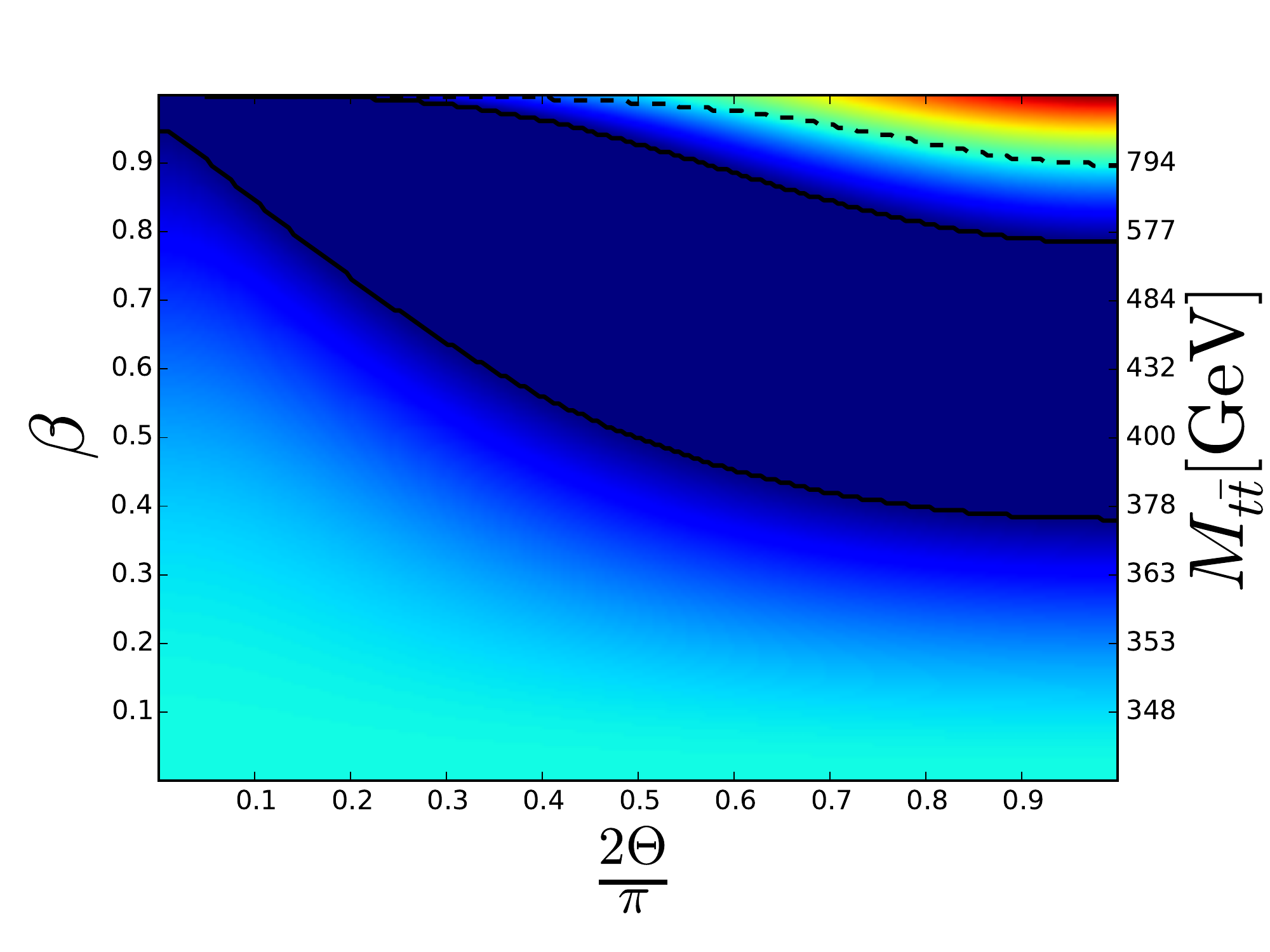}} & \stackinset{l}{18pt}{t}{12pt}{\textcolor{white}{(g)}}{\includegraphics[width=0.5\columnwidth]{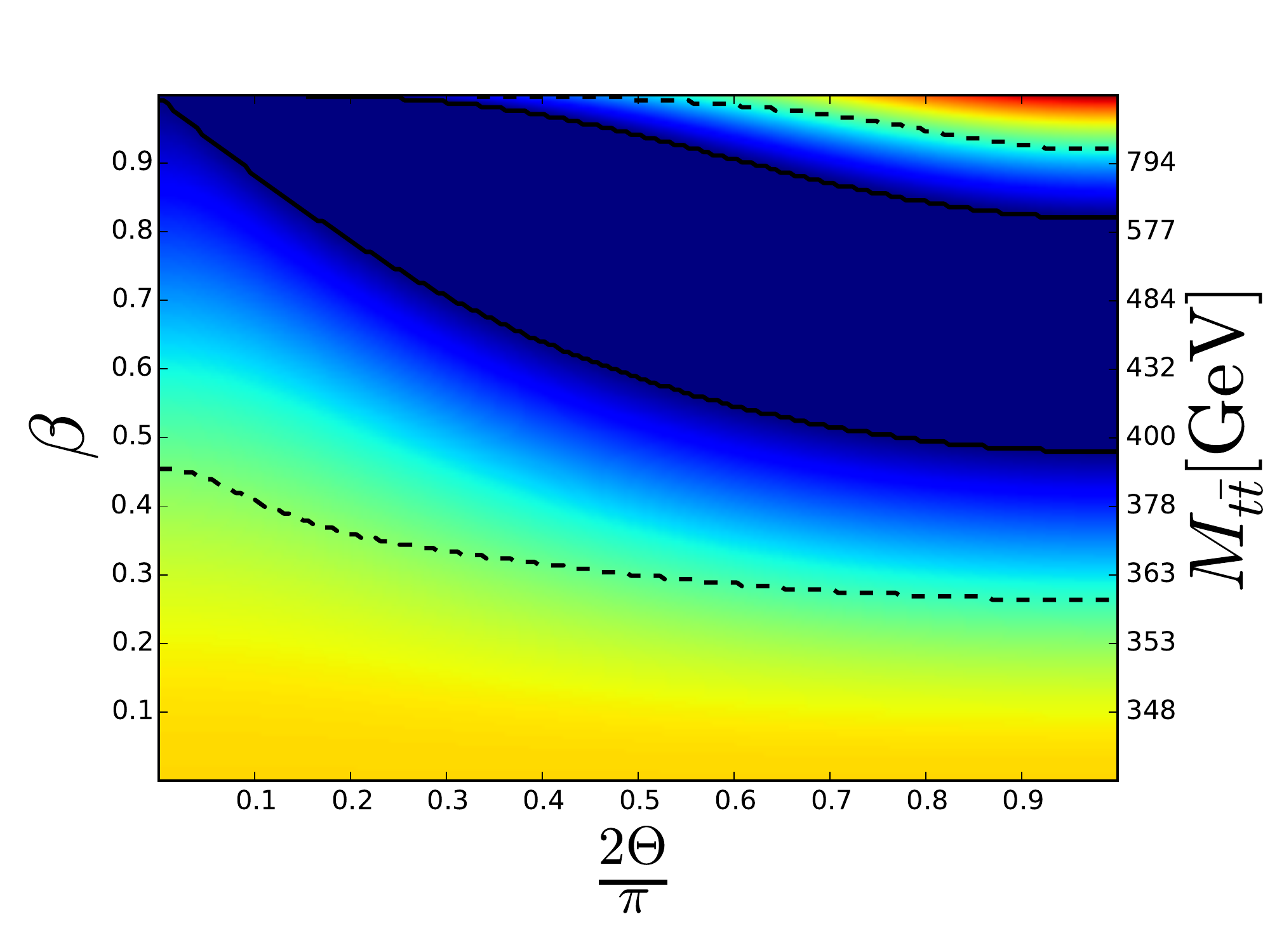}} &  
    \stackinset{l}{18pt}{t}{12pt}{\textcolor{white}{(h)}}
    {\includegraphics[width=0.5\columnwidth]{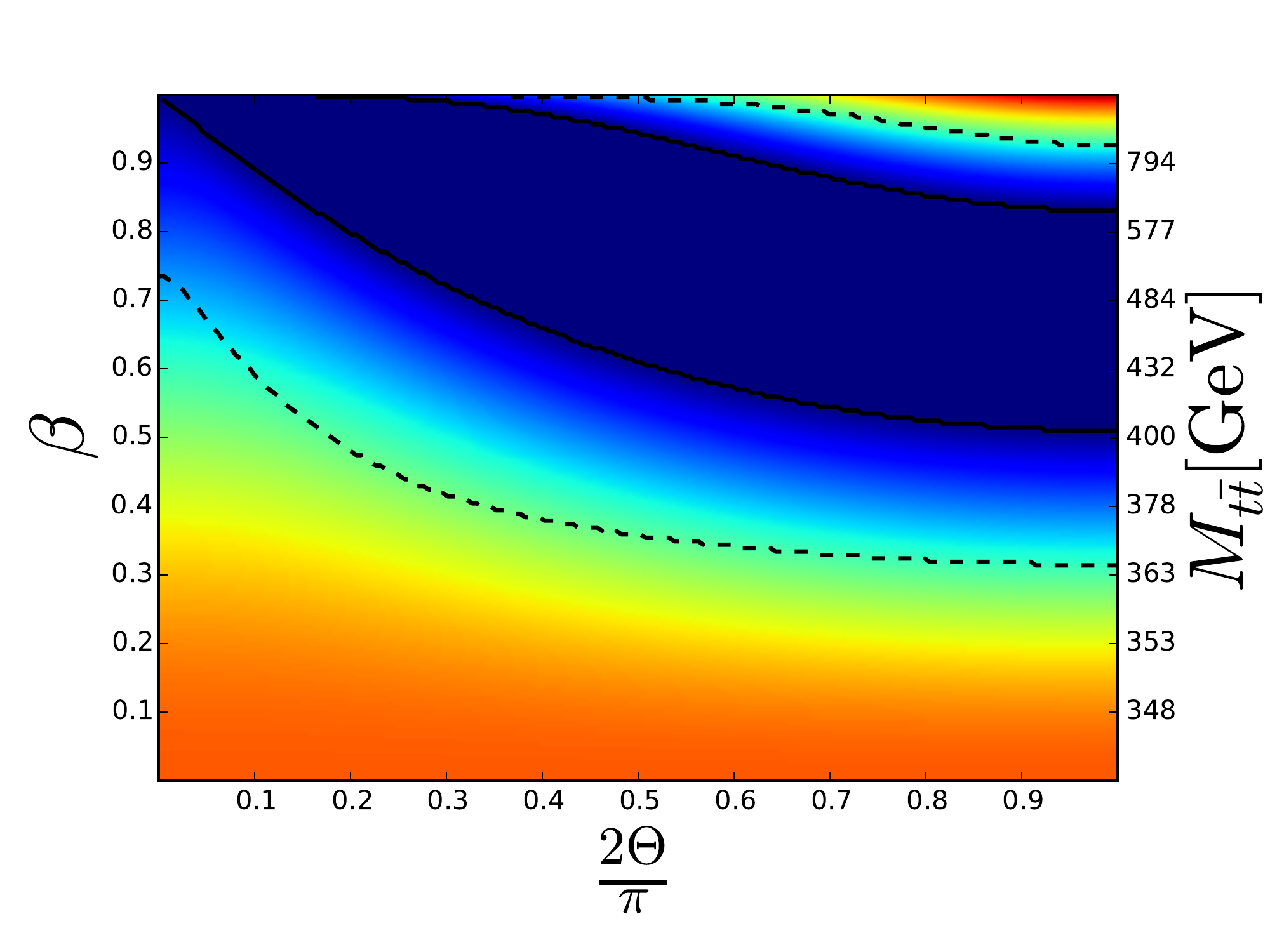}}\\
    \stackinset{l}{18pt}{t}{12pt}{\textcolor{white}{(i)}}{\includegraphics[width=0.5\columnwidth]{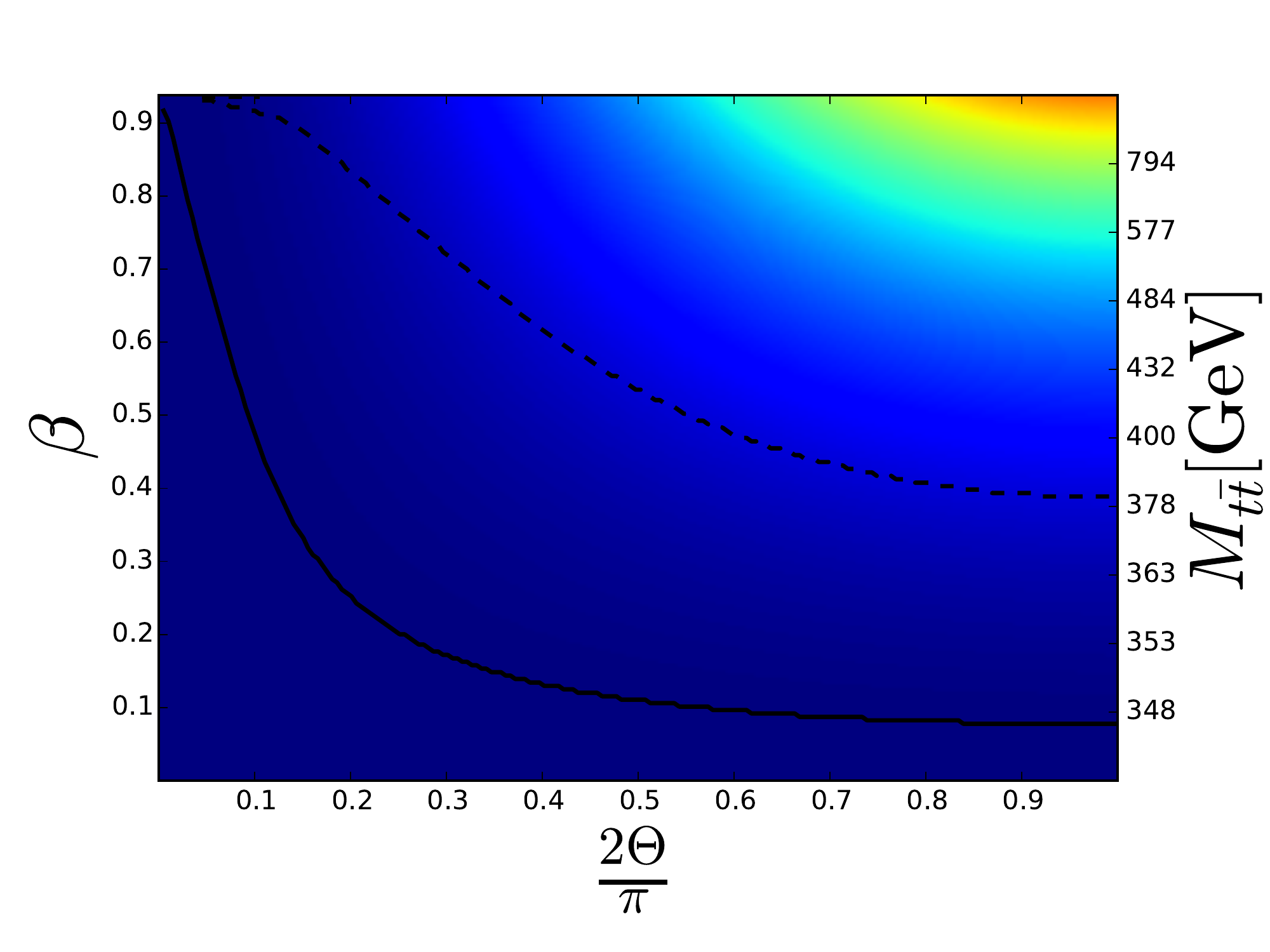}} &
    \stackinset{l}{18pt}{t}{12pt}{\textcolor{white}{(j)}}
    {\includegraphics[width=0.5\columnwidth]{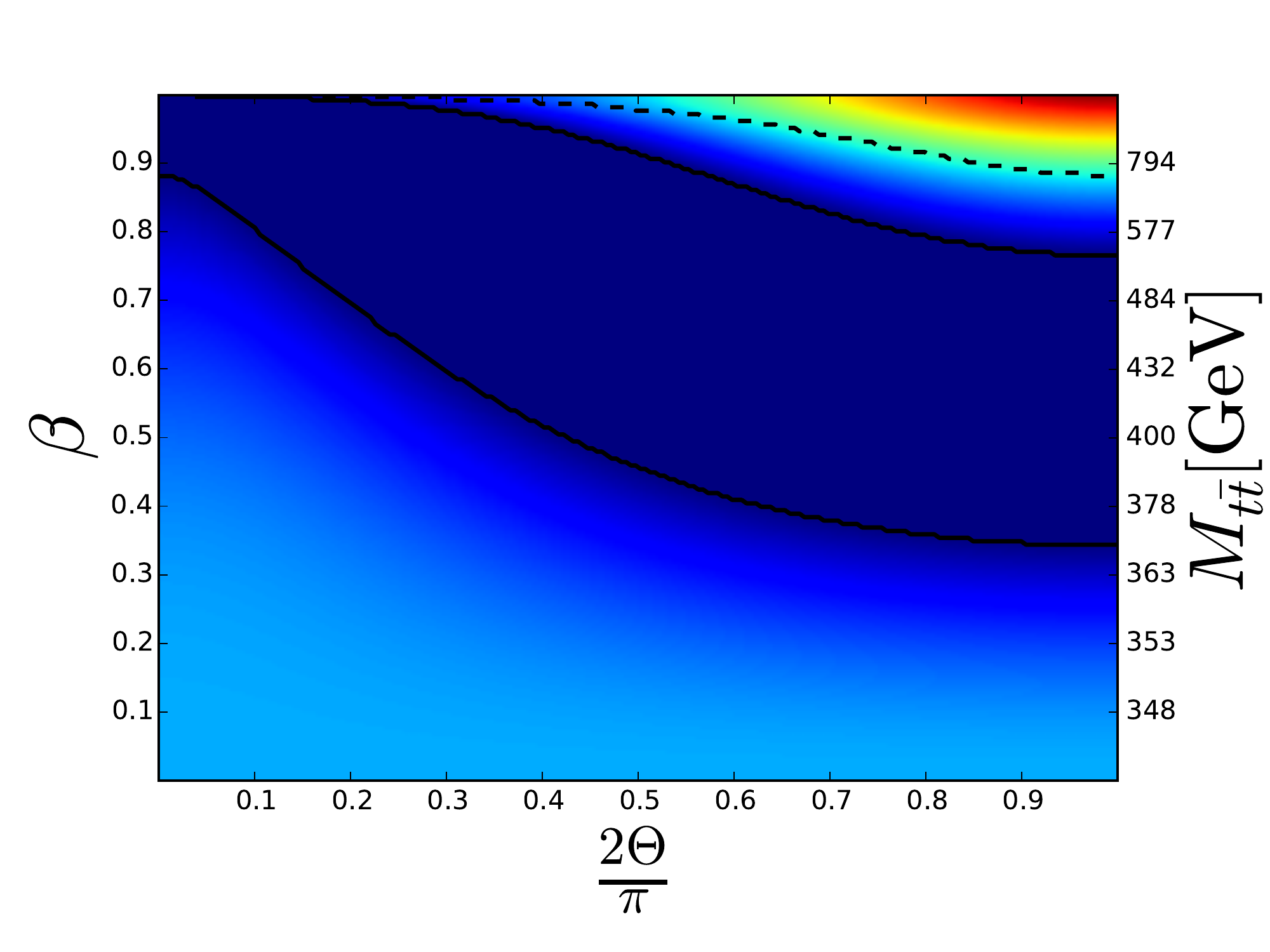}} & \stackinset{l}{18pt}{t}{12pt}{\textcolor{white}{(k)}}{\includegraphics[width=0.5\columnwidth]{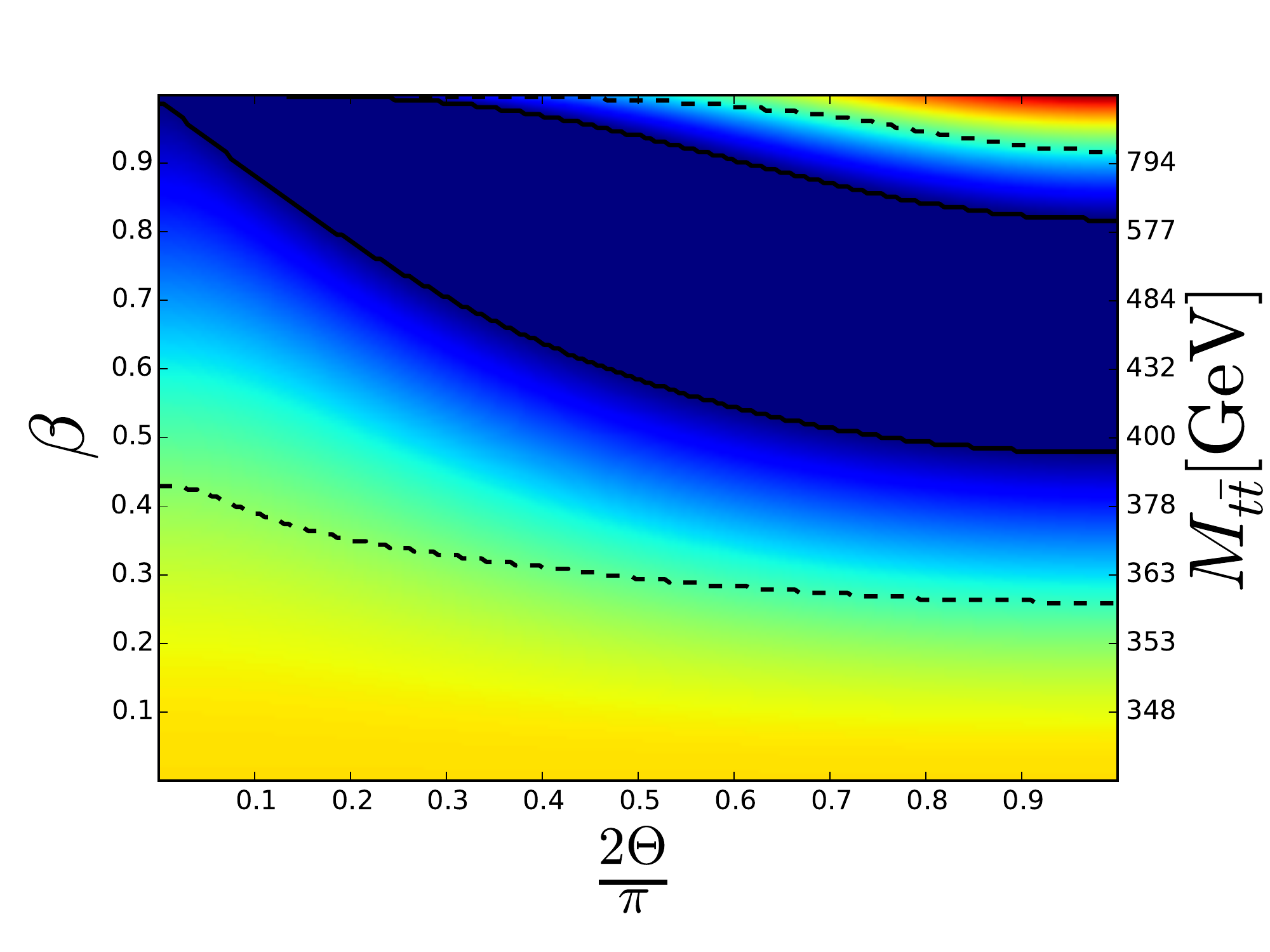}} &  
    \stackinset{l}{18pt}{t}{12pt}{\textcolor{white}{(l)}}
    {\includegraphics[width=0.5\columnwidth]{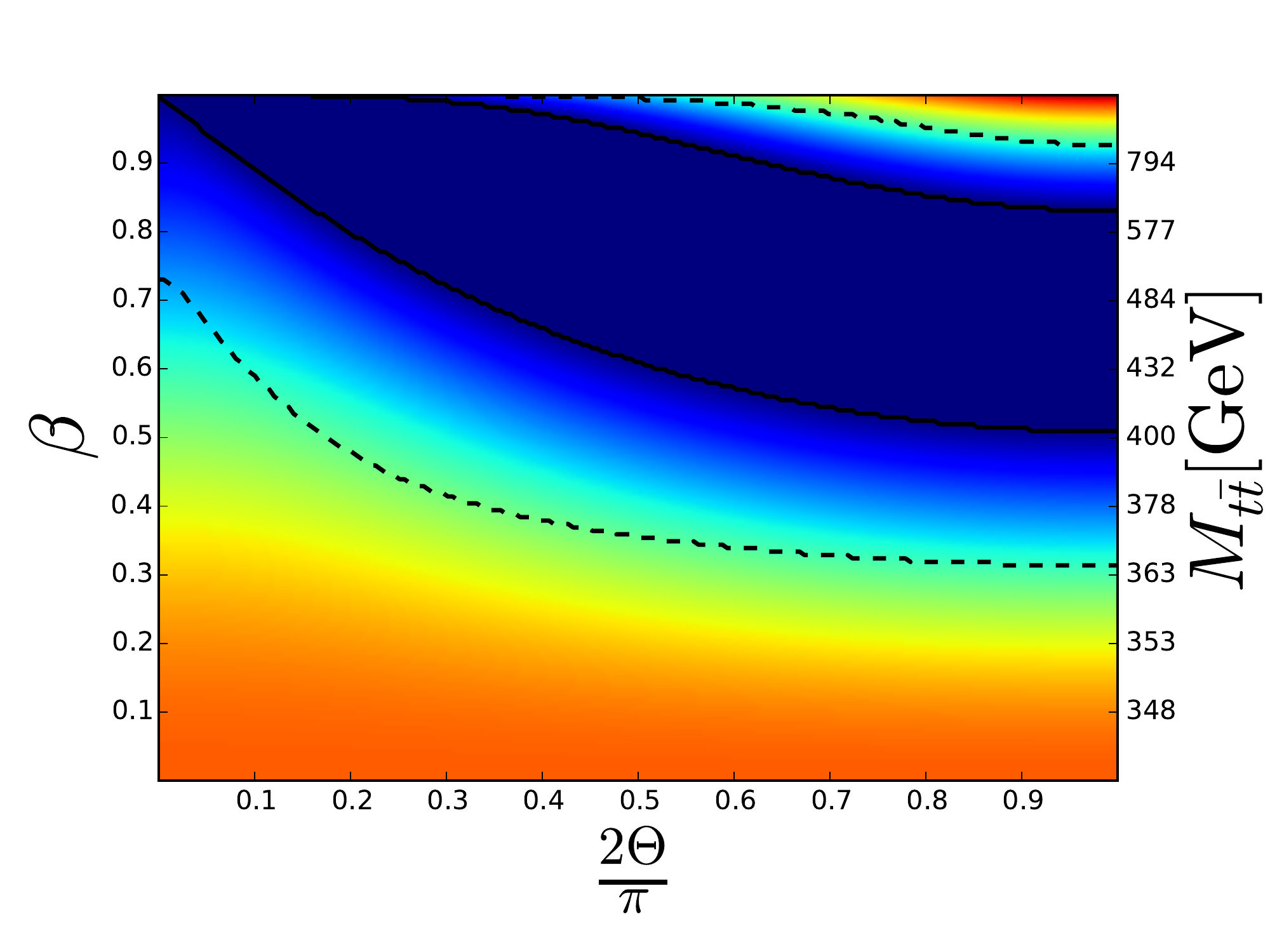}}
\end{tabular}
\caption{Same as Fig.~\ref{fig:ConcurrenceFundamental} but for the $t\bar{t}$ quantum state $\rho(M_{t\bar{t}},\hat{k},\sqrt{s})$ resulting from realistic hadronic processes. Upper row: Toy model where the probabilities are constant in whole phase space, $w_{I}(M_{t\bar{t}},\hat{k},\sqrt{s})=w_I$.  (a) $w_{gg}=0.2$. (b) $w_{gg}=0.4$.(c) $w_{gg}=0.6$.(d) $w_{gg}=0.8$. Middle row: $t\bar{t}$ production from $pp$ collisions. (e) $\sqrt{s}=1~\textrm{TeV}$. (f) $\sqrt{s}=10~\textrm{TeV}$. (g) $\sqrt{s}=30~\textrm{TeV}$. (h) $\sqrt{s}=100~\textrm{TeV}$. Lower row: $t\bar{t}$ production from $p\bar{p}$ collisions. (i)-(l) Same values of $\sqrt{s}$ as (e)-(h), respectively.} 
\label{fig:ConcurrenceMIX}
\end{figure*}

Regarding the actual spin quantum state, since $\rho^{I}=R^{I}/4\tilde{A}^I$, then
\begin{equation}\label{eq:partonicstates}
\rho(M_{t\bar{t}},\hat{k},\sqrt{s})=\sum_{I=q\bar{q},gg} w_I(M_{t\bar{t}},\hat{k},\sqrt{s})\rho^{I}(M_{t\bar{t}},\hat{k})
\end{equation}
Similar expressions can be written for the spin polarizations and spin correlations $\mathbf{B}^{\pm}(M_{t\bar{t}},\hat{k},\sqrt{s}),\mathbf{C}(M_{t\bar{t}},\hat{k},\sqrt{s})$ of $\rho(M_{t\bar{t}},\hat{k},\sqrt{s})$ in terms of their partonic counterparts, computed from Eqs. (\ref{eq:LOSpinCorrelationsqq}), (\ref{eq:LOSpinCorrelationsgg}). The weights $w_I$ are obtained from the luminosities as
\begin{equation}\label{eq:partonicweights}
w_I(M_{t\bar{t}},\hat{k},\sqrt{s})=\frac{L_{I}(M_{t\bar{t}},\sqrt{s})\tilde{A}^I(M_{t\bar{t}},\hat{k})}{\sum_{J}L_{J}(M_{t\bar{t}},\sqrt{s})\tilde{A}^J(M_{t\bar{t}},\hat{k})} 
\end{equation}
and represent the probability of production of each individual partonic quantum state $\rho^{I}(M_{t\bar{t}},\hat{k})$, satisfying $w_{q\bar{q}}+w_{gg}=1$. They can be simply understood as the probability of occurrence of the initial state $I$, $L_I$, multiplied by the probability of producing a $t\bar{t}$ pair from $I$, proportional to the differential cross-section, in turn proportional to the coefficient $\tilde{A}^I(M_{t\bar{t}},\hat{k})$. 

We stress that the nature of the specific hadron collision only enters in the calculations through the probabilities $w_I(M_{t\bar{t}},\hat{k},\sqrt{s})$, which are PDF dependent. Hence, the LO QCD $q\bar{q}$ and $gg$ processes are the building blocks of any realistic hadronic production process, whose role is reduced to change the amount of mixing between them. From a more general perspective, we can regard the production spin density matrices $R^{I}(M_{t\bar{t}},\hat{k})$ and the luminosities $L_{I}(M_{t\bar{t}},\sqrt{s})$ as inputs from the theory of high-energy physics, from which we can compute the physical spin density matrices $\rho^I(M_{t\bar{t}},\hat{k})$ and the probabilities $w_I(M_{t\bar{t}},\hat{k},\sqrt{s})$. Once done, we are simply left with a typical problem in quantum information involving the convex sum of two-qubit states, where the usual techniques of the field can be applied.

In order to understand better how the mixture between $q\bar{q}$ and $gg$ channels affects the $t\bar{t}$ quantum state $\rho(M_{t\bar{t}},\hat{k},\sqrt{s})$, we first use a toy model where both partonic probabilities are constant, $w_I(M_{t\bar{t}},\hat{k},\sqrt{s})=w_I$. We gradually introduce $gg$ processes in a pure $q\bar{q}$ reaction in upper row of Fig.~\ref{fig:ConcurrenceMIX}. For a small amount of $gg$ processes, the effect of mixing is reduced to shrink the entangled region towards the high $p_T$ region. By further increasing the amount of $gg$ processes, at some point entangled $t\bar{t}$ states also emerge at threshold. We can actually compute from Eq.~(\ref{eq:Deltatt}) the critical value $w^{\textrm{PH}}_{c}$ at which entanglement appears, since
\begin{equation}
    \Delta(\beta=0,\Theta)=\frac{w_{gg}+|3w_{gg}-1|-1}{2}
\end{equation}
For $w_{gg}>1/3$, $\Delta(0,\Theta)=2w_{gg}-1$, so 
\begin{equation}
    w^{\textrm{PH}}_{c}=\frac{1}{2}
\end{equation}
Hence, in order to have entanglement close to threshold, we need at least $50\%$ of the $t\bar{t}$ pairs to be produced through gluon fusion. Regarding the violation of the CHSH inequality, the corresponding critical value $w^{\textrm{CH}}_{c}$ is computed by switching to the off-diagonal basis where both correlation matrices are diagonal ($\rho^{gg}$ at threshold is a spin singlet, invariant under rotations). This gives
\begin{equation}
    \mu(\beta=0,\Theta)=2w^2_{gg}-1
\end{equation}
from where we find
\begin{equation}
    w^{\textrm{CH}}_{c}=\frac{1}{\sqrt{2}}>w^{\textrm{PH}}_{c}
\end{equation}

Once this simple model is understood, we switch to real hadron processes. We analyze $t\bar{t}$ production from $pp$ collisions for different c.m. energies in central row of Fig.~\ref{fig:ConcurrenceMIX}. By direct comparison with upper row, we observe that for low c.m. energies $q\bar{q}$ processes dominate. However, the amount of $gg$ processes increases with the c.m. energy, eventually dominating the $t\bar{t}$ production mechanism. A similar calculation is shown in the lower row of Fig.~\ref{fig:ConcurrenceMIX} for $p\bar{p}$ collisions where, although $q\bar{q}$ contributions are stronger at low energies, $gg$ dominance is again recovered for sufficiently high c.m. energies, and both types of collisions converge to a similar quantum state. 

\begin{figure*}[t!]
\begin{tabular}{@{}cc@{}}
    \includegraphics[width=\columnwidth]{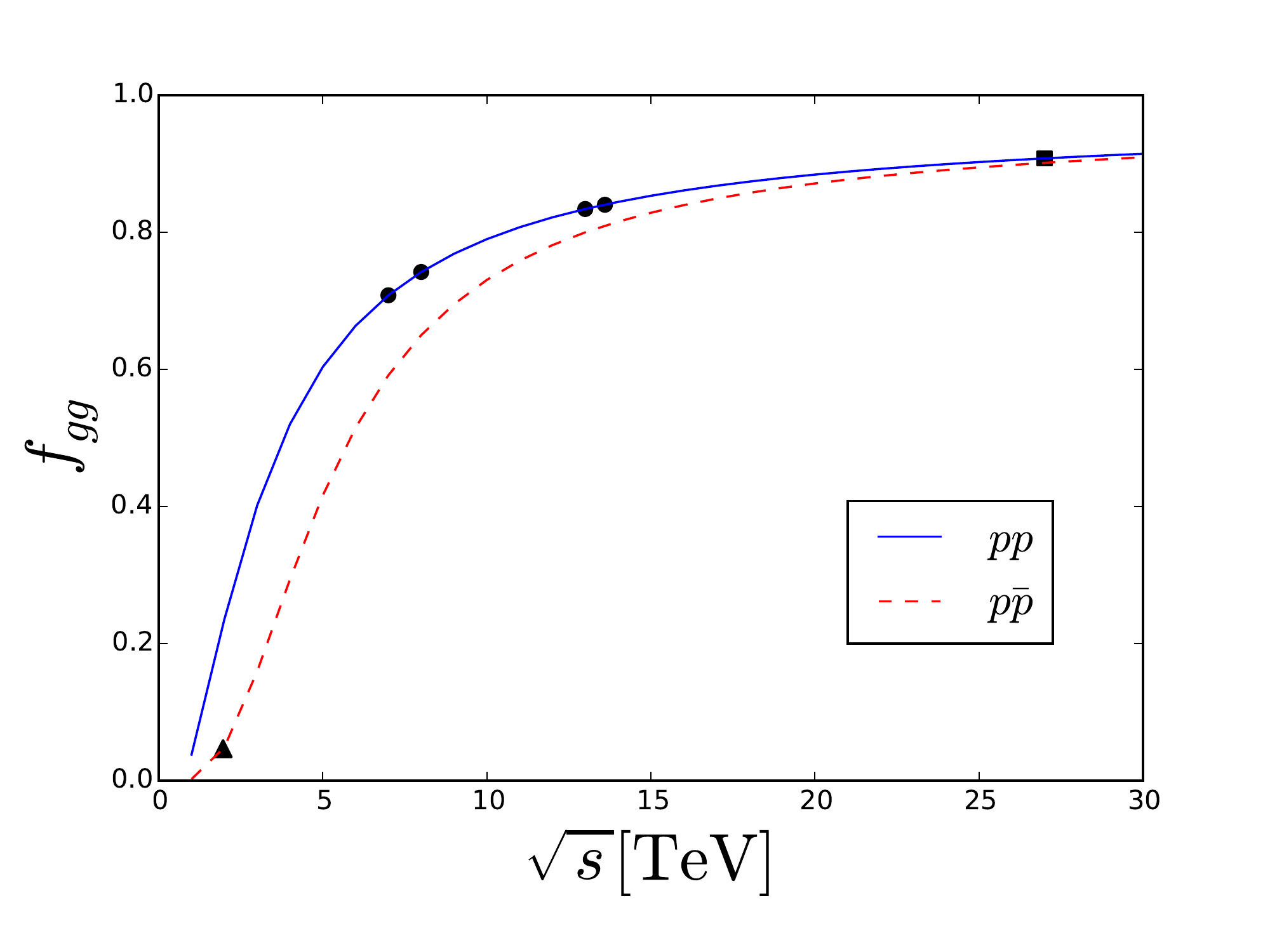} &
    \includegraphics[width=\columnwidth]{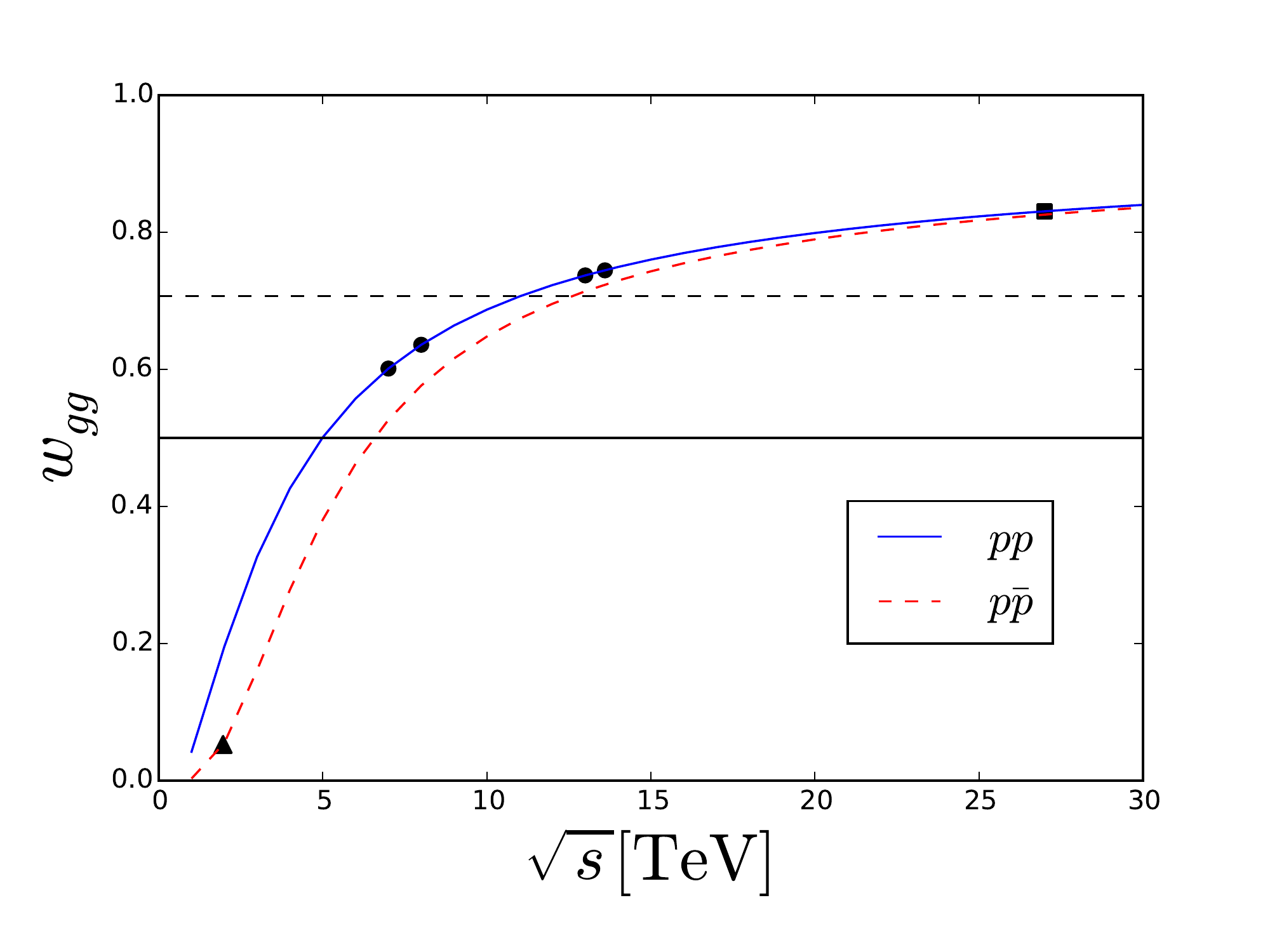}
\end{tabular}
\caption{Study of the dependence of $t\bar{t}$ production mechanism with the hadron c.m. energy $\sqrt{s}$ for $pp$ collisions (solid blue line) and for $p\bar{p}$ collisions (dashed red line), corresponding to the LHC and Tevatron, respectively. Black markers indicate the values for c.m. energies of actual high-energy colliders. Circles: Run~1 ($\sqrt{s}=7,8~\textrm{TeV}$), Run~2 ($\sqrt{s}=13~\textrm{TeV}$) and Run~3 ($\sqrt{s}=13.6~\textrm{TeV}$) of the LHC. Squares: Possible upgrade of the LHC ($\sqrt{s}=27~\textrm{TeV}$). Triangles: Tevatron ($\sqrt{s}=1.96~\textrm{TeV}$). Left: Gluon fraction $f_{gg}$. Right: $gg$ probability at threshold $w_{gg}(\beta=0)$. Horizontal black lines mark the critical values $w^{\textrm{PH}}_{gg,c}=1/2$ for entanglement (solid), and $w^{\textrm{CH}}_{gg,c}=1/\sqrt{2}$ for CHSH violation (dashed).}
\label{fig:Criticality}
\end{figure*}

We can further quantify the amount of mixing of $q\bar{q}$ and $gg$ processes in the total $t\bar{t}$ production by computing the relative contribution of each channel to the total cross-section from Eq.~(\ref{eq:CrossSectionDifferentialReal}):
\begin{equation}
    f_I\equiv \frac{\int\limits^{\sqrt{s}}_{2m_t}\mathrm{d}M_{t\bar{t}}\int\mathrm{d}\Omega~L_I\frac{\mathrm{d}\sigma^I}{\mathrm{d}\Omega}}{\int\limits^{\sqrt{s}}_{2m_t}\mathrm{d}M_{t\bar{t}}\int\mathrm{d}\Omega~\frac{\mathrm{d}\sigma}{\mathrm{d}\Omega\mathrm{d}M_{t\bar{t}}}}
\end{equation}
where we stress that the maximum invariant mass $M_{t\bar{t}}$ achievable by the $t\bar{t}$ pair is precisely the c.m. energy of the hadron pair, $M_{t\bar{t}}=\sqrt{s}$. The fraction of $gg$ processes $f_{gg}$ for both $pp$ and $p\bar{p}$ collisions is represented in left Fig.~\ref{fig:Criticality}, where its predicted increase with the c.m. energy indicated by the entanglement analysis is clearly observed, as well as the stronger contribution of $q\bar{q}$ processes for $p\bar{p}$ collisions with respect to $pp$ at low energies. At high energies, both hadronic processes converge to the same gluon fraction $f_{gg}$. This analysis suggests that entanglement measurements can be also used to understand the underlying structure of a certain high-energy process without direct knowledge of it.

A related plot is that of right Fig.~\ref{fig:Criticality}, in which the threshold values of the gluon probability $w_{gg}(\beta=0,\sqrt{s})$ are represented (there is no angular dependence at threshold). We observe that entanglement [$w_{gg}(0,\sqrt{s})>w^{\textrm{PH}}_{c}=1/2$] can be achieved in $pp$ collisions for small c.m. energies $\sqrt{s}\gtrsim 5~\textrm{TeV}$, below current LHC energies. In particular, the relevant experimental values $\sqrt{s}=7,8~\textrm{TeV}$~\cite{Aad2012,Chatrchyan2013,Aad2014mfk}, $\sqrt{s}=13~\textrm{TeV}$~\cite{Sirunyan2019,Aaboud2019hwz} and $\sqrt{s}=13.6~\textrm{TeV}$~\cite{Fartoukh:2790409} are marked by circles, corresponding to the c.m. energies of Run~1, Run~2 and the ongoing Run~3 of the LHC, respectively. The expected value for a possible upgrade of the LHC is $\sqrt{s}=27~\textrm{TeV}$~\cite{FCC:2018bvk}, marked by a square. In addition, the c.m. energy for $pp$ collisions in the Future Circular Collider (FCC), $\sqrt{s}=100~\textrm{TeV}$~\cite{Benedikt2020}, is represented explicitly in Fig.~\ref{fig:ConcurrenceMIX}h. With respect to CHSH violation [$w_{gg}(0,\sqrt{s})>w^{\textrm{CH}}_{c}=1/\sqrt{2}$], we observe that we need larger energies $\sqrt{s}\gtrsim 10~\textrm{TeV}$, with Run~2 and Run~3 only slightly above this limit. For $p\bar{p}$ collisions, even threshold entanglement cannot be observed since the c.m. energies required are well above the maximum energy achieved at the Tevatron, $\sqrt{s}=1.96~\textrm{TeV}$~\cite{Aaltonen2010,Abazov2011ka,Abazov2015psg}, marked by a triangle. 

From these plots, we conclude that the Tevatron and the LHC provide experimental realizations to a very good approximation of the two main paradigms of $t\bar{t}$ production through QCD, $q\bar{q}$ and $gg$ reactions.  

\section{Experimental observables}\label{sec:totalquantumstate}

\subsection{Experimental motivation}\label{subsec:Experiment}

How are all these spin magnitudes translated into actual observables that can be measured in high-energy colliders? This is where the unique properties of the top quark enter in place. As explained in the Introduction, the top quark is the most massive particle of the Standard Model. This property makes that even its large width $\Gamma_t\simeq 1~\textrm{GeV}$ is still narrow when compared to its mass $m_t\simeq 173~\textrm{GeV}$. In turn, this large width results in a very short lifetime $\tau=1/\Gamma_t\sim 10^{-25} \textrm{s}$. This implies that a $t\bar{t}$ pair decays so fast that any other process, such as hadronisation (with a time scale $\sim 10^{-23} \textrm{s}$) or spin decorrelation (with a time scale $\sim 10^{-21} \textrm{s}$), cannot affect their spin quantum state, whose information is directly translated into the properties of the decay products. This is what makes the top quark so special, as any other quark hadronizes before decaying, losing its spin information in the process. On the other hand, if the top quark did not decay, its spin state could not be extracted because detectors only measure the momentum of the arriving particles. 

Quantitatively, we describe the decay of a top quark to some final state $F$ in terms of a decay spin density matrix, defined in the top rest frame as
\begin{equation}\label{eq:DecaySpinDensityMatrixF}
    \Gamma^F_{\alpha'\alpha}\equiv\bra{F}T\ket{t\alpha}\bra{t\alpha'}T^{\dagger}\ket{F}
\end{equation}
where $\ket{t\alpha}$ is the top state with zero momentum and spin $\alpha$, $\ket{F}$ is the final state after the decay, and $T$ is here the \textit{on-shell} $T$-matrix determining the amplitude of the decay process. A similar decay spin density matrix $\bar{\Gamma}^{\bar{F}}_{\beta'\beta}$ can be defined for the decay of the antitop quark to a final state $\bar{F}$. 

The idea is to describe the production and decay
of a $t\bar{t}$ pair using the production and decay spin density matrices. Specifically, we consider a $t\bar{t}$ pair with fixed c.m. energy and momentum produced from a hadron collision, whose quantum state is specified by the matrix $R(M_{t\bar{t}},\hat{k},\sqrt{s})$. In the so-called narrow-width approximation, valid since $\Gamma_t\ll m_t$~\cite{Bernreuther1994}, the differential cross-section characterizing the decay of a $t\bar{t}$ pair to two final states $F\bar{F}$ is proportional to~\cite{Bernreuther2004,Bernreuther:2010ny}
\begin{align}\label{eq:LeptonicDifferentialCrossSectionTotal}
 \nonumber \mathrm{d}\sigma_{F\bar{F}} &\sim \sum_{\alpha\alpha',\beta\beta'}R_{\alpha\beta,\alpha'\beta'}(M_{t\bar{t}},\hat{k},\sqrt{s})\Gamma^F_{\alpha'\alpha}\bar{\Gamma}^{\bar{F}}_{\beta'\beta}\\
 &=\textrm{tr}\left[\left(\Gamma^F\otimes\bar{\Gamma}^{\bar{F}}\right) R(M_{t\bar{t}},\hat{k},\sqrt{s}) \right]
\end{align}
Thus, the information about the quantum state of the $t\bar{t}$ pair, encoded in $R(M_{t\bar{t}},\hat{k},\sqrt{s})$, is contained in the cross-section characterizing the decay products of the final states $F\bar{F}$.

\begin{figure}[t]
  \centering
     \includegraphics[width=0.7\columnwidth]{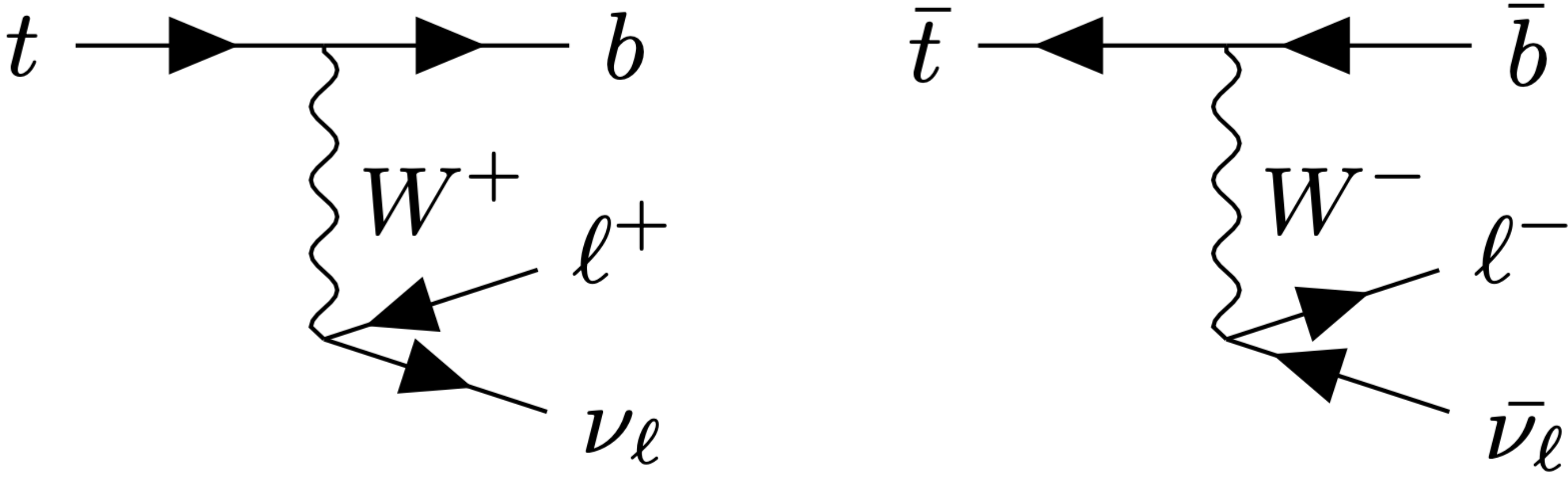} 
     \caption{Representative Feynman diagrams for a leptonic electroweak decay of a top (left) and antitop (right) quark.}
     \label{fig:FeynmanDecay}
\end{figure}

We can retrieve this information from final states including a lepton $\ell$, which can be either an electron or a muon, $\ell=e,\mu$. Specifically, we consider the electroweak leptonic decay of both the top and the antitop quark:
\begin{align}\label{eq:ttdecays}
    t&\rightarrow b+W^+\rightarrow b+\ell^+ + \nu_\ell,\\
\nonumber \bar{t}&\rightarrow \bar{b}+W^-\rightarrow \bar{b}+\ell^{-}+\bar{\nu}_\ell
\end{align}
where $F=b\ell^+\nu_\ell$, with $\ell^+$ the antilepton and $\nu_\ell$ its associated neutrino, and $\bar{F}$ its conjugate,
$\bar{F}=\bar{b}\ell^-\bar{\nu}_\ell$. A diagrammatic representation of the top/antitop decay is provided in Fig.~\ref{fig:FeynmanDecay}.

The expression for the decay spin density matrices vastly simplifies if we integrate out all the degrees of freedom of the final states $F\bar{F}$ except for the lepton directions. In that case, due to the rotational invariance in the top (antitop) rest frames, the resulting decay spin density matrices $\Gamma_\ell$ $(\bar{\Gamma}_\ell)$ can only be of the form~\cite{Baumgart2013}
\begin{equation}\label{eq:DecaySpinDensityMatrix}
    \Gamma_\ell \propto \frac{I_2+\kappa_\ell (\hat{\mathbf{\ell}}_{+}\cdot \mathbf{\sigma})}{2},~
    \bar{\Gamma}_\ell\propto\frac{I_2+\bar{\kappa}_\ell (\hat{\mathbf{\ell}}_{-}\cdot \mathbf{\sigma})}{2}
\end{equation}
$\hat{\mathbf{\ell}}_{\pm}$ being the antilepton (lepton) directions in each one of the parent top (antitop) rest frames, and $\kappa_\ell=-\bar{\kappa}_\ell\simeq 1$ the so-called spin analyzing powers of the leptons. 

As a result, the cross-section $\sigma_{\ell\bar{\ell}}$ characterizing the $\ell^+\ell^-$ angular distribution reads
\begin{align}\label{eq:LeptonicCrossSectionTotal}
 \nonumber &\frac{\mathrm{d}\sigma_{\ell\bar{\ell}}}{\mathrm{d} \Omega_+\mathrm{d}\Omega_-\mathrm{d}M_{t\bar{t}}\mathrm{d}\Omega_{\hat{k}}} \propto\frac{\alpha^2_s\beta}{M^2_{t\bar{t}}} \textrm{tr}\left[(\Gamma_\ell\otimes \bar{\Gamma}_\ell) R \right]\\ &\propto\frac{\alpha^2_s\beta}{M^2_{t\bar{t}}}\left[\tilde{A}+\tilde{\mathbf{B}}^{+}\cdot\hat{\mathbf{\ell}}_{+}-\tilde{\mathbf{B}}^{-}\cdot\hat{\mathbf{\ell}}_{-}
-\hat{\mathbf{\ell}}_{+}\cdot \tilde{\mathbf{C}} \cdot\hat{\mathbf{\ell}}_{-}\right]
\end{align}
where $\Omega_{\pm}$, $\Omega_{\hat{k}}$ are the solid angles associated to $\hat{\mathbf{\ell}}_{\pm}$, $\hat{k}$. The differential elements $\mathrm{d}M_{t\bar{t}}\mathrm{d}\Omega_{\hat{k}}$ arise because we are considering the decay of a $t\bar{t}$ pair with fixed energy and direction. The total angular differential cross-section describing the leptons arising from the $t\bar{t}$ decay is then obtained by integration over all possible top directions and c.m. energies:
\begin{equation}\label{eq:LeptonicCrossSectionTotalIntegrated}
 \frac{\mathrm{d}\sigma_{\ell\bar{\ell}}}{\mathrm{d} \Omega_+\mathrm{d}\Omega_-}= \int\limits^{\sqrt{s}}_{2m_t}\mathrm{d}M_{t\bar{t}}\int\mathrm{d}\Omega~\frac{\mathrm{d}\sigma_{\ell\bar{\ell}}}{\mathrm{d} \Omega_+\mathrm{d}\Omega_-\mathrm{d}M_{t\bar{t}}\mathrm{d}\Omega_{\hat{k}}}
\end{equation}
This implies that the normalized angular differential cross-section characterizing the dileptonic decay is 
\begin{equation}\label{eq:LeptonicCrossSectionNormalized}
\frac{1}{\sigma_{\ell\bar{\ell}}}\frac{\mathrm{d}\sigma_{\ell\bar{\ell}}}{\mathrm{d}\Omega_{+}\mathrm{d}\Omega_{-}}=\frac{1+\mathbf{B}^{+}\cdot\hat{\mathbf{\ell}}_{+}-\mathbf{B}^{-}\cdot\hat{\mathbf{\ell}}_{-}
-\hat{\mathbf{\ell}}_{+}\cdot \mathbf{C} \cdot\hat{\mathbf{\ell}}_{-}}{(4\pi)^2}
\end{equation}
where the vectors $\mathbf{B}^{\pm}$ and the matrix $\mathbf{C}$ are the integrated values of the top/antitop spin polarizations and the spin correlation matrix, respectively. Therefore, we can measure all the integrated spin information of a $t\bar{t}$ pair by fitting the angular distribution of the leptonic decay products. 

In particular, from the kinematic reconstruction of each event, the momenta of the $t\bar{t}$ and $\ell^+\ell^-$ pairs can be determined. By defining the polar angle with respect to a certain direction given by the unit vector $\hat{u}_i$, $\cos\theta^i_{\pm}\equiv \hat{\mathbf{\ell}}_{\pm}\cdot \hat{u}_i$, and after integrating over the azimuthal angles, we find that the spin polarizations can be easily obtained from a linear fit of the angular distributions for each individual lepton as
\begin{align}\label{eq:polarizations}
\frac{1}{\sigma_{\ell\bar{\ell}}} \frac{\mathrm{d}\sigma_{\ell\bar{\ell}}}{\mathrm{d}\cos \theta_{\pm}^i} &= \frac{1}{2}(1\pm B_i^{\pm} \cos \theta^i_\pm)
\end{align}
The spin correlations for a pair of directions $i,j$ can be measured from the distribution of the product $x_{ij}\equiv \cos\theta^i_{+}\cos\theta^j_{-}$, given by integration of Eq.~(\ref{eq:LeptonicCrossSectionNormalized}):
\begin{equation}\label{eq:correlationsproduct}
\frac{1}{\sigma_{\ell\bar{\ell}}} \frac{\mathrm{d}\sigma_{\ell\bar{\ell}}}{\mathrm{d}x_{ij}}=\frac{1}{2}\left[1-C_{ij} x_{ij}\right]\ln\frac{1}{|x_{ij}|}
\end{equation}
Another measurable magnitude of interest is the trace of the correlation matrix, which can be directly extracted from
\begin{equation}\label{eq:DCrossSection}
\frac{1}{\sigma_{\ell\bar{\ell}}} \frac{\mathrm{d}\sigma_{\ell\bar{\ell}}}{\mathrm{d} \cos \varphi}=\frac{1}{2}(1-D\cos\varphi),~D=\frac{\textrm{tr}[\mathbf{C}]}{3}    \end{equation}
where $\varphi$ is the angle between the lepton directions in each one of their parent top and antitop rest frames, $\cos\varphi=\hat{\mathbf{\ell}}_{+}\cdot \hat{\mathbf{\ell}}_{-}$.

The determination of the momentum of the $t\bar{t}$ pair also implies that their spins can be characterized in any orthonormal basis, e.g. in the helicity basis considered so far. The measurement of the $t\bar{t}$ spin polarizations and spin correlations through this technique is well-established, and has already been performed by the CMS collaboration at the LHC~\cite{Sirunyan2019}, where $\mathbf{B}^{\pm},\mathbf{C}$ were obtained in the helicity basis, with no restrictions on $t\bar{t}$ phase space. However, no entanglement signature was provided by these measurements.

\subsection{Integrated expectation values}

The formalism of the previous subsection focused on integrated values over all $(M_{t\bar{t}},\hat{k})$ phase space. However, as explained, the $t\bar{t}$ momenta can be experimentally reconstructed, allowing to restrict the contributions to the integral of Eq.~(\ref{eq:LeptonicCrossSectionTotal}) to specific regions $\Pi$ of phase space. In that case, the measured spin polarizations and spin correlations would correspond to integrated values just within the region $\Pi$. This allows to further explore the $t\bar{t}$ properties in search of richer physics. For instance, we can take $\Pi$ within the regions of phase space where entanglement is present (delimited by the black lines in Figs.~\ref{fig:ConcurrenceFundamental},~\ref{fig:ConcurrenceMIX}). 

Ideally, one would restrict $\Pi$ to a small bin around a given value $(M_{t\bar{t}},\hat{k})$, since that would allow to study the individual spin quantum states $\rho(M_{t\bar{t}},\hat{k},\sqrt{s})$. However, in actual experiments, the signal needs to be integrated in order to collect a sufficient number of events for achieving statistically significant measurements. Therefore, an adaptation of the formalism developed in the previous sections is required. 

We begin by noting that the total quantum state $\rho_{\textrm{T}}(\sqrt{s})$ describing the $t\bar{t}$ pairs produced from a certain hadronic process is obtained by generalizing Eq.~(\ref{eq:QuantumStateRSpin}) as
\begin{widetext}
\begin{equation}\label{eq:QuantumStateTotaltt}
\rho_{\textrm{T}}(\sqrt{s})=\frac{1}{\sigma}\sum_{\alpha\beta,\alpha'\beta'}\int\limits^{\sqrt{s}}_{2m_t}\mathrm{d}M_{t\bar{t}}\int\mathrm{d}\Omega~\frac{\mathrm{d}\sigma}{\mathrm{d}\Omega\mathrm{d}M_{t\bar{t}}}\rho_{\alpha\beta,\alpha'\beta'}(M_{t\bar{t}},\hat{k},\sqrt{s})\frac{\ket{M_{t\bar{t}}\hat{k}\alpha\beta}\bra{M_{t\bar{t}}\hat{k}\alpha'\beta'}}{\braket{M_{t\bar{t}}\hat{k}|M_{t\bar{t}}\hat{k}}}
\end{equation}
If we consider now some observable $O$ diagonal in momentum space, with matrix elements $O_{\alpha'\beta',\alpha\beta}(M_{t\bar{t}},\hat{k})=\bra{M_{t\bar{t}}\hat{k}\alpha'\beta'}O\ket{M_{t\bar{t}}\hat{k}\alpha\beta}/\braket{M_{t\bar{t}}\hat{k}|M_{t\bar{t}}\hat{k}}$, its expectation value $\braket{O}=\textrm{tr}[O\rho_{\textrm{T}}(\sqrt{s})]$ is simply computed as
\begin{equation}\label{eq:ExpectationValueTotaltt}
\braket{O}=\frac{1}{\sigma}\int\limits^{\sqrt{s}}_{2m_t}\mathrm{d}M_{t\bar{t}}\int\mathrm{d}\Omega~\frac{\mathrm{d}\sigma}{\mathrm{d}\Omega\mathrm{d}M_{t\bar{t}}}\textrm{tr}\left[O(M_{t\bar{t}},\hat{k}) \rho(M_{t\bar{t}},\hat{k},\sqrt{s})\right]
\end{equation}
This expectation value is readily understood as the average of the expectation values of $O$ for all possible quantum states $\rho(M_{t\bar{t}},\hat{k},\sqrt{s})$ of the $t\bar{t}$ pair, with a probability proportional to the differential cross-section of the process, normalized by the total cross-section for $t\bar{t}$ production.

The above equation can be simply rewritten in terms of the $R$-matrix as
\begin{equation}\label{eq:ExpectationValueR}
\braket{O}= \frac{\int\limits^{\sqrt{s}}_{2m_t}\mathrm{d}M_{t\bar{t}}\int\mathrm{d}\Omega~\frac{\alpha^2_s\beta}{M^2_{t\bar{t}}}\mathrm{tr}\left[O(M_{t\bar{t}},\hat{k})R(M_{t\bar{t}},\hat{k},\sqrt{s})\right]}
{\int\limits^{\sqrt{s}}_{2m_t}\mathrm{d}M_{t\bar{t}}\int\mathrm{d}\Omega~\frac{\alpha^2_s\beta}{M^2_{t\bar{t}}}\mathrm{tr}\left[R(M_{t\bar{t}},\hat{k},\sqrt{s})\right]}
\end{equation}

From these considerations, it is immediate to compute the reduced quantum state $\rho_{\textrm{T},\Pi}(\sqrt{s})$ that describes the $t\bar{t}$ pair within a certain region $\Pi$ of phase space by projecting only onto the relevant states $\ket{M_{t\bar{t}}\hat{k}}$ contained in $\Pi$:

\begin{equation}\label{eq:QuantumStateTotalPi}
\rho_{\textrm{T},\Pi}(\sqrt{s})=\frac{1}{\sigma_\Pi}\sum_{\alpha\beta,\alpha'\beta'}\int_\Pi\mathrm{d}\Omega\mathrm{d}M_{t\bar{t}}~\frac{\mathrm{d}\sigma}{\mathrm{d}\Omega\mathrm{d}M_{t\bar{t}}}\rho_{\alpha\beta,\alpha'\beta'}(M_{t\bar{t}},\hat{k},\sqrt{s})\frac{\ket{M_{t\bar{t}}\hat{k}\alpha\beta}\bra{M_{t\bar{t}}\hat{k}\alpha'\beta'}}{\braket{M_{t\bar{t}}\hat{k}|M_{t\bar{t}}\hat{k}}}
\end{equation}
\end{widetext}
with the partition function $\sigma_\Pi$ being the $t\bar{t}$ cross-section in the region $\Pi$
\begin{equation}
    \sigma_\Pi\equiv\int_\Pi\mathrm{d}\Omega\mathrm{d}M_{t\bar{t}}~\frac{\mathrm{d}\sigma}{\mathrm{d}\Omega\mathrm{d}M_{t\bar{t}}} 
\end{equation}

Similar relations can be obtained for the expectation value $\braket{O}_\Pi$ of any observable $O$ in the region $\Pi$ by restricting the integrals of Eqs. (\ref{eq:ExpectationValueTotaltt}), (\ref{eq:ExpectationValueR}) to the region $\Pi$.

\begin{figure}[tb!]\includegraphics[width=0.8\columnwidth]{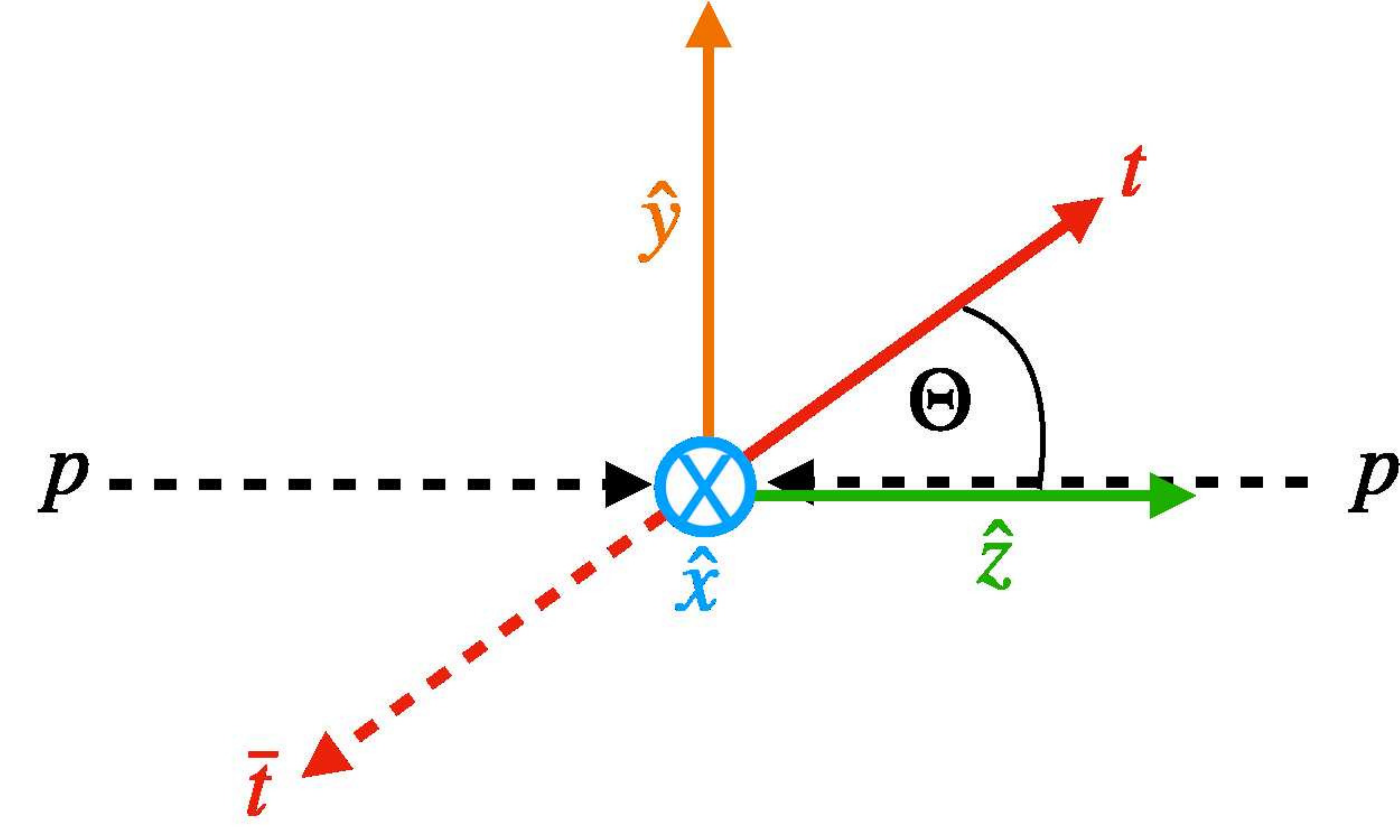}
\centering
     \caption{Orthonormal beam basis, defined in the c.m. frame. The $\hat{z}$ vector points along the direction of the initial hadron beam, $\hat{z}=\hat{p}$, while $\hat{x},\hat{y}$ point fixed transverse directions.} 
     \label{fig:BeamBasis}
\end{figure}

Furthermore, we can define a genuine two-qubit quantum state by taking the trace in momentum space
\begin{align}\label{eq:QuantumStateTotalTwoQubit}
\nonumber \rho_\Pi(\sqrt{s})&\equiv\textrm{tr}_{M_{t\bar{t}}\hat{k}}\left[\rho_{\textrm{T},\Pi}(\sqrt{s})\right]\\
&=\sum_{\alpha\beta,\alpha'\beta'}\rho_{\Pi,\alpha\beta,\alpha'\beta'}\ket{\alpha\beta}\bra{\alpha'\beta'}
\end{align}
In matrix notation,
\begin{align}\label{eq:TotalDensityMatrix}
\nonumber &\rho_\Pi(\sqrt{s})=\frac{1}{\sigma_\Pi}\int_\Pi\mathrm{d}\Omega\mathrm{d}M_{t\bar{t}}~\frac{\mathrm{d}\sigma}{\mathrm{d}\Omega\mathrm{d}M_{t\bar{t}}}\rho(M_{t\bar{t}},\hat{k},\sqrt{s})\\
&=\frac{\sum_I\int_\Pi\mathrm{d}\Omega\mathrm{d}M_{t\bar{t}}~\frac{\alpha^2_s\beta}{M^2_{t\bar{t}}} L_{I}(M_{t\bar{t}},\sqrt{s})R^I(M_{t\bar{t}},\hat{k})}
{\sum_{I}\int_\Pi\mathrm{d}\Omega\mathrm{d}M_{t\bar{t}}~\frac{\alpha^2_s\beta}{M^2_{t\bar{t}}}L_{I}(M_{t\bar{t}},\sqrt{s})4\tilde{A}^I(M_{t\bar{t}},\hat{k})}
\end{align}
As a two-qubit quantum state, $\rho_\Pi$ is determined by the respective integrated spin polarizations and spin correlations $\mathbf{B}_{\Pi}^{\pm},\mathbf{C}_{\Pi}$, which can be measured from the cross-section of the dileptonic decay, Eqs. (\ref{eq:polarizations}), (\ref{eq:correlationsproduct}). However, since $\rho_\Pi$ is already the result of an integration in phase space, the orthonormal basis used to compute $\mathbf{B}_{\Pi}^{\pm},\mathbf{C}_{\Pi}$ cannot depend on $(M_{t\bar{t}},\hat{k})$, which invalidates the use of the helicity or the diagonal basis. A natural choice is then the beam basis $\{\hat{x},\hat{y},\hat{z}\}$, where $\hat{z}=\hat{p}$ points along the initial beam, and $\hat{x},\hat{y}$ point transverse directions to the beam, where all directions are fixed in the c.m. frame~\cite{Bernreuther2004}. The beam basis is represented in Fig.~\ref{fig:BeamBasis}. 

If one inserts the density matrices $\rho(M_{t\bar{t}},\hat{k},\sqrt{s})$ computed in the helicity or the diagonal basis in Eq. (\ref{eq:TotalDensityMatrix}), the resulting integrated density matrix, which we denote as $\bar{\rho}_\Pi(\sqrt{s})$, does not represent an actual spin quantum state of the $t\bar{t}$ pair, Eq. (\ref{eq:QuantumStateTotalTwoQubit}), even though it satisfies the usual properties of a density matrix as it is a convex sum of physical density matrices (i.e., it is a non-negative Hermitian matrix with unit trace). The coefficients $\mathbf{B}_\Pi^{\pm},\mathbf{C}_\Pi$ characterizing this \textit{fictitious} quantum state $\bar{\rho}_\Pi(\sqrt{s})$ are precisely the expectation values in $\Pi$ of the spin polarizations and the spin correlations in the basis chosen for the integration. Nevertheless, since both the set of separable and Bell-local states are convex, any signature of entanglement/CHSH violation in the fictitious quantum state $\bar{\rho}_\Pi(\sqrt{s})$ implies entanglement/CHSH violation in some of the physical substates $\rho(M_{t\bar{t}},\hat{k},\sqrt{s})$, and so in the total spin-momentum quantum state $\rho_{\textrm{T},\Pi}$.

The upshot of the discussion of this subsection is that we can aim at obtaining entanglement signatures by applying the previous criteria to the corresponding integrated expectation values in selected regions $\Pi$ of phase space where the $t\bar{t}$ pair is entangled. Moreover, we can analyze directly if the \textit{physical} two-qubit quantum state $\rho_\Pi$ itself is entangled. In fact, by measuring the coefficients $\mathbf{B}_{\Pi}^{\pm},\mathbf{C}_{\Pi}$ we can perform the full quantum tomography of $\rho_\Pi$.

Due to its experimental simplicity and availability of analytical results, leading to a better understanding of the physics involved, in the following we take the region $\Pi$ as $\Pi=\Sigma \times S^{2}$. That is, we integrate over all possible top directions so we only have to specify the cuts in the invariant mass spectrum that delimit the region $\Sigma$. 

\begin{figure*}[tb!]
\begin{tabular}{@{}cc@{}}
    \includegraphics[width=\columnwidth]{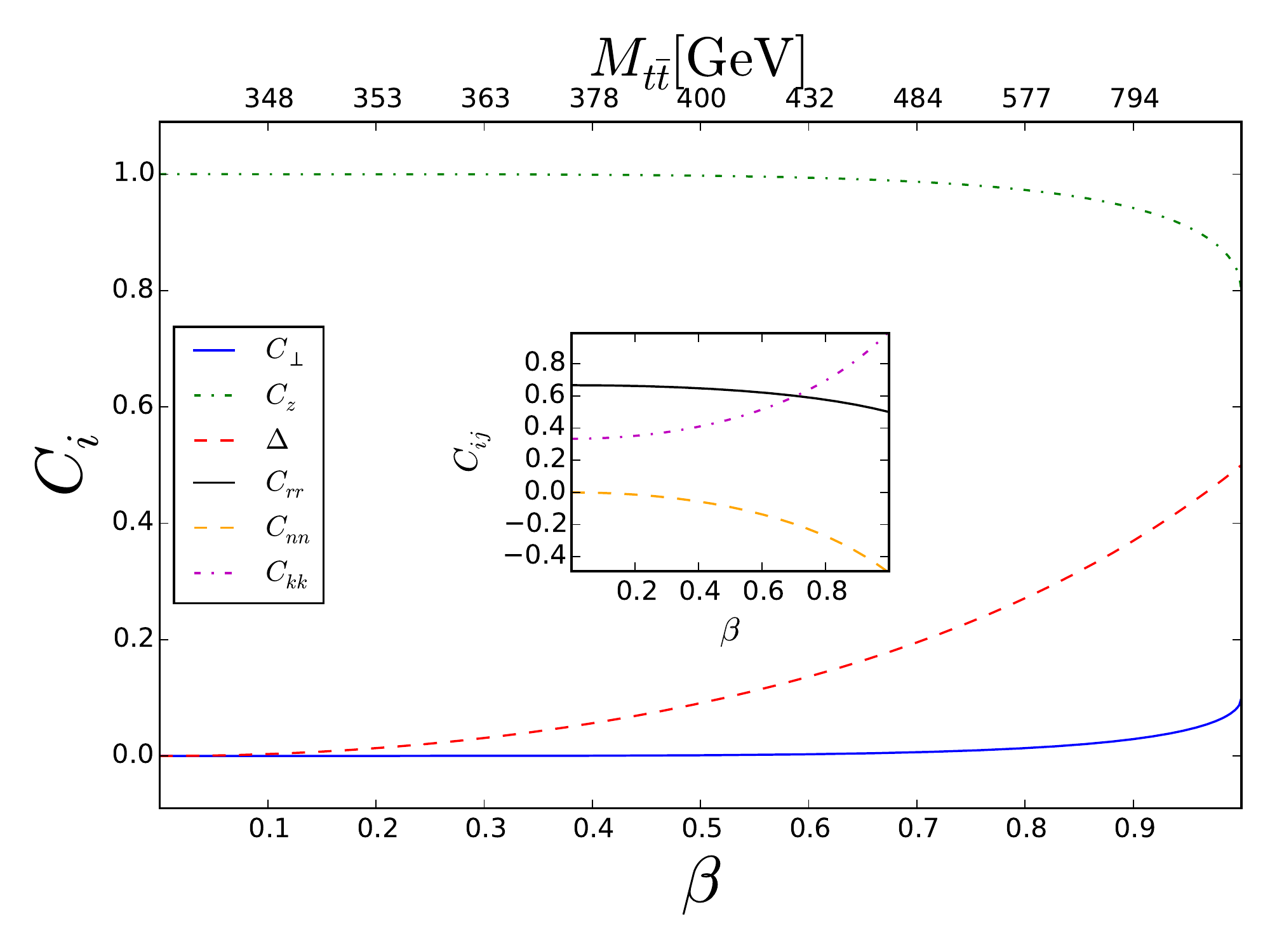} &
    \includegraphics[width=\columnwidth]{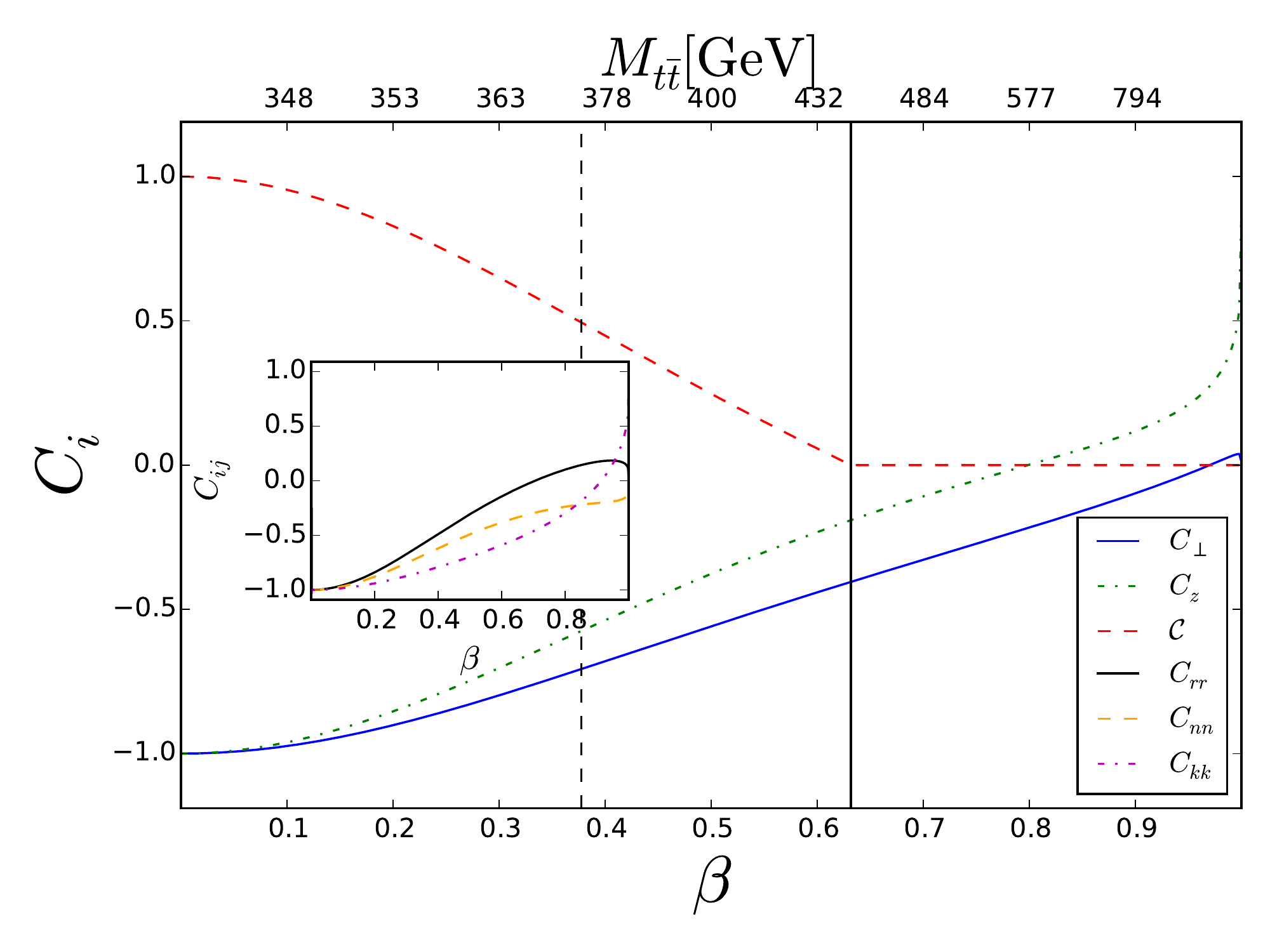}
\end{tabular}
\caption{Angular-averaged spin correlations as a function of $\beta$. Main plot shows $C^{I}_{\perp}$ (solid blue) and $C^I_z$ (dashed-dotted green). Inset shows
$C^I_{rr}$ (solid black), $C^I_{nn}$ (dashed orange), and $C^I_{kk}$ (dashed-dotted purple). Left: $I=q\bar{q}$. Dashed red line is $\Delta_{\Omega}^{q\bar{q}}$. Right: $I=gg$. Dashed red line is $\mathcal{C}[\rho^{gg}]=\max(\delta_{\Omega}^{gg},0)$. Vertical black lines represent the critical values $\beta^{\textrm{CH}}_{c}$ (dashed) and $\beta^{\textrm{PH}}_{c}$ (solid) above which there is no CHSH violation and entanglement, respectively.}
\label{fig:CorrelationsAngular}
\end{figure*}

\subsection{Angular integration}\label{subsec:angularint}

We compute first the distributions resulting from averaging over the angular coordinates. We start with the $R$-matrix for each partonic initial state $I$:
\begin{align}\label{eq:CorrelationAzimuthAveraged}
R^I_{\Omega}(M_{t\bar{t}})&=\frac{1}{4\pi}\int\mathrm{d}\Omega~R^I(M_{t\bar{t}},\hat{k})
\end{align}
The invariance under rotations around the beam axis implies that, in the beam basis, the correlation matrix describing $R^I_{\Omega}(M_{t\bar{t}})$ after azimuthal integration is diagonal, $\tilde{C}^I_{ij}=\delta_{ij}\tilde{C}^I_j$, with $\tilde{C}^I_{x}=\tilde{C}^I_{y}\equiv \tilde{C}^I_{\perp}$. Thus, the matrices $R^I_{\Omega}(M_{t\bar{t}})$ are characterized by just $3$ parameters that can be computed analytically from Eqs. (\ref{eq:LOSpinCorrelationsqq}), (\ref{eq:LOSpinCorrelationsgg}): $\tilde{A}^I(M_{t\bar{t}}),\tilde{C}^I_{\perp}(M_{t\bar{t}})$ and $\tilde{C}^I_z(M_{t\bar{t}})$, where $\tilde{A}^I(M_{t\bar{t}})$ is the angular average of $\tilde{A}^I(M_{t\bar{t}},\hat{k})$. Actual spin density matrices $\rho_{\Omega}^{I}(M_{t\bar{t}})$ are obtained from $R^I(M_{t\bar{t}})$ by normalization, $\rho_{\Omega}^{I}(M_{t\bar{t}})=R_{\Omega}^I(M_{t\bar{t}})/4\tilde{A}^I(M_{t\bar{t}})$.

With respect to entanglement, due to the symmetry around the beam axis of $\rho^I_{\Omega}$, the Peres-Horodecki criterion is equivalent now to $\delta_{\Omega}^I>0$ [see Eq.~(\ref{eq:PeresHorodeckiAxialLO}) and ensuing discussion], with
\begin{equation} \label{eq:PeresHorodeckiRotation}
\delta_{\Omega}^I\equiv\frac{-C^I_{z}+|2C^I_{\perp}|-1}{2},
\end{equation}
from where the concurrence reads $\mathcal{C}[\rho^I_{\Omega}]=\max(\delta^I_{\Omega},0)$.

Regarding the CHSH violation, $\rho^I_{\Omega}$ can achieve it \textit{iff}  $\mu^I_\Omega(M_{t\bar{t}})>0$, where now
\begin{equation}
    \mu^I_\Omega\equiv \max \left\{2\left(C^{I}_{\perp}\right)^2,\left(C^{I}_{\perp}\right)^2+\left(C^{I}_z\right)^2\right\}-1
\end{equation}

The angular average of the expectation values in the helicity and the diagonal basis are computed similarly:
\begin{align}\label{eq:ExpectationValueRAngular}
\nonumber &\tilde{C}^I_{ij}(M_{t\bar{t}})
\equiv\frac{1}{4\pi}\int\mathrm{d}\Omega~\tilde{C}^I_{ij}(M_{t\bar{t}},\hat{k})
\\&C^I_{ij}(M_{t\bar{t}})=\frac{\tilde{C}^I_{ij}(M_{t\bar{t}})}{\tilde{A}^I(M_{t\bar{t}})}
\end{align}
We note that the matrix $\tilde{C}^I_{ij}(M_{t\bar{t}})$ is also diagonal in the helicity basis after the angular averaging since $\tilde{C}^I_{kr}(M_{t\bar{t}})$ vanishes at LO due to the odd parity under inversion of the coefficient $\tilde{C}^I_{kr}(M_{t\bar{t}},\hat{k})$.

Moreover, a complementary entanglement signature can be obtained from $\Delta^I_{\Omega}(M_{t\bar{t}})>0$, with
\begin{equation}
    \Delta^I_{\Omega}\equiv\frac{-C_{nn}^{I}(M_{t\bar{t}})+|C_{kk}^{I}(M_{t\bar{t}})+C_{rr}^{I}(M_{t\bar{t}})|-1}{2},
\end{equation}
since, if all substates $\rho^I(M_{t\bar{t}},\hat{k})$ for fixed $M_{t\bar{t}}$ were separable, $\Delta^I(M_{t\bar{t}},\hat{k})\leq 0$ and so would its angular average $\braket{\Delta^I(M_{t\bar{t}})}_\Omega$, which satisfies $\Delta^I_{\Omega}(M_{t\bar{t}})\leq\braket{\Delta^I(M_{t\bar{t}})}_\Omega$. Hence, $\Delta^I_{\Omega}(M_{t\bar{t}})> 0$ necessarily implies that some of the substates $\rho^I(M_{t\bar{t}},\hat{k})$ must be entangled. Another way to put it is that $\Delta^I_{\Omega}(M_{t\bar{t}})>0$ is an entanglement criterion for a fictitious quantum state $\bar{\rho}^I_\Omega(M_{t\bar{t}})$ [see discussion after Eq. (\ref{eq:TotalDensityMatrix})]. Similarly, one can obtain a criterion for CHSH violation $\mathcal{B}^I_{\Omega}>2$ that results from the angular integration of $\mathcal{B}[\rho^I(M_{t\bar{t}},\hat{k})]$, see Eq. (\ref{eq:CHSH}).

Technical details about the analytical calculation of the angular integrals are provided in Appendix~\ref{app:AngularAveraging}.

\subsubsection{$q\bar{q}$ processes}
We obtain for $R^{q\bar{q}}_\Omega(M_{t\bar{t}})$ that
\begin{align}\label{eq:AngularAveragedCorrelationsqq}
\tilde{A}^{q\bar{q}}(M_{t\bar{t}})&=\frac{1}{9}\left[1-\frac{\beta^2}{3}\right]\\
\nonumber \tilde{C}_{\perp}^{q\bar{q}}(M_{t\bar{t}})&=\frac{2}{135}f(\beta)\\
\nonumber \tilde{C}_{z}^{q\bar{q}}(M_{t\bar{t}})&=\frac{1}{9}\left[1-\frac{\beta^2}{3}-\frac{4}{15}f(\beta)\right]\\
\nonumber f(\beta)&\equiv\frac{\left(1-\sqrt{1-\beta^2}\right)^2}{2}
\end{align}
while for the angular averages of the helicity spin correlations we find
\begin{align}\label{eq:AngularAveragedCorrelationsHelicityqq}
\tilde{C}^{q\bar{q}}_{rr}(M_{t\bar{t}})&=\frac{(2-\beta^2)}{27}\\
\nonumber \tilde{C}^{q\bar{q}}_{nn}(M_{t\bar{t}})&=-\frac{\beta^2}{27}\\
\nonumber \tilde{C}^{q\bar{q}}_{kk}(M_{t\bar{t}})&=\frac{1+\beta^2}{27}
\end{align}

These results imply that the quantum state $\rho_{\Omega}^{q\bar{q}}$ is completely separable as
\begin{equation}
\delta_{\Omega}^{q\bar{q}}(M_{t\bar{t}})=\frac{-1+\frac{\beta^2}{3}+\frac{4}{15}f(\beta)}{1-\frac{\beta^2}{3}}<0
\end{equation}

We note that this only means that the angular-averaged state $\rho_{\Omega}^{q\bar{q}}$ is separable, but not that entanglement disappears after the angular average. Indeed,  
\begin{equation}
\Delta_{\Omega}^{q\bar{q}}(M_{t\bar{t}})=\frac{\beta^2}{3-\beta^2}\geq 0
\end{equation}
which means that entanglement is still present in the whole energy range (except exactly at threshold, $\beta=0$). Since in the diagonal basis $C_{+}(M_{t\bar{t}},\hat{k})=1$, its angular average also satisfies $C_{+}(M_{t\bar{t}})=1$, and thus there is CHSH violation for any non-zero $\beta$, $\mathcal{B}^{q\bar{q}}_{\Omega}(\beta)> 2$. Left Fig.~\ref{fig:CorrelationsAngular} displays all the angular-averaged magnitudes for $I=q\bar{q}$.

\subsubsection{$gg$ processes}
We obtain for $R_\Omega^{gg}(M_{t\bar{t}})$ that

\begin{widetext}
\begin{align}\label{eq:AngularAveragedCorrelations}
\tilde{A}^{gg}(M_{t\bar{t}})&=
\frac{1}{192}\left[-59+31\beta^2+(66-36\beta^2+2\beta^4)\frac{\textrm{atanh}(\beta)}{\beta}\right]\\
\nonumber \tilde{C}^{gg}_{\perp}(M_{t\bar{t}})&=
\frac{1-\beta^2}{192}\left[9-16\frac{\textrm{atanh}(\beta)}{\beta}\right]+g(\beta)\\
\nonumber 
\tilde{C}^{gg}_{z}(M_{t\bar{t}})&=\frac{1}{192}\left[-109+49\beta^2+(102-72\beta^2+2\beta^4)\frac{\textrm{atanh}(\beta)}{\beta}\right]-2g(\beta)\\
\nonumber g(\beta)&\equiv\frac{f(\beta)}{96\beta^4}\left[49-\frac{149}{3}\beta^2+\frac{24}{5}\beta^4-(49-66\beta^2+17\beta^4)\frac{\textrm{atanh}(\beta)}{\beta}\right]
\end{align}
while for the helicity spin correlations we find
\begin{align}\label{eq:AngularAveragedCorrelationsHelicity}
\tilde{C}^{gg}_{rr}(M_{t\bar{t}})&=-\frac{1}{192}
\left[87-31\beta^2+66\frac{\frac{\textrm{atanh}(\beta)}{\beta}-1}{\beta^2}-
\left(102-38\beta^2+2\beta^4\right)\frac{\textrm{atanh}(\beta)}{\beta}\right]\\
\nonumber \tilde{C}^{gg}_{nn}(M_{t\bar{t}})&=-\frac{1}{192}
\left[41-31\beta^2-\left(34-36\beta^2+2\beta^4\right)\frac{\textrm{atanh}(\beta)}{\beta}\right]\\
\nonumber \tilde{C}^{gg}_{kk}(M_{t\bar{t}})&=-\frac{1}{192}
\left[-37+31\beta^2-66\frac{\frac{\textrm{atanh}(\beta)}{\beta}-1}{\beta^2}+
\left(66-34\beta^2+2\beta^4\right)\frac{\textrm{atanh}(\beta)}{\beta}\right]
\end{align}

We note that $C^{gg}_{\perp}(M_{t\bar{t}})<0$ for all the energies of interest (the sign crossover is produced only in the ultrarelativistic limit $\beta\approx 0.970$). This implies that for practical purposes we can take $\delta_{\Omega}^{gg}(M_{t\bar{t}})$ as
\begin{equation}
\delta_{\Omega}^{gg}=-\frac{1+\textrm{tr}[\mathbf{C}^{gg}]}{2}=\frac{75-31\beta^2-\left(68-38\beta^2+2\beta^4\right)\frac{\textrm{atanh}(\beta)}{\beta}}{-59+31\beta^2+(66-36\beta^2+2\beta^4)\frac{\textrm{atanh}(\beta)}{\beta}}
\end{equation}
\end{widetext}
From this expression, we compute the critical top velocity $\beta^{\textrm{PH}}_{c}\approx 0.632$ below which the state $\rho^{gg}_{\Omega}$ is still entangled, with the associated critical mass being $M^{\textrm{PH}}_{c}\approx 446~\mathrm{GeV}$. 

On the other hand, since in the entangled region $|C^{gg}_{\perp}|>|C^{gg}_{z}|$, $\rho^{gg}_{\Omega}$ can violate the CHSH inequality \textit{iff}
\begin{equation}\label{eq:CHSHAngular}
    \mu^{gg}_{\Omega}=2\left(C^{gg}_{\perp}\right)^2-1>0
\end{equation}
This condition is satisfied below the critical value $\beta^{\textrm{CH}}_{c}\approx 0.378<\beta^{\textrm{PH}}_{c}$, corresponding to a critical mass $M^{\textrm{CH}}_{c}\approx 374~\mathrm{GeV}$.

We can also reproduce these results from the spin correlations in the helicity basis. Indeed, since 
\begin{widetext}
\begin{equation}
    C^{gg}_{kk}+C^{gg}_{rr}=-\frac{50-\left(36-4\beta^2\right)\frac{\textrm{atanh}(\beta)}{\beta}}{-59+31\beta^2+(66-36\beta^2+2\beta^4)\frac{\textrm{atanh}(\beta)}{\beta}}
\end{equation}
we have that $C^{gg}_{kk}+C^{gg}_{rr}<0$ for $\beta<\beta_\Delta\approx 0.864$, and thus, in the energy range where the state is entangled, $\delta_{\Omega}^{gg}=\Delta_{\Omega}^{gg}=-(1+\textrm{tr}[\mathbf{C}^{gg}])/2$. In fact, this result can be obtained in any orthonormal basis due to the rotational invariance of the trace, $\textrm{tr}[\mathbf{C}^{gg}]=2C^{gg}_{\perp}+C^{gg}_z=C^{gg}_{rr}+C^{gg}_{nn}+C^{gg}_{kk}$, reflecting the symmetry of the spin-singlet state.

However, entanglement is again lost at high energies for $\rho^{gg}_{\Omega}$: even if we use the full $\Delta_{\Omega}^{gg}$,
\begin{equation}
    \Delta_{\Omega}^{gg}=\frac{50-31\beta^2-\left(50-36\beta^2+2\beta^4\right)\frac{\textrm{atanh}(\beta)}{\beta}+\left|25-\left(18-2\beta^2\right)\frac{\textrm{atanh}(\beta)}{\beta}\right|}{-59+31\beta^2+(66-36\beta^2+2\beta^4)\frac{\textrm{atanh}(\beta)}{\beta}}
\end{equation}
we do not get any entanglement signature, in contrast to $q\bar{q}$ processes. Right Fig.~\ref{fig:CorrelationsAngular} displays all the angular-averaged magnitudes for $I=gg$.
\end{widetext}

\subsubsection{General considerations}

We analyze the global results of Fig.~\ref{fig:CorrelationsAngular} for both processes in the light of the 2D plots of Fig.~\ref{fig:ConcurrenceFundamental} for $\rho^{I}(M_{t\bar{t}},\hat{k})$. The entanglement loss in  $\rho_{\Omega}^I(M_{t\bar{t}})$ for both quark and gluon processes arises due to the statistical average over all possible top directions. However, close to threshold, gluon fusion produces a $t\bar{t}$ pair in a spin singlet [see Eq.~(\ref{eq:singletgg})], invariant under rotations, and thus unaffected by the angular average, keeping the entanglement. 

Nevertheless, an entanglement signature can be recovered for $q\bar{q}$ processes by averaging $\Delta^{q\bar{q}}$, since the helicity basis changes accordingly its orientation to produce a constructive sum of the spin correlations. The same technique does not work for the $gg$ channel at high energies because there is no entanglement close to forward production $\Theta=0$, which spoils the high-$p_T$ entanglement when averaging over all top directions.

\begin{figure*}[tb!]
\begin{tabular}{@{}cc@{}}
    \stackinset{l}{31pt}{t}{22pt}{\textcolor{white}{(a)}}
    {\includegraphics[width=\columnwidth]{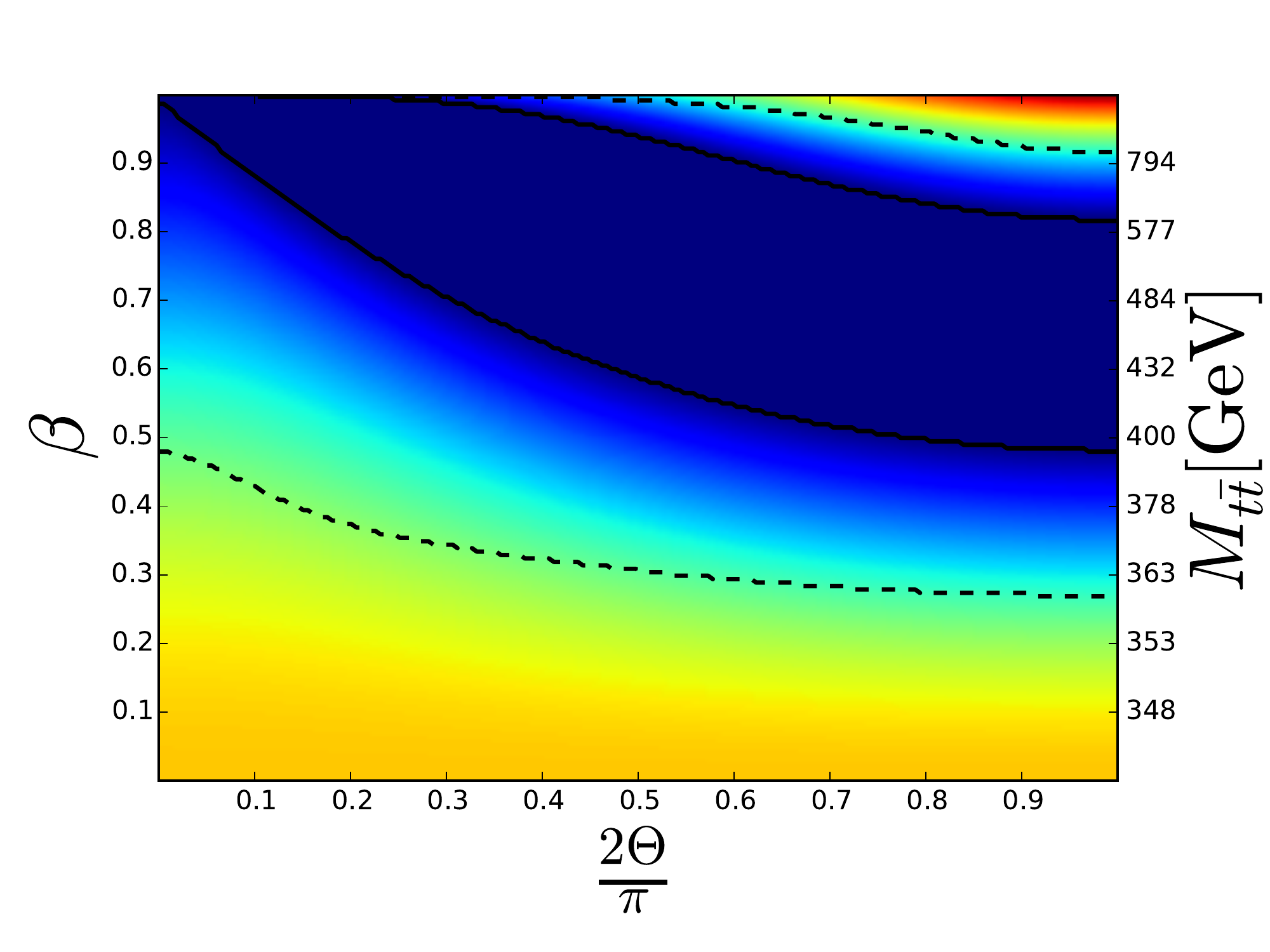}} &  ~~\stackinset{l}{37pt}{t}{25pt}{(b)}
    {\includegraphics[width=\columnwidth]{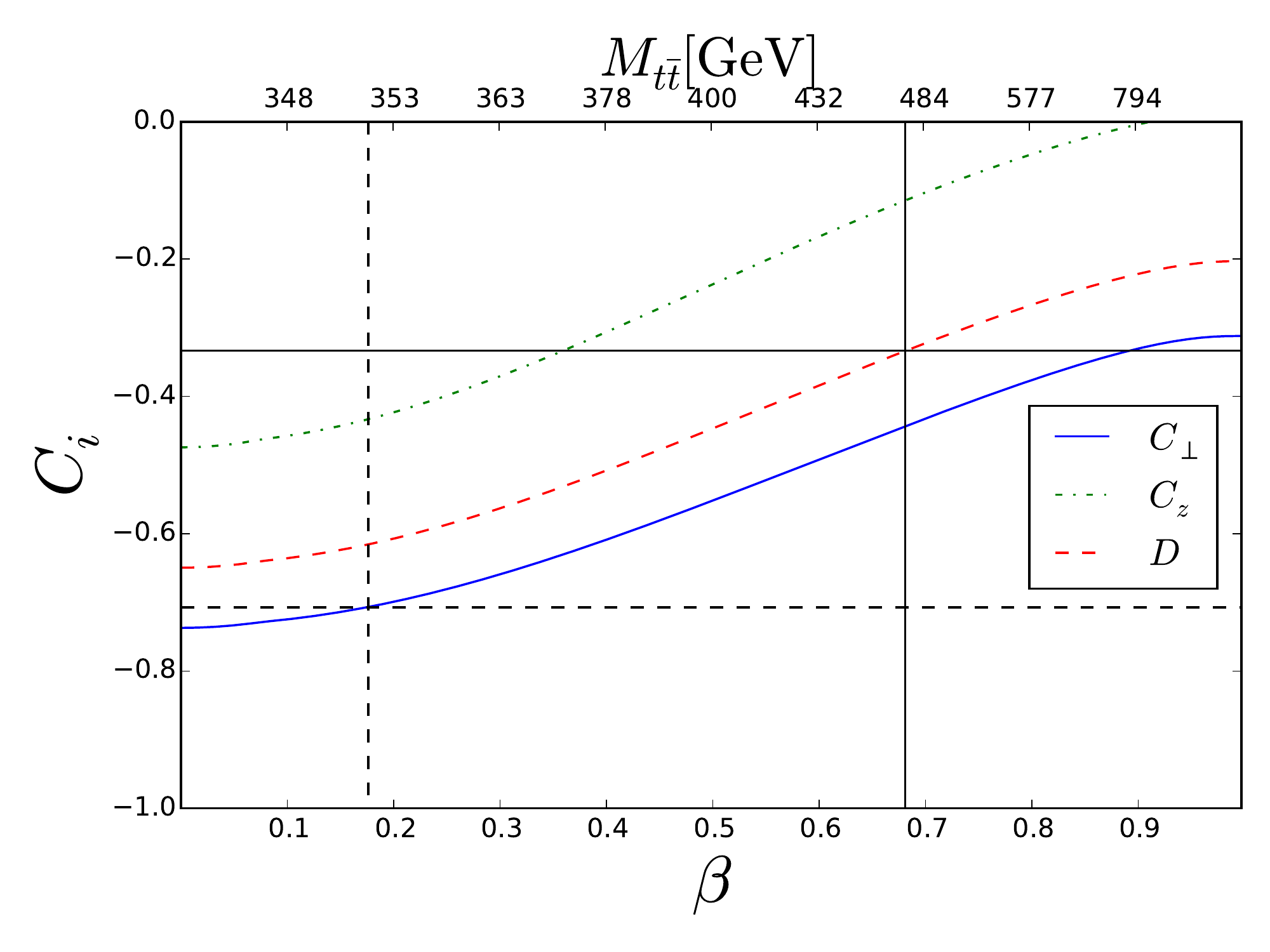}}
    \\ \stackinset{l}{31pt}{t}{22pt}{\textcolor{white}{(c)}}
    {\includegraphics[width=\columnwidth]{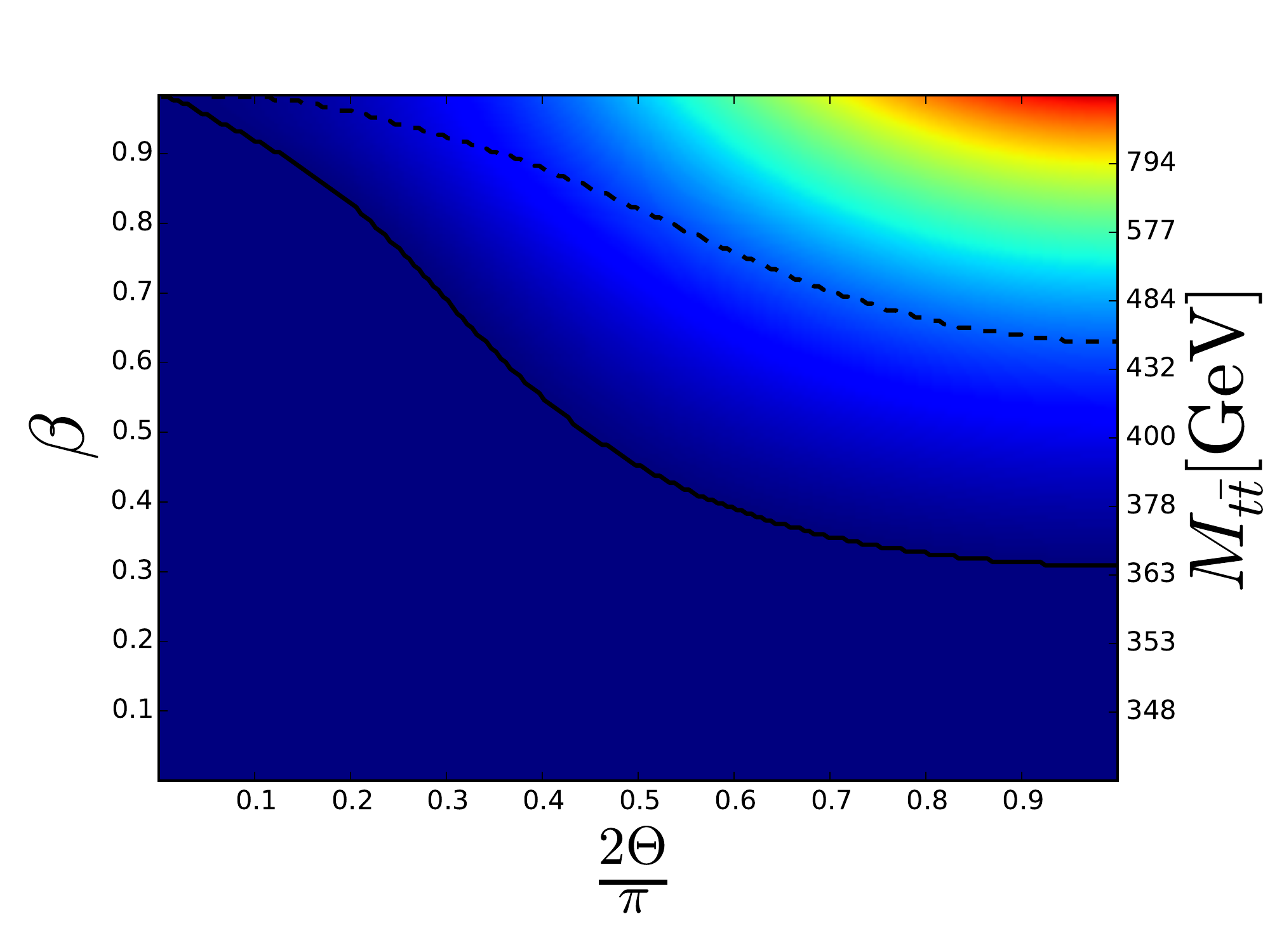}} & \,\,\stackinset{l}{39pt}{t}{25pt}{(d)}
    {\includegraphics[width=\columnwidth]{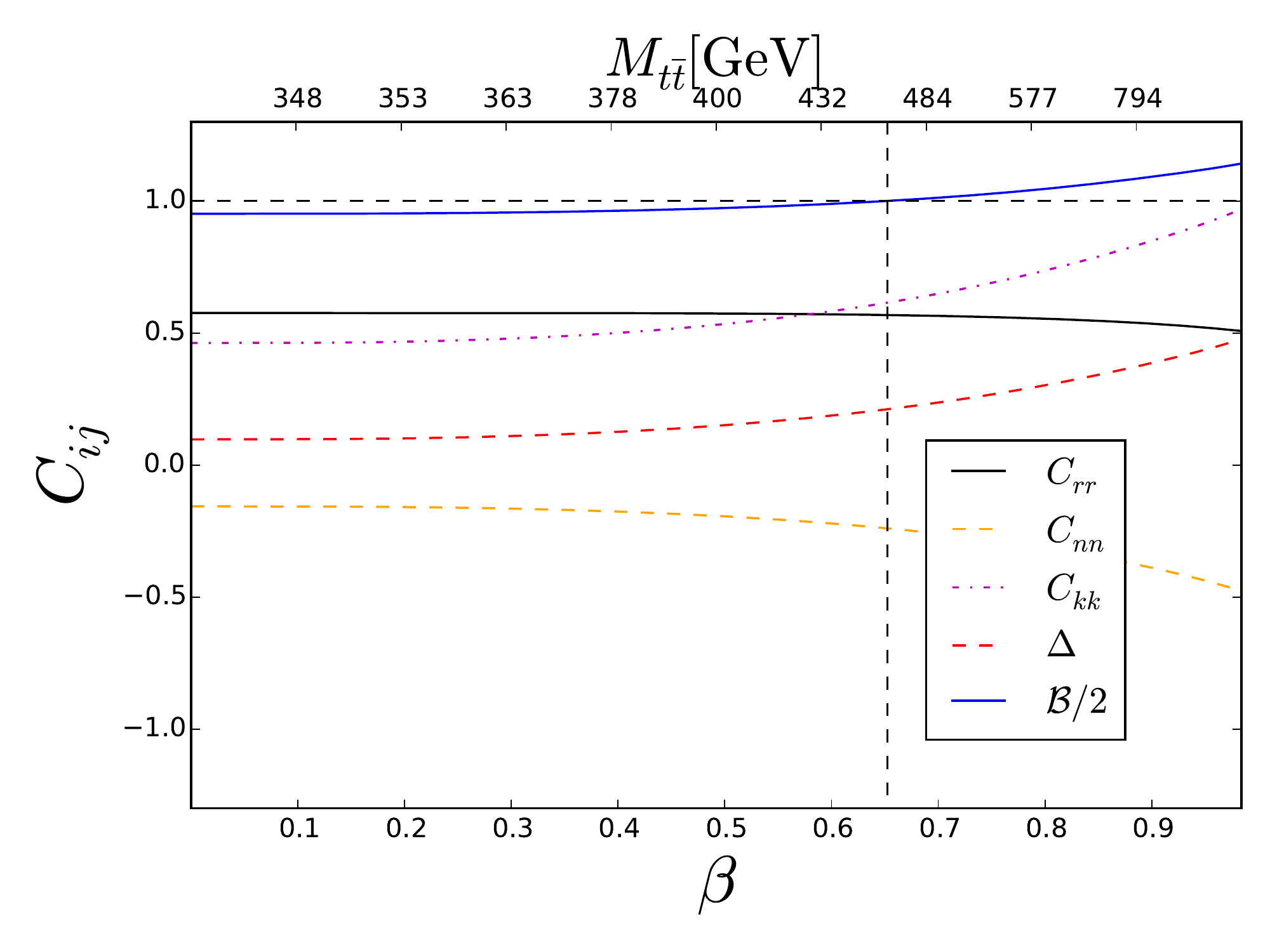}}\\
\end{tabular}
\caption{Upper row: Analysis of $t\bar{t}$ production at the LHC for $\sqrt{s}=13$~TeV. (a) 2D plot of the concurrence as in Figs.~\ref{fig:ConcurrenceFundamental},~\ref{fig:ConcurrenceMIX}. (b) Integrated spin correlations $C_{\perp}$ (solid blue), $C_z$ (dashed dotted green) and $D$ (dashed red) for $\rho(M_{t\bar{t}},\sqrt{s})$. The horizontal and vertical solid black lines signal the entanglement limit $D=-1/3$ while dashed ones signal the CHSH violation limit $|C_{\perp}|=1/\sqrt{2}$.  Lower row: Analysis of $t\bar{t}$ production at the Tevatron for $\sqrt{s}=2$~TeV. (c) Same as (a). (d) Integrated correlations in the spin helicity basis: $C_{rr}$ (solid black), $C_{nn}$ (dashed orange), and $C_{kk}$ (dashed-dotted purple). Dashed red  and solid blue lines show the integrated value of $\Delta$ and $\mathcal{B}/2$, respectively. Horizontal and vertical dashed black lines signal the CHSH violation limit $\mathcal{B}/2=1$.}
\label{fig:CorrelationsIntegrated}
\end{figure*}

\subsection{Mass integration}\label{subsec:massint}

With the help of the angular-averaged $R$-matrix, it is quite simple to compute expectation values in the region $\Pi$ of phase space by integrating in the mass range specified by $\Sigma$. Since mass integration involves luminosity functions (which we stress that do not depend on the $t\bar{t}$ direction), one has necessarily to consider specific hadron processes. 

The angular average of the total matrix $R(M_{t\bar{t}},\hat{k},\sqrt{s})$ describing a realistic hadronic $t\bar{t}$ production process for fixed c.m. energy is 
\begin{equation}
    R_{\Omega}(M_{t\bar{t}},\sqrt{s})=\sum_{I=q\bar{q},gg} L_I(M_{t\bar{t}},\sqrt{s})R_{\Omega}^{I}(M_{t\bar{t}})
\end{equation}
The actual spin density matrix $\rho_{\Omega}(M_{t\bar{t}},\sqrt{s})=R_{\Omega}(M_{t\bar{t}},\sqrt{s})/4\tilde{A}(M_{t\bar{t}},\sqrt{s})$ is computed from its partonic counterparts
$\rho_{\Omega}^{I}(M_{t\bar{t}})$ as in Eq.~(\ref{eq:partonicstates}):
\begin{equation}\label{eq:partonicstatesangular}
\rho_{\Omega}(M_{t\bar{t}},\sqrt{s})=\sum_{I=q\bar{q},gg} w_I(M_{t\bar{t}},\sqrt{s})\rho_{\Omega}^{I}(M_{t\bar{t}})
\end{equation}
where the probabilities $w_I(M_{t\bar{t}},\sqrt{s})$ are now
\begin{equation}\label{eq:partonicweightsangular}
w_I(M_{t\bar{t}},\sqrt{s})=\frac{L_{I}(M_{t\bar{t}},\sqrt{s})\tilde{A}^I(M_{t\bar{t}})}{\sum_{J}L_{J}(M_{t\bar{t}},\sqrt{s})\tilde{A}^J(M_{t\bar{t}})} 
\end{equation}

Finally, by Eq.~(\ref{eq:TotalDensityMatrix}), the density matrix $\rho_\Pi$ reads in terms of these angular-averaged magnitudes as
\begin{align}\label{eq:TotalDensityMatrixMass}
\nonumber \rho_\Pi(\sqrt{s})&=\frac{\sum_I\int_\Sigma\mathrm{d}M_{t\bar{t}}~\frac{\alpha^2_s\beta}{M^2_{t\bar{t}}} L_{I}(M_{t\bar{t}},\sqrt{s})R_{\Omega}^I(M_{t\bar{t}})}
{\sum_{I}\int_\Sigma\mathrm{d}M_{t\bar{t}}~\frac{\alpha^2_s\beta}{M^2_{t\bar{t}}}L_{I}(M_{t\bar{t}},\sqrt{s})4\tilde{A}^I(M_{t\bar{t}})}\\
&=\frac{1}{\sigma_\Sigma}\int_\Sigma\mathrm{d}M_{t\bar{t}}~\frac{\mathrm{d}\sigma}{\mathrm{d}M_{t\bar{t}}}\rho_{\Omega}(M_{t\bar{t}},\sqrt{s})
\end{align}
with
\begin{equation}\label{eq:TotalDensityMatrixMass}
\frac{\mathrm{d}\sigma}{\mathrm{d}M_{t\bar{t}}}=\int\mathrm{d}\Omega~\frac{\mathrm{d}\sigma}{\mathrm{d}\Omega\mathrm{d}M_{t\bar{t}}},~\sigma_\Sigma=\int_\Sigma\mathrm{d}M_{t\bar{t}}~\frac{\mathrm{d}\sigma}{\mathrm{d}M_{t\bar{t}}}
\end{equation}
Similar considerations apply to the mass integration of the angular-averaged expectation values, such as the spin correlations in the helicity basis. 

Because of their illustrative character and their high-experimental relevance, we focus on two particular hadronic processes: $p\bar{p}$ collisions at $\sqrt{s}=2~\textrm{TeV}$ (very close to the actual value $\sqrt{s}=1.96~\textrm{TeV}$ at the Tevatron), and $pp$ collisions at $\sqrt{s}=13~\textrm{TeV}$ (corresponding to the c.m. energy of Run~2 at the LHC). As shown in Fig.~\ref{fig:Criticality}, at the Tevatron $q\bar{q}$ processes dominate while at the LHC $gg$ processes do. The 2D plot of their concurrences is depicted in left column of Fig.~\ref{fig:CorrelationsIntegrated}. We focus on integrating the signal in the two relevant regions for entanglement: close to threshold and at high $p_T$.

\subsubsection{Threshold analysis}

Close to threshold, the angular results of the previous subsection predict that only the $gg$ channel gives rise to entangled $t\bar{t}$ pairs. Therefore, we restrict to the LHC example and take $\Sigma=[2m_t,M_{t\bar{t}}]$, which means that only events in the window $[2m_t,M_{t\bar{t}}]$ are selected, so
\begin{align}\label{eq:TotalDensityMatrixMass}
\rho_\Pi(\sqrt{s})&=\rho(M_{t\bar{t}},\sqrt{s})\\
\nonumber &\equiv\frac{1}{\sigma(M_{t\bar{t}})}\int^{M_{t\bar{t}}}_{2m_t}\mathrm{d}M~\frac{\mathrm{d}\sigma}{\mathrm{d}M}\rho_{\Omega}(M,\sqrt{s})
\end{align}
with $\sigma(M_{t\bar{t}})$ the total integrated cross-section in the same mass window. Since it is computed in terms of the angular-averaged substates $\rho_{\Omega}(M_{t\bar{t}},\sqrt{s})$, the quantum state $\rho(M_{t\bar{t}},\sqrt{s})$ is also characterized by its integrated transverse and longitudinal spin correlations $C_{\perp},C_{z}$, represented in Fig.~\ref{fig:CorrelationsIntegrated}b. The necessary and sufficient condition for $\rho(M_{t\bar{t}},\sqrt{s})$ to be entangled is that the integrated value $\delta(M_{t\bar{t}},\sqrt{s})$ satisfies $\delta(M_{t\bar{t}},\sqrt{s})>0$. 
Remarkably, this implies that a directly measurable entanglement witness $W$, which satisfies $W<0$ only for entangled states~\cite{Terhal_2000}, is provided by the observable $D$ of Eq.~(\ref{eq:DCrossSection}),
\begin{equation}\label{eq:Witness}
    W\equiv D+1/3
\end{equation}
The concurrence is also readily computed from $D$ as $\mathcal{C}[\rho]=\max(-1-3D,0)/2$. The integrated value of $D$ is represented in Fig.~\ref{fig:CorrelationsIntegrated}b, along with the critical value $\beta^{\textrm{PH}}_{c}(\sqrt{s})$ (marked by horizontal and vertical solid black lines) below which $W<0$.

Regarding the CHSH violation, its presence can be also signaled by just one parameter, $C_{\perp}(M_{t\bar{t}},\sqrt{s})$, as given by the condition $\mu(M_{t\bar{t}},\sqrt{s})=2\left[C_{\perp}(M_{t\bar{t}},\sqrt{s})\right]^2-1>0$. The critical value $\beta^{\textrm{CH}}_{c}(\sqrt{s})$ below which there is CHSH violation, $|C_{\perp}(\beta^{\textrm{CH}}_{c},\sqrt{s})|=1/\sqrt{2}$, is marked in Fig.~\ref{fig:CorrelationsIntegrated}b by horizontal and vertical dashed black lines.

\begin{figure}[tb!]\includegraphics[width=\columnwidth]{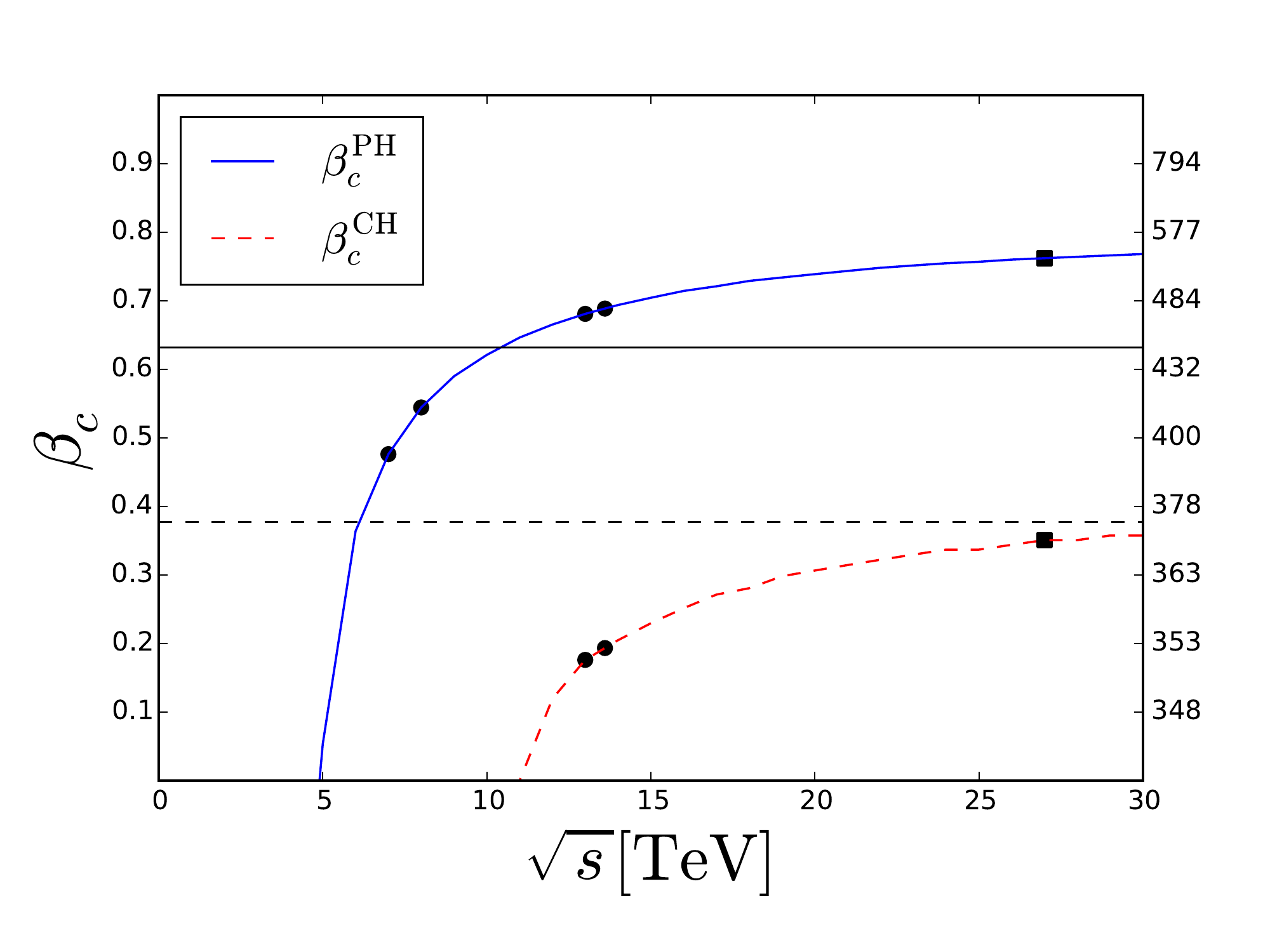}
     \caption{Critical values $\beta_c^{\textrm{PH}}(\sqrt{s})$ (solid blue) and $\beta_c^{\textrm{CH}}(\sqrt{s})$ (dashed red) below which entanglement and CHSH violation can be observed, respectively, close to threshold as a function of the c.m. energy $\sqrt{s}$ for $pp$ collisions. The horizontal lines mark the critical values $\beta_c^{\textrm{CH}}$ (dashed) and $\beta_c^{\textrm{PH}}$ (solid) for $\rho^{gg}_{\Omega}$. Black markers indicate values for c.m. energies at the LHC. Circles: Run~1 ($\sqrt{s}=7,8~\textrm{TeV}$), Run~2 ($\sqrt{s}=13~\textrm{TeV}$) and Run~3 ($\sqrt{s}=13.6~\textrm{TeV}$). Squares: Possible upgrade of the LHC ($\sqrt{s}=27~\textrm{TeV}$).}
     \label{fig:CriticalityLHC}
\end{figure}

We compute the critical values $\beta_c^{\textrm{CH}}(\sqrt{s})\leq \beta_c^{\textrm{PH}}(\sqrt{s})$ as a function of the c.m. energy and represent them in Fig.~\ref{fig:CriticalityLHC}. We compare these critical values with those for the angular-averaged substates $\rho^{gg}_{\Omega}(M_{t\bar{t}})$, $\beta_{c}^{\textrm{CH}}\approx 0.378<\beta_{c}^{\textrm{PH}}\approx 0.632$ (horizontal dashed and solid lines). We observe that, for sufficiently large c.m. energies, the critical value $\beta_c^{\textrm{PH}}(\sqrt{s})$ exceeds its angular-averaged counterpart, even though $\rho(M_{t\bar{t}},\sqrt{s})$ contains some mixing with $q\bar{q}$ processes that reduces the entanglement (see lower part of right Fig.~\ref{fig:ConcurrenceFundamental} and Fig.~\ref{fig:CorrelationsIntegrated}a). This exceeding arises from the fact that the total quantum state $\rho(M_{t\bar{t}},\sqrt{s})$ is a convex sum of the substates $\rho^I_{\Omega}(M_{t\bar{t}})$. Therefore, one needs higher energies to include a sufficient amount of separable states in order to dilute the entanglement. Nevertheless, if the integration window $\Sigma$ was entirely placed in the region of separability of $\rho^I_{\Omega}(M_{t\bar{t}})$, no entanglement would be observed. In contrast, for the CHSH violation, $\beta_c^{\textrm{CH}}(\sqrt{s})$ is always below its angular-averaged counterpart, at least within the range of energies considered. This is because of the critical effect introduced by the mixing with $q\bar{q}$ processes at threshold, see Fig.~\ref{fig:Criticality}, which cannot be overcome by the opposite effect of mass integration due to the lower size of the CHSH-violating window. Indeed, for the considered energy of Run~2, $\sqrt{s}=13~\textrm{TeV}$, the weight of the $gg$ channel at threshold is barely above the critical value $1/\sqrt{2}$, Fig.~\ref{fig:Criticality}b.

\subsubsection{High-$p_T$ analysis}

For high $p_T$ at the Tevatron, entanglement is not present in the two-qubit quantum state $\rho_\Pi$ but is instead signaled by the integrated value of $\Delta_{\Omega}$.  Indeed, Fig.~\ref{fig:CorrelationsIntegrated}c suggests to choose now the mass window $\Sigma$ as $[M_{t\bar{t}},\sqrt{s}]$: 
\begin{equation}\label{eq:DeltaIntegrated}
\Delta(M_{t\bar{t}},\sqrt{s})=\frac{1}{\bar{\sigma}(M_{t\bar{t}})}\int^{\sqrt{s}}_{M_{t\bar{t}}}\mathrm{d}M~\frac{\mathrm{d}\sigma}{\mathrm{d}M}\Delta_{\Omega}(M,\sqrt{s})
\end{equation}
with $\bar{\sigma}(M_{t\bar{t}})=\sigma-\sigma(M_{t\bar{t}})$. We define in a similar fashion the integrated value of the CHSH violation, $\mathcal{B}(M_{t\bar{t}},\sqrt{s})$.

The values of $\Delta(M_{t\bar{t}},\sqrt{s})$ and $\mathcal{B}(M_{t\bar{t}},\sqrt{s})/2$, along with the integrated helicity spin correlations, are represented in Fig.~\ref{fig:CorrelationsIntegrated}d. Interestingly, even though the $t\bar{t}$ pairs produced close to threshold are not entangled (see lower Fig.~\ref{fig:CorrelationsIntegrated}c), the integration in the mass range makes $\Delta(M_{t\bar{t}},\sqrt{s})>0$ in the whole mass spectrum, similarly to the increase of the critical value of $\beta_c^{\textrm{PH}}(\sqrt{s})$ for $gg$ processes. However, this does not apply to the CHSH violation, which can only be observed now at high energies $\beta> \beta^{\textrm{CH}}_c\approx 0.652$.

Within the angular-averaged scheme considered here, no entanglement can be detected for $gg$ processes at high $p_T$.  We note, however, that entanglement and CHSH violation can be observed at the LHC for high $p_T$ if one also introduces a cut in the top direction, as suggested by Fig.~\ref{fig:CorrelationsIntegrated}a and implemented in Refs.~\cite{Fabbrichesi2021,Severi2022}.

\section{Experimental remarks}\label{sec:experimental}

We complement here the theoretical predictions for observables of the previous section with some experimental remarks about the actual measurement scheme in high-energy colliders, and the potential statistical significance of the discovery. Nevertheless, a dedicated analysis of the expected experimental sensitivity of each prediction is beyond the scope of the present work.

\subsection{Entanglement detection}

The detection of entanglement is more delicate than it could be naively expected from Figs.~\ref{fig:ConcurrenceFundamental},~\ref{fig:ConcurrenceMIX} since, even though entanglement is present in a wide region of phase space, one has to select carefully the phase space regions where the signal is integrated in order to obtain an entanglement signature. In fact, the recent measurement of the CMS collaboration at the LHC with the data of Run~2~\cite{Sirunyan2019}, with no restrictions in phase space, yielded no entanglement signature, $D=-0.237\pm 0.011>-1/3$.

Our analysis reveals that there are two main regions where entanglement can be in principle detected: close to threshold and for high transverse momentum.

\subsubsection{Threshold}

Close to threshold, as shown in Fig.~\ref{fig:Criticality}, the only experiment with high enough c.m. energy to give rise to entanglement is the LHC. The idea here is to measure $D$ from the cross-section of Eq.~(\ref{eq:DCrossSection}), applying an upper cut in the mass spectrum for the integrated signal. Although our calculation is restricted to LO, in general, the criterion $W=D+1/3<0$ still provides a sufficient condition for the entanglement of $\rho(M_{t\bar{t}},\sqrt{s})$ without any assumption on its specific form [see Eq.~(\ref{eq:DeltaTraceGeneralized})].

The entanglement witness $W<0$ can be rewritten as the violation of a Cauchy-Schwarz inequality, a typical entanglement signature in other fields such as quantum optics, condensed matter, or analog Hawking radiation~\cite{Walls2008,Wolk2014,deNova2014,deNova:2015lva}. This is proven by using the $P$-representation of Eq.~(\ref{eq:PRepresentation2Qubit}). If a state is separable, $P(\mathbf{n}_A,\mathbf{n}_B)>0$, and then
\begin{align}
\nonumber |\textrm{tr}~\mathbf{C}|&=\left|\braket{\mathbf{\sigma}\cdot \mathbf{\sigma}}\right|\\
\nonumber &=\left|\int\mathrm{d}\Omega_A\mathrm{d}\Omega_B~P(\mathbf{n}_A,\mathbf{n}_B)\mathbf{n}_A\cdot\mathbf{n}_B\right|\\
\nonumber &\leq \int\mathrm{d}\Omega_A\mathrm{d}\Omega_B~P(\mathbf{n}_A,\mathbf{n}_B)\left|\mathbf{n}_A\cdot\mathbf{n}_B\right|\\
&\leq \int\mathrm{d}\Omega_A\mathrm{d}\Omega_B~P(\mathbf{n}_A,\mathbf{n}_B)=1
\end{align}
Hence, the directly measurable observable $D$ represents in the range of values $-1\leq D<-1/3$ a genuine non-classical feature, which can be qualitatively understood from the fact that the classical average of the scalar product of two vectors with unit length is never larger than one, $3|D|=|\textrm{tr}~\mathbf{C}|=|\braket{\mathbf{\sigma}\cdot \mathbf{\sigma}}|\leq 1$. A similar entanglement criterion based on the trace of the correlation matrix was derived for Heisenberg spin chains~\cite{Schliemann2003}.

A dedicated analysis of the entanglement discovery was performed in Refs.~\cite{Afik2021,Severi2022}, finding that high-statistical significance can be expected, potentially more than $5$ statistical deviations ($5\sigma$), the standard candle for discovery in high-energy physics.

Figures~\ref{fig:Criticality},~\ref{fig:CriticalityLHC} show that the effect of increasing the c.m. energy saturates for large energies. Therefore, we do not expect that future hadron colliders with higher collision energies, such as possible upgrades of the LHC or the FCC, will strongly increase the entanglement in $t\bar{t}$ production. However, we do expect that a larger data collection will enhance the sensitivity of the measurements by reducing the statistical uncertainties. For example, the LHC is expected to collect about $20-30$ times more events with respect to the currently recorded data~\cite{ZurbanoFernandez:2020cco}, a possible upgrade of the LHC to access higher energies (HE-LHC) is expected to collect about $100$ times more events~\cite{FCC:2018bvk}, and the FCC is expected to collect about $140-220$ times more events~\cite{FCC:2018vvp}. Another possibility to increase entanglement is to enhance the contribution from the $gg$ channel by further rejecting events from the $q\bar{q}$ channel, as recently proposed in Ref.~\cite{Aguilar2022}.

\subsubsection{High $p_T$}

For high $p_T$, both partonic processes $q\bar{q}$ and $gg$ give rise to the same triplet state due to orbital angular momentum dominance. Nevertheless, because the procedure used here averages over all possible top directions, only $q\bar{q}$ processes yield an entanglement signature in our scheme. A detailed study of the statistical significance of an entanglement detection for high $p_T$ at the LHC has been already provided in Ref.~\cite{Severi2022}, predicting also a potential $5\sigma$ discovery.

Regarding the Tevatron, entanglement is detected from the integrated value of $\Delta$, which is in general a sufficient condition for the presence of entanglement [see Eq.~(\ref{eq:DeltaGeneralized})]. There are important conceptual differences between the detection of entanglement close to threshold (involving the measurement of $D$) and that at high $p_T$ (involving the measurement of $\Delta$). First, close to threshold, entanglement is detected from one single magnitude, $D$, while $\Delta$ requires the measurement of $3$ magnitudes, which are the diagonal spin correlations in the helicity or the diagonal basis (we note that a recent work has proposed to measure $\Delta$ also from a single parameter~\cite{Aguilar2022}). Second, $D+1/3<0$ reveals that the \textit{physical} two-qubit quantum state $\rho_\Pi$ is entangled, while $\Delta>0$ signals entanglement in the \textit{fictitious} quantum state $\bar{\rho}_\Pi$, and so in the total spin-momentum quantum state $\rho_{\textrm{T},\Pi}$.

At the Tevatron, the relatively low c.m. energy and data recorded yielded high uncertainties on spin-correlation observables~\cite{Aaltonen2010,Abazov2011ka,Abazov2015psg}. Therefore, by considering typical uncertainties of the carried measurements, we do not expect in principle to find there an entanglement signature with high-statistical significance.

\subsubsection{Lorentz invariance}

Finally, we would like to discuss the possible role of the reference frame choice on the entanglement detection. While our calculations assume that the $t\bar{t}$ momenta for each production process are well-defined and therefore their spin-entanglement is Lorentz invariant, in realistic situations their wave function is necessarily described by wave packets with finite width, so their entanglement is no longer independent of the reference frame. However, within the current experimental scheme, the $t\bar{t}$ spins are by definition measured in their respective rest frames, since these are precisely the frames where the lepton directions $\mathbf{\ell}_{\pm}$ are defined in Eq.~(\ref{eq:DecaySpinDensityMatrix}).

\subsection{Quantum tomography}

A protocol for the quantum tomography of the $t\bar{t}$ pair was developed in Ref.~\cite{Afik2021}, based on the fact that the experimental procedure of Section~\ref{subsec:Experiment} allows to measure the integrated value of the spin correlations and spin polarizations. Thus, by implementing a cut in the invariant mass spectrum, one can extract all the parameters characterizing the total quantum state $\rho(M_{t\bar{t}},\sqrt{s})$ of Eq.~(\ref{eq:TotalDensityMatrixMass}). As explained, such a reconstruction requires the use of a fixed orthonormal basis in space, like the beam basis.

We extend here the proposed quantum tomography protocol to more general situations. First, we note that the protocol is not restricted to the state $\rho(M_{t\bar{t}},\sqrt{s})$, but it is rather valid for any arbitrary integrated quantum state $\rho_\Pi$ by measuring all the $15$ parameters $B^{\pm}_{i},C_{ij}$ in the beam basis. We can further simplify the process if we integrate over the azimuth around the beam axis, since then rotational invariance is recovered and thus, at LO, only $2$ parameters are needed for quantum tomography: the integrated transverse and longitudinal spin correlations $C_{\perp},C_{z}$. In general, by only assuming symmetry around the beam axis, the quantum tomography of the $t\bar{t}$ pair requires the measurement of just $4$ parameters, $B^{\pm}_{z}, C_{\perp}, C_z$, with $B^{\pm}_{z}$ the spin polarizations along the beam axis. Nevertheless, from the recent CMS measurement~\cite{Aaboud2019hwz}, the values of these polarizations are still expected to be quite small. A summary of the parameters needed for the quantum tomography of the $t\bar{t}$ pair is presented in Table~\ref{table}. 

\begin{table}[h]
\begin{tabular}[c]{|c|c|c|}
\hline
Assumption & Coefficients & \#Parameters\\
\hline
Symmetry + LO & $C_{\perp},C_{z}$ & 2 \\
\hline
Symmetry & $B^{\pm}_{z},C_{\perp},C_{z}$ & 4 \\
\hline
None & $B^{\pm}_{i},C_{ij}$ & 15 \\
\hline
\end{tabular}
\caption{Summary of the parameters needed to be measured in order to perform the quantum tomography of the $t\bar{t}$ pair for different assumptions on the form of $\rho_\Pi$. ``Symmetry'' denotes symmetry around the beam axis, obtained after integration over the azimuth in the beam basis.}
\label{table}
\end{table}

Moreover, since the quantum tomography protocol only depends on the decay properties of the $t\bar{t}$ pair and not on the nature of the production process, it can be extended to any $t\bar{t}$ production mechanism, such as positron-electron ($e^{+}e^{-}$) collisions. Furthermore, even though we have focused on the case of a dileptonic decay of the $t\bar{t}$ pair,  Eqs. (\ref{eq:DecaySpinDensityMatrix}), (\ref{eq:LeptonicCrossSectionTotal}) are valid for any pair of detectable decay products by just replacing $\mathbf{\ell}_{\pm}$ and $\kappa_{\ell},\bar{\kappa}_{\ell}$ by the corresponding flight directions in the parent rest frames and spin analyzing powers. We conclude by noting that the proposed quantum tomography protocol for the $t\bar{t}$ pair goes beyond the general approach to high-energy processes presented in Ref.~\cite{Martens:2017cvj}.

\subsection{CHSH violation}
Since entanglement is a necessary condition for the violation of CHSH inequalities, CHSH violations are generally expected to be measured with lower statistical significance than entanglement~\cite{Severi2022}. This is clearly seen by examining  Fig.~\ref{fig:CriticalityLHC}, where we see that the upper critical value for a CHSH violation is well below that for entanglement. Remarkably, in a similar fashion to entanglement, the CHSH violation close to threshold is also inferred from the value of just one parameter, $C_{\perp}$. Nevertheless, one can always measure the full CHSH violation in a more conventional way through Eq.~(\ref{eq:CHSHOriginalSpin}). Due to the rotational invariance in the perpendicular plane to the beam, the experimental scheme is similar to the usual case of Bell inequalities with spin-singlet states, and one chooses four vectors $\mathbf{a}_1,\mathbf{a}_2,\mathbf{b}_1,\mathbf{b}_2$ in that plane satisfying $\mathbf{a}_1\cdot \mathbf{a}_2=\mathbf{b}_1\cdot \mathbf{b}_2=0$ and $\mathbf{a}_1\cdot \mathbf{b}_1=\mathbf{a}_2\cdot \mathbf{b}_2=\mathbf{a}_2\cdot \mathbf{b}_1=-\mathbf{a}_1\cdot \mathbf{b}_2=1/\sqrt{2}$. Since the correlation matrix should be isotropic in this plane, this experimental scheme is equivalent to measure $|C_{\perp}|>1/\sqrt{2}$. We note that our choice of spin directions is fixed \textit{a priori}, so no statistical bias is introduced~\cite{Severi2022}.

Regarding the statistical significance of the CHSH violation close to threshold, we expect it to be much lower than that of entanglement. However, even if a significant detection is not achievable with the current Run~2 data, it could be performed with the increased c.m. energies (as seen from Fig.~\ref{fig:CriticalityLHC}) and larger amount of data provided by future runs and/or colliders. At the Tevatron, since entanglement detection already seems quite challenging, we do not expect to achieve a statistically significant observation of a CHSH violation.

As for the case of entanglement, the scheme presented here does not allow to observe CHSH violation at the LHC at high $p_T$. Nevertheless, a thorough analysis of the potential detection of CHSH violation in this regime and its statistical significance is presented in Refs.~\cite{Fabbrichesi2021,Severi2022}.

So far, we have discussed the violation of the CHSH inequality of Eq.~(\ref{eq:CHSHOriginalSpin}). However, a measurement of a genuine violation of Bell theorem implies much more than just measuring some linear combination of spin correlations above certain critical value, like in the case of entanglement. In order to fully rule out a local hidden-variable model, one needs to perform a loophole-free test that ensures that the experimental setup satisfies all the hypotheses of Bell theorem. Loophole-free violations of Bell inequalities were measured for the first time in 2015~\cite{Hensen:2015ccp,Giustina2015}, and only in 2018 the so-called ``free will'' loophole was completely closed~\cite{BigBell2018}.

Consequently, a careful analysis of the measurement process in a collider is in order. Due to its illustrative character, we examine the paradigmatic case of Run~2 at the LHC. Run~2 consisted of colliding a bunch of protons every $25$~ns, yielding typically 60-80 collisions per bunch~\cite{ATLAS:2020esi}. The products of these collisions are recorded by different detectors surrounding the $pp$ interaction point that trace their direction and momentum/energy. This process goes on for several years (Run~1 went from 2009 to 2012, and Run~2 went from 2015 to 2018) to collect sufficient amount of statistics.

As a result, in our specific case, the experiments of Alice and Bob should be regarded as the detector recording of the $\ell^+\ell^-$ pair resulting from a $t\bar{t}$ decay. First, one needs to ensure that the detection events are causally disconnected. This in principle could be done, since the lepton pair is typically ultrarelativistic due to the large top mass, and in addition causally-connected detection events can be rejected after reconstruction of the lepton momenta. 

However, a major problem arises due to the fact that the direction of the spin measurements cannot be controlled, since we only detect the lepton directions. In fact, technically, one does not even have access to the $t\bar{t}$ spins, but rather infers their expectation values since they are correlated with the lepton directions. We stress that Eq.~(\ref{eq:DecaySpinDensityMatrix}) is only valid after integrating out all the remaining degrees of freedom of the decay products, and cannot be used on an event by event basis. In other words: there is not a measurement setting that yields a $\pm 1$ in each detection event or that can be controlled by Alice or Bob. Therefore, the free-will loophole cannot be closed and even the specific measurement setting of the CHSH inequality is not achievable; only once the spin correlation matrix has been extracted from the fit of the angular differential cross-section of the decay products, one can aim at measuring a CHSH violation. Moreover, many of the events are not useful for the analysis, which gives rise to the so-called detection loophole. 

The emergence of all these loopholes is quite natural because high-energy colliders were not specifically designed for testing Bell inequalities. As a result, in a high-energy collider, we can only detect a \textit{weak} violation of the CHSH inequality, in the sense that some loopholes can never be closed. Related discussions on the validity of Bell tests in high-energy colliders can be found in Refs.~\cite{Fabbrichesi2021,Severi2022,Barr:2021zcp}.

\section{Conclusions and outlook}\label{sec:conclusions}

In this work, we have provided the general framework to study the quantum state of $t\bar{t}$ pairs produced in QCD processes. We have discussed that, due to the nature of the measurement scheme, the most general quantum state that can be probed in a collider is fully determined by the production spin density matrix. This has to be necessarily a mixed state since a) only momentum measurements are carried out in a collider and b) one has to average over the internal degrees of freedom (like spin or color) of the initial state.

We have analyzed $t\bar{t}$ production for the most elementary QCD partonic reactions, extending the work of Ref.~\cite{Afik2021} to also include the analysis of CHSH violation. For $q\bar{q}$ processes, entanglement and CHSH violation are equivalent conditions, and are present in whole phase space. For $gg$ processes, entangled $t\bar{t}$ pairs are produced at threshold in a spin-singlet state, and at high $p_T$ in a spin-triplet state, both violating the CHSH inequality. Remarkably, all these features can be essentially understood in terms of basic conservation laws of angular momentum, without invoking the specific details of QCD interactions.

We have shown that, at least at LO, any $t\bar{t}$ production from a real hadronic process can be written in terms of these basic building blocks through the luminosity functions, which determine the probability of a certain partonic reaction at a given c.m. energy in terms of PDF. In particular, we have focused on $pp$ and $p\bar{p}$ collisions, which are those carried out at the LHC and the Tevatron, respectively. We have performed a detailed analysis on how the $t\bar{t}$ quantum state depends on the c.m. energy of the collisions, finding that at low energies the $q\bar{q}$ channel dominates, with the $gg$ contribution increasing with the c.m. energy for both types of collisions. For the LHC, the $gg$ channel already dominates at the energies of its first run. For the Tevatron, however, due to its relatively low-energy operating point, $q\bar{q}$ processes dominate. Thus, both colliders represent a perfect example of each elementary QCD process.

Thinking about potential experimental realizations, we have proposed a number of realistic observables for the characterization of the $t\bar{t}$ quantum state that provide signatures of entanglement and CHSH violation. Interestingly, at the LHC these signatures are obtained by the measurement of a single magnitude: for entanglement, the trace of the correlation matrix is a good entanglement witness, as already predicted in Ref.~\cite{Afik2021}, while CHSH violation can be signaled by measuring the spin correlations in the transverse plane to the beam.

Finally, we analyze in detail the experimental implementation of these ideas. We explicitly show that an entanglement measurement at the LHC represents the violation of a Cauchy-Schwarz inequality. Regarding the Tevatron, we find that a statistically significant observation of entanglement seems quite challenging due to the relatively large expected uncertainties. 

We also extend the quantum tomography protocol developed in Ref.~\cite{Afik2021} to more general quantum states, arguing that it can be applied in general to any $t\bar{t}$ quantum state. In particular, if one averages over the beam axis, rotational symmetry is recovered and the quantum tomography can be implemented from the measurement of just $4$ parameters, related to the transverse and longitudinal spin correlations, and to the longitudinal spin polarizations. Furthermore, since the protocol does not depend on the specific production process, it can be extended to any $t\bar{t}$ production mechanism, like electroweak production from $e^+e^-$ collisions.

Regarding the CHSH violation, since it is a stronger requirement than entanglement, its statistical significance is expected to be lower. Moreover, we argue that, due to the nature of the detection process, only weak violations of Bell inequalities can be measured in a high-energy collider, since some loopholes, like those related to the free-will or to the detection efficiency, cannot be closed. This is not surprising: after all, high-energy colliders were not designed to test Bell inequalities.

From a quantum information perspective, top quarks allow to export fundamental concepts in quantum information, such as entanglement, CHSH violation or quantum tomography, to the high-energy field. This opens the prospect of using high-energy colliders to study quantum information problems at the highest-energy scale available. The genuine relativistic behavior, the exotic character of the interactions and symmetries involved, and the fundamental nature of this environment make it especially attractive for such purpose. Indeed, the detection of entanglement or CHSH violation in $t\bar{t}$ pairs would represent their highest-energy detections ever, many orders of magnitude above standard laboratory setups. Another interesting experiment is the implementation of the quantum tomography of the $t\bar{t}$ pair, which could be used for instance to measure quantum discord~\cite{Ollivier2001,Afik2022Discord}. 

A very interesting perspective is provided by future colliders, where $t\bar{t}$ pairs are expected to be produced from $e^{+}e^{-}$ collisions, such as the FCC-$ee$~\cite{Blondel:2019jmp}.
In particular, in future linear $e^{+}e^{-}$ colliders such as the International Linear Collider (ILC)~\cite{Barklow:2015tja} and the the Compact Linear Collider (CLIC)~\cite{CLIC:2016zwp,CLICdp:2018cto}, spin degrees of the initial $e^{+}e^{-}$ state can be controlled to a large extent, in contrast to the QCD production discussed in this work. These colliders are then revealed as quite promising scenarios to study quantum information problems, where the techniques of this work can be straightforwardly adapted. Alternative candidates to top quarks for the study of quantum information problems are gauge bosons arising from Higgs decays and $\tau$ leptons~\cite{Barr:2021zcp,Barr:2022wyq,Fabbrichesi2022}.

From the high-energy perspective, the introduction of quantum information concepts can provide new relevant observables in the field. For example, entanglement measurements can help to understand the underlying mechanism of a certain production process. A very intriguing extension of this work is to explore New Physics beyond the Standard Model by measuring the quantum state of the $t\bar{t}$ pair, comparing the experimental results with the predictions of the different theories available. Indeed, some recent works already address the possibility of using entanglement to trace signatures of New Physics~\cite{Aoude2022,Fabbrichesi2022}.

We conclude by stressing that the work has been fully developed within a genuine quantum information framework, since once the production spin density matrix and the luminosity functions are computed by the theory of high-energy physics, all the calculations are reduced to study convex combinations of two-qubit quantum states using standard tools of quantum information theory. Our work then provides a simple and general approach for non-particle physicists to the quantum information aspects of high-energy colliders, a fascinating arena to study the foundations of quantum mechanics.

\acknowledgements

We thank E.~Madge, M.~Vos, F. Sols, G.~Sierra and J.~J.~Garc\'ia-Ripoll for valuable comments. JRMdN acknowledges funding from European Union's Horizon 2020 research and innovation programme under the Marie Sk\l{}odowska-Curie grant agreement No 847635, and also from Grant FIS2017-84368-P from Spain's MINECO.

\appendix

\section{$P$-representation and coherent states for qudits}\label{app:Coherent}

We review here the main properties of the $P$-representation for qudits introduced in Ref.~\cite{Giraud2008}. We describe a Hilbert space of dimension $N$ as an irreducible representation of $\text{SU(2)}$ with angular momentum $j$ such that $N=2j+1$. The qudit basis is relabeled as $\ket{j,m}$, with $m=-j,-j+1\ldots,j$ the eigenvalues of the $z$-component of the angular momentum, $J_z\ket{j,m}=m\ket{j,m}$. Angular momentum coherent states $\ket{j,\mathbf{\hat{n}}}$ are defined as those simultaneously eigenstates of $\mathbf{J}^2$ and $\mathbf{n}\cdot \mathbf{J}$, i.e., $\mathbf{J}^2\ket{j,\mathbf{\hat{n}}}=j(j+1)\ket{j,\mathbf{\hat{n}}}$ and $\mathbf{n}\cdot \mathbf{J}\ket{j,\mathbf{\hat{n}}}=j\ket{j,\mathbf{\hat{n}}}$, where we work in the usual orthonormal basis in spherical coordinates $\{\mathbf{\hat{n}},\mathbf{\hat{n}}_\theta,\mathbf{\hat{n}}_\phi\}$:
\begin{align}\label{eq:Sphericalbasis}
\nonumber \mathbf{\hat{n}}&=[\sin \theta \cos\phi,\sin \theta \sin\phi,\cos \theta]\\ \mathbf{\hat{n}}_\theta&=[\cos \theta \cos\phi,\cos \theta \sin\phi,-\sin \theta]\\
\nonumber \mathbf{\hat{n}}_\phi&=[-\sin\phi,\cos\phi,0]
\end{align}
One can easily compute $\ket{j,\mathbf{\hat{n}}}$ from the ladder operators 
$J_{\pm,\mathbf{\hat{n}}}=\mathbf{\hat{n}}_{\theta}\cdot\mathbf{J}\pm i \mathbf{\hat{n}}_{\phi}\cdot\mathbf{J}$ as $J_{+,\mathbf{\hat{n}}}\ket{j,\mathbf{\hat{n}}}=0$, finding
\begin{align}\label{eq:CoherentStateDefinitive}
\nonumber \ket{j,\mathbf{\hat{n}}}&=\sum^{2j}_{n=0}\sqrt{\binom{2j}{n}}\left[\cos\frac{\theta}{2}\right]^{2j-n}\left[\sin\frac{\theta}{2}\right]^{n}\\
&\times e^{-i(j-n)\phi}\ket{j, j-n}
\end{align}
As the more usual coherent states, these states form an overcomplete basis of the Hilbert space,
\begin{equation}\label{eq:Identity}
\frac{2j+1}{4\pi}\int \mathrm{d}\Omega\ket{j,\mathbf{\hat{n}}}\bra{j,\mathbf{\hat{n}}}=\sum_m\ket{j,m}\bra{j,m}=I_{2j+1}
\end{equation}
The $P$-representation for a quantum state $\rho$ is defined from these coherent states as
\begin{equation}\label{eq:PRepresentation}
    \rho=\int \mathrm{d}\Omega~P(\mathbf{\hat{n}})\ket{j,\mathbf{\hat{n}}}\bra{j,\mathbf{\hat{n}}},~\int \mathrm{d}\Omega~P(\mathbf{\hat{n}})=1
\end{equation}
This definition does not completely fix $P(\mathbf{\hat{n}})$. In order to check it, we consider its expansion in spherical harmonics
\begin{align}
    P(\mathbf{\hat{n}})&=\sum^{\infty}_{\ell=0}\sum^{\ell}_{m=-\ell} p_{\ell m}Y^{m}_\ell(\mathbf{\hat{n}})\\
    \nonumber Y^{m}_\ell(\mathbf{\hat{n}})&=\sqrt{\frac{2\ell+1}{4\pi}\frac{(\ell-m)!}{(\ell+m)!}}\sin^m\theta P_\ell^m(\cos\theta) e^{im\phi}
\end{align}
where $P_\ell^m(\cos\theta)$ are the associated Legendre polynomials. Normalization of $P(\mathbf{\hat{n}})$ implies $p_{00}=1/\sqrt{4\pi}$. Since spherical harmonics form an orthonormal basis, when inserting Eq.~(\ref{eq:CoherentStateDefinitive}) in Eq.~(\ref{eq:PRepresentation}) we realize that only the coefficients $p_{\ell m}$ with $\ell\leq 2j$ are involved, while those for $\ell>2j$ remain undetermined. Thus, the most general expression for $P(\mathbf{\hat{n}})$ is
\begin{equation}
    P(\mathbf{\hat{n}})=\frac{1}{4\pi}+\sum^{2j}_{\ell=1}\sum^{\ell}_{m=-\ell} p_{\ell m}Y^{m}_\ell(\mathbf{\hat{n}})+Q_{2j}f(\mathbf{\hat{n}})
\end{equation}
with $Q_{2j}$ the projector onto $\ell>2j$ and $f(\mathbf{\hat{n}})$ an arbitrary angular function. Since $P(\mathbf{\hat{n}})$ is a real function, the number of real parameters determining the coefficients $\{p_{\ell m}\}^{2j}_{\ell=1}$ is precisely $(2j+1)^2-1=4j(j+1)$, the number of parameters fixing an arbitrary density matrix in a Hilbert space of dimension $2j+1$. Therefore, any density matrix can be described in terms of a certain function $P(\mathbf{\hat{n}})$, where those admitting a non-negative $P$-function represent classical states described by a conventional probability distribution.

We now focus on the case of qubits, where $j=1/2$ and the coherent states are the usual spin states
\begin{align}\label{eq:CoherentState1/2}
    \ket{\mathbf{\hat{n}}}&=\cos\frac{\theta}{2}e^{-i\frac{\phi}{2}}\ket{\uparrow}+\sin\frac{\theta}{2}e^{i\frac{\phi}{2}}\ket{\downarrow}
\end{align}
where $\ket{\uparrow},\ket{\downarrow}$ are the spin states along the $z$-axis, $\ket{\frac{1}{2}\pm \frac{1}{2}}$. Since $2j=1$, only spherical harmonics up to $\ell=1$ in the expansion of $P(\mathbf{\hat{n}})$ are involved in the expression of $\rho$. This can be straightforwardly checked by noticing in Eq.~(\ref{eq:PRepresentationQubit}) that
\begin{equation}
    \ket{\mathbf{n}}\bra{\mathbf{n}}=\frac{1+\mathbf{n}\cdot \mathbf{\sigma}}{2}
\end{equation}
Moreover, if we choose the $i=x,y,z$ basis for the spherical harmonics $Y^{1}_i(\mathbf{\hat{n}})=\sqrt{\frac{3}{4\pi}} \hat{n}_i$, we simply find that the $B_i$ coefficients of Eq.~(\ref{eq:GeneralQubitQuantumState}) are
\begin{equation}\label{eq:coefficientsxyzQubit}
B_i=\sqrt{\frac{4\pi}{3}}p_{1i}
\end{equation}
and thus
\begin{equation}
    P(\mathbf{\hat{n}})=\frac{1}{4\pi}\left[1+3(\mathbf{\hat{n}}\cdot\mathbf{B})\right]+Q_1f(\mathbf{\hat{n}})
\end{equation}
For one qubit, a non-negative $P(\mathbf{\hat{n}})$ can always be found since $\rho$ is in fact an incoherent mixture of the two spin states along the direction of the Bloch vector $\mathbf{B}$.

The two-qubit case is straightforwardly adapted from the one qubit case. In particular, the most general expression for $P(\mathbf{\hat{n}}_A,\mathbf{\hat{n}}_B)$, defined through Eq.~(\ref{eq:PRepresentation2Qubit}), is now
\begin{align}
    \nonumber P(\mathbf{\hat{n}}_A,\mathbf{\hat{n}}_B)&=\frac{1}{(4\pi)^2}[1+3\mathbf{B}^{+}\cdot\mathbf{\hat{n}}_A+3\mathbf{B}^{-}\cdot\mathbf{\hat{n}}_B\\
    &+9\mathbf{\hat{n}}_A\cdot\mathbf{C}\cdot\mathbf{\hat{n}}_B]+Q_{1}\otimes Q_{1} f(\mathbf{\hat{n}}_A,\mathbf{\hat{n}}_B)
\end{align}
Entangled states are thus states for which no $f(\mathbf{\hat{n}}_A,\mathbf{\hat{n}}_B)$ can be found such that $P(\mathbf{\hat{n}}_A,\mathbf{\hat{n}}_B)$ is non-negative.\\

\section{Specific entanglement criteria for $t\bar{t}$ pairs}\label{app:criteria}

We derive here some adapted results for $t\bar{t}$ quantum states from the general entanglement criteria discussed in Section~\ref{subsec:Hilbert}. For that purpose, we write the full matrix form of the general expression for a density matrix $\rho$ in a $2\times 2$ Hilbert space, Eq.~(\ref{eq:GeneralBipartiteStateRotations}):
\begin{widetext}
\begin{equation}\label{eq:GeneralBipartiteStateExplicit}
\resizebox{\hsize}{!}{$\rho=\frac{1}{4}\left[\begin{array}{cccc}
1+B^{+}_3+B^{-}_{3}+C_{33} & B^{-}_1+C_{31}-i(B^{-}_2+C_{32}) & B^{+}_1+C_{13}-i(B^{+}_2+C_{23}) & C_{11}-C_{22}-i(C_{12}+C_{21}) \\
B^{-}_1+C_{31}+i(B^{-}_2+C_{32})& 1+B^{+}_3-B^{-}_{3}-C_{33} & C_{11}+C_{22}+i(C_{12}-C_{21}) & B^{+}_1-C_{13}-i(B^{+}_2-C_{23}) \\
B^{+}_1+C_{13}+i(B^{+}_{2}+C_{23}) & C_{11}+C_{22}+i(C_{21}-C_{12}) & 1-B^{+}_3+B^{-}_{3}-C_{33} & B^{-}_1-C_{31}-i(B^{-}_{2}-C_{32})\\
C_{11}-C_{22}+i(C_{21}+C_{12}) & B^{+}_1-C_{13}+i(B^{+}_2-C_{23}) & B^{-}_{1}-C_{31}+i(B^{-}_{2}-C_{32}) & 1-B^{+}_3-B^{-}_{3}+C_{33}\\
\end{array}\right]$}
\end{equation}
\end{widetext}
where we are following the usual convention for the Pauli matrices $\sigma_i$. For the analysis of the Peres-Horodecki criterion, we compute the partial transpose of $\rho$ with respect to the second subsystem, $\rho^{\rm{T_2}}$, which amounts to transpose its four $2\times 2$ blocks, namely:
\begin{widetext}
\begin{equation}\label{eq:GeneralBipartiteStateExplicitTranspose}
\resizebox{\hsize}{!}{$\rho^{\rm{T_2}}=\frac{1}{4}\left[\begin{array}{cccc}
1+B^{+}_3+B^{-}_{3}+C_{33} & B^{-}_1+C_{31}+i(B^{-}_2+C_{32}) & B^{+}_1+C_{13}-i(B^{+}_2+C_{23}) & C_{11}+C_{22}+i(C_{12}-C_{21}) \\
B^{-}_1+C_{31}-i(B^{-}_2+C_{32})& 1+B^{+}_3-B^{-}_{3}-C_{33} & C_{11}-C_{22}-i(C_{12}+C_{21}) & B^{+}_1-C_{13}-i(B^{+}_2-C_{23}) \\
B^{+}_1+C_{13}+i(B^{+}_{2}+C_{23}) & C_{11}-C_{22}+i(C_{21}+C_{12}) & 1-B^{+}_3+B^{-}_{3}-C_{33} & B^{-}_{1}-C_{31}+i(B^{-}_{2}-C_{32}) \\
C_{11}+C_{22}+i(C_{21}-C_{12}) & B^{+}_1-C_{13}+i(B^{+}_2-C_{23}) & B^{-}_1-C_{31}-i(B^{-}_{2}-C_{32}) & 1-B^{+}_3-B^{-}_{3}+C_{33}\\
\end{array}\right]$}
\end{equation}
The Peres-Horodecki criterion states that $\rho^{\rm{T_2}}$ is non-negative \textit{iff}  $\rho$ is separable. A simpler version of this statement is obtained by considering vectors with only first and fourth component, which gives a reduced quadratic form
\begin{equation}\label{eq:GeneralBipartiteStateExplicitTranspose}
\rho_{C}\equiv\left[\begin{array}{cc}
1+B^{+}_3+B^{-}_{3}+C_{33} & C_{11}+C_{22}+i(C_{12}-C_{21}) \\
C_{11}+C_{22}+i(C_{21}-C_{12}) & 1-B^{+}_3-B^{-}_{3}+C_{33}\\
\end{array}\right]
\end{equation}
\end{widetext}
The non-negative character of $\rho^{\rm{T_2}}$ implies $\det \rho_{C}\geq 0$. Thus, $\det \rho_{C}<0$ is a \textit{sufficient} condition of entanglement since it implies that $\rho^{\rm{T_2}}$ is not positive semi-definite. In turn, $\det \rho_{C}<0$ can be rewritten as $\mathcal{P}>0$, with
\begin{align}
\nonumber \mathcal{P}&\equiv(B^{+}_3+B^{-}_3)^2+(C_{11}+C_{22})^2\\
&+(C_{21}-C_{12})^2-(1+C_{33})^2
\end{align}
Furthermore, 
\begin{equation}
\mathcal{P}\geq (C_{11}+C_{22})^2-(1+C_{33})^2\equiv \tilde{\mathcal{P}},
\end{equation}
so $\tilde{\mathcal{P}}>0$ provides an even simpler entanglement criterion. Specifically, since $1+C_{33}\geq0$, $\tilde{\mathcal{P}}>0$ \textit{iff}
\begin{equation}
|C_{11}+C_{22}|>1+C_{33}
\end{equation}
Thus, the entanglement signatures used in the main text
\begin{equation}\label{eq:DeltaGeneralized}
\Delta\equiv\frac{-C_{33}+|C_{11}+C_{22}|-1}{2}>0
\end{equation}
as well as
\begin{equation}\label{eq:DeltaTraceGeneralized}
W=D+\frac{1}{3}\equiv \frac{\textrm{tr}[\mathbf{C}]}{3}+\frac{1}{3}<0
\end{equation}
are sufficient conditions for entanglement in general, valid for arbitrary quantum states in $2\times 2$ bipartite Hilbert spaces.

We now address the specific case of $t\bar{t}$ production through LO QCD, characterized by unpolarized quantum states (the so-called $T$-states~\cite{Horodecki1996}), $B^{+}_i=B^{-}_i=0$, and a symmetric correlation matrix, $C_{ij}=C_{ji}$. The latter condition implies that the correlation matrix can be diagonalized after the appropriated rotation,  $C=\textrm{diag}[C_1,C_2,C_3]$,  which reduces Eq.~(\ref{eq:GeneralBipartiteStateExplicit}) to
\begin{equation}\label{eq:GeneralBipartiteStateExplicitDiagonal}
\resizebox{\hsize}{!}{$\rho=\frac{1}{4}\left[\begin{array}{cccc}
1+C_{3} & 0 & 0 & C_{1}-C_{2} \\
0 & 1-C_{3} & C_{1}+C_{2} & 0 \\
0  & C_{1}+C_{2} & 1-C_{3} & 0\\
C_{1}-C_{2} & 0 & 0 & 1+C_{3}\\
\end{array}\right]$}
\end{equation}
It is easy to see that
\begin{equation}\label{eq:PhysicalityRho}
\pm C_{3}+|C_{1}\pm C_{2}|-1\leq 0
\end{equation}
by demanding $\rho$ to be a physical state described by a non-negative density matrix. This complements the Peres-Horodecki criterion, which says that the state is entangled \textit{iff}
\begin{equation}\label{eq:EntanglementPeresHorodeckiDiagonal}
\pm C_{3}+|C_{1}\mp C_{2}|-1>0
\end{equation}
By combining both conditions, we find that Peres-Horodecki is equivalent to
\begin{align}\label{eq:EntanglementPeresHorodeckiDiagonalSign}
-C_3+|C_1+C_2|-1&>0,~C_3\leq 0\\
\nonumber C_3+|C_1-C_2|-1&>0,~C_3\geq 0
\end{align}
where the first line is just the condition $\Delta>0$ of Eq.~(\ref{eq:DeltaGeneralized}).

For a $T$-state, the concurrence can be also analytically computed from its definition (\ref{eq:Concurrence}), since in that case $\tilde{\rho}=\rho$. Hence, $\sqrt{\sqrt{\rho}\tilde{\rho}\sqrt{\rho}}=\rho$, and $\lambda_i$ are the eigenvalues of $\rho$,  $\lambda_i=\frac{1}{4}(1+C_3\pm|C_{1}-C_{2}|)$,~$\frac{1}{4}(1-C_3\pm|C_{1}+C_{2}|)$. Moreover, as $\rho$ has unit trace,
\begin{equation}
\sum_i \lambda_i=1,
\end{equation}
and the concurrence is simply $\mathcal{C}[\rho]=\max(2\lambda_1-1,0)$, with $2\lambda_1-1=(\pm C_{3}+|C_{1}\mp C_{2}|-1)/2$. By noting the analogy with Eq.~(\ref{eq:EntanglementPeresHorodeckiDiagonal}), we can write the concurrence directly in terms of the Peres-Horodecki criterion:
\begin{align}\label{eq:ConcurrenceSign}
\nonumber \mathcal{C}[\rho]&=\frac{1}{2}\max[-C_3+|C_1+C_2|-1,0],~C_3\leq 0\\
 \mathcal{C}[\rho]&=\frac{1}{2}\max[C_3+|C_1-C_2|-1,0],\,~~~C_3\geq 0
\end{align}

For the study of the angular-averaged quantum states $\rho_{\Omega}$, we consider the particular case in which there is invariance under rotations around a certain direction, chosen along $\sigma_3$ and labeled as the $z$-axis. In that case, the spin polarizations must be longitudinal, $B^{\pm}_{i}=B^{\pm}_{z}\delta_{i3}$, and the correlation matrix is diagonal, $C_{ij}=\delta_{ij}C_j$, with eigenvalues $C_{1}=C_{2}=C_{\perp}$ and $C_{3}=C_{z}$. The Peres-Horodecki criterion is then equivalent to 
\begin{equation}\label{eq:PeresHorodeckiAxial}
4C^2_{\perp}+(B^{+}_{z}+B^{-}_{z})^2-(1+C_z)^2>0
\end{equation}
If the state is unpolarized,
\begin{equation}\label{eq:PeresHorodeckiAxialLO}
\delta\equiv\frac{-C_z+2|C_{\perp}|-1}{2}>0
\end{equation}
is a necessary and sufficient condition for entanglement, while for $B^{\pm}_z\neq 0$, $\delta>0$ is just a sufficient condition.

\section{Parton distribution functions}\label{app:PDF}

We further explain here the concept of PDF and detail how the luminosity functions $L^{I}(M_{t\bar{t}},\sqrt{s})$ of Eq.~(\ref{eq:Rtotal}) are computed. The PDF $N_{\pi}(x)$ determines the probability of originating a parton $\pi$ from the corresponding hadron with an energy fraction $0 \leq x \leq 1$. For instance, in the case of $pp$ collisions, a $t\bar{t}$ pair is originated from the interaction between a parton $\pi$ from one of the protons and the corresponding antiparton $\bar{\pi}$ from the other proton. The probability of producing a state $I=\pi\bar{\pi}$ with partonic c.m. energy $\hat{s}=x_1x_2 s$ is $N_{\pi}(x_1)N_{\bar{\pi}}(x_2)$. In the case of $I=q\bar{q}$, $\pi$ can be either some light quark, $\pi=u,d,c,s,b$, or its corresponding antiparticle $\pi=\bar{u},\bar{d},\bar{c},\bar{s},\bar{b}$. In the case of $I=gg$, $\pi$ is a gluon. Since the partonic c.m. frame is that of the $t\bar{t}$ pair, the c.m. energy is the same, $M_{t\bar{t}}=\sqrt{\hat{s}}$, and thus
we can compute the luminosity function as
\begin{align}
\nonumber L_I(M_{t\bar{t}},s)&=\sum_{\pi}\int_{0}^{1}\int_{0}^{1}\mathrm{d}x_1\mathrm{d}x_2~\delta(M_{t\bar{t}}-\sqrt{\hat{s}})\\
&\times N_{\pi}(x_1)N_{\bar{\pi}}(x_2)
\end{align}
where the sum over the index $\pi$ runs over all the possible partons giving rise to the state $I$. Changing to dimensionless variables $x=\sqrt{x_1x_2}=M_{t\bar{t}}/\sqrt{s}$, $t=\sqrt{x_1/x_2}$ gives
\begin{equation}\label{eq:LuminosityPDF}
L_I(M_{t\bar{t}},s)=\sum_{\pi}\frac{2x}{\sqrt{s}}\int_{x}^{\frac{1}{x}}\frac{\mathrm{d}t}{t}~N_{\pi}(xt)N_{\bar{\pi}}\left(\frac{x}{t}\right)
\end{equation}
For the case of $p\bar{p}$ collisions, we simply replace one of the proton PDF above by the antiproton PDF $\bar{N}_\pi(x)$, which is that of the proton but interchanging partons with antipartons, $\bar{N}_\pi(x)=N_{\bar{\pi}}(x)$.

The luminosity integral (\ref{eq:LuminosityPDF}) is computed numerically, using the PDF values proportioned by the NNPDF30LO PDF set~\cite{Ball2014}. No significant change is found if other PDF sets are used.

\section{Angular averaging}\label{app:AngularAveraging}

We outline here the main technical details about the computation of the angular integrals presented in Section~\ref{subsec:angularint}.

\subsubsection{$q\bar{q}$ processes}

For computing the angular averages involved in the $q\bar{q}$ channel, the polar integrals in $\Theta$ can be computed either as polynomials in $t=\cos\Theta$ or in terms of 
\begin{equation}
F_n\equiv\int^{\frac{\pi}{2}}_0\mathrm{d}\Theta~\sin^n\Theta=\frac{n-1}{n}I_{n-2}=\frac{\Gamma\left(\frac{n+1}{2}\right)}{\Gamma\left(\frac{n+2}{2}\right)}\frac{\sqrt{\pi}}{2}
\end{equation}
with $\Gamma(x)$ the usual Euler gamma function.

The integral giving rise to the CHSH violation $\mathcal{B}^{q\bar{q}}_{\Omega}$ is much more delicate. We sketch here the main steps for its computation. First, we change variable as $\beta\cos\theta=\sqrt{z}$, so 
\begin{align}\label{eq:CHSHIntegralAngular}
    \nonumber &\frac{1}{2}\int^{\pi}_{0}\mathrm{d}\Theta~\sin\Theta\sqrt{(2-\beta^2\sin^2\Theta)^2+\beta^4\sin^4\Theta}\\
    &=\frac{1}{\sqrt{2}\beta}
    \int^{\beta^2}_{0}\mathrm{d}z~\sqrt{\frac{(z+1-\beta^2)^2+1}{z}} 
\end{align}
This last integral can be recurrently rewritten after integrating by parts as
\begin{align}
&\int^{\beta^2}_{0}\mathrm{d}z~\sqrt{\frac{(z+1-\beta^2)^2+1}{z}}=\frac{2\sqrt{2}}{3}\beta\\
&\nonumber +\frac{2}{3}\left[1+(1-\beta^2)^2\right]E_0(1-\beta^2)\\
&\nonumber +\frac{2}{3}(1-\beta^2) E_1(1-\beta^2)
\end{align}
with
\begin{equation}
    E_n(x)\equiv \int^{1-x}_0\mathrm{d}z~\frac{z^n}{\sqrt{z\left[(z+x)^2+1\right]}}
\end{equation}
This integral can be in turn expressed in terms of elliptic integrals:
\begin{align}
E_n(x)&=2\int^{\sqrt{1-x}}_0\mathrm{d}y~\frac{y^{2n}}{\sqrt{(y^2+x)^2+1}}\\
    &=a^{2n-1}\int^{1}_{\tau}\mathrm{d}t~\frac{\left(\frac{1-t}{1+t}\right)^{2n}}{\sqrt{\left(At^2+B\right)\left(A+Bt^2\right)}}\nonumber
\end{align}
where 
\begin{align}
t&\equiv \frac{a-y}{a+y},~\tau=\frac {a-\sqrt{1-x}}{a+\sqrt{1-x}} \\
A&\equiv \frac{1}{2}\left[1+\frac{b}{a}\right],~B\equiv\frac{1}{2}\left[1-\frac{b}{a}\right] \nonumber\\
a&=\sqrt[4]{x^2+1},~b=\sqrt{\frac{\sqrt{x^2+1}-x}{2}}  \nonumber
\end{align}
The last step is changing to angular variables as
\begin{align}
t&=\sqrt{k'}\tan\phi,~k'\equiv \frac{B}{A}\\
\nonumber \phi_1&=\arctan \frac{\tau}{\sqrt{k'}},~\phi_2=\arctan \frac{1}{\sqrt{k'}}
\end{align}
which yields
\begin{align}
    &E_n(x)=\frac{a^{2n-1}}{A}\int^{\phi_2}_{\phi_1}\mathrm{d}\phi~\frac{\left(\frac{1-\sqrt{k'}\tan\phi}{1+\sqrt{k'}\tan\phi}\right)^{2n}}{\sqrt {1-k^2\sin^2\phi}}\\
    &=\frac{a^{2n-1}}{A}\sum^{2n}_{m=0}\binom{2n}{m}(-1)^m2^m\eta_m(\phi_2,\phi_1,\sqrt{k'},k^2) \nonumber
\end{align}
In the above equation, $k^2=1-k'^2$ is the modulus of the elliptic integral, and $\eta_n(\phi_2,\phi_1,\alpha,k^2)\equiv \eta_n(\phi_2,\alpha,k^2)-\eta_n(\phi_1,\alpha,k^2)$, with
\begin{equation}
\eta_n(\varphi,\alpha,k^2)\equiv \int^{\varphi}_{0}\frac{\mathrm{d}\phi}{(1+\alpha \tan\phi)^n\sqrt {1-k^2\sin^2\phi}}
\end{equation}
The integrals $\eta_n(\varphi,\alpha,k^2)$ can be analitically obtained in terms of elliptic functions. For instance,
\begin{equation}
    \eta_0(\varphi,\alpha,k^2)= \int^{\varphi}_{0}\frac{\mathrm{d}\phi}{\sqrt {1-k^2\sin^2\phi}}\equiv F(\varphi,k^2)
\end{equation}
is just the usual incomplete elliptic integral of the first kind, which through the relation $u=F(\phi,k^2)$ defines the Jacobi elliptic functions 
\begin{equation}
    \textrm{sn}\, u=\sin \phi,~\textrm{cn}\, u=\cos \phi,~\textrm{tn}\, u=\tan \phi
\end{equation}
These functions allow to rewrite
\begin{equation}
    \eta_n(\varphi,\alpha,k^2)= \int^{F(\varphi,k^2)}_{0}\frac{\mathrm{d}u}{(1+\alpha\,\textrm{tn}\, u)^n}
\end{equation}
This integral can be computed analytically through recurrence relations. Nevertheless, the values of $\eta_n(\varphi,\alpha,k^2)$ for $n=1,2$, needed for the computation of $E_1(x)$, are too cumbersome to write them here; we refer the interested reader to Ref.~\cite{Byrd1971} for its complete expressions. For any practical purpose, we find that direct numerical integration of Eq. (\ref{eq:CHSHIntegralAngular}) is faster and simpler.

\subsubsection{$gg$ production}

For $gg$ processes, it can be seen that all the involved integrals can be put in terms of $K_{n,2}(\beta)$, with
\begin{equation}
K_{n,m}(x)\equiv\int^{x}_{-x}\mathrm{d}z~\frac{z^{2n}}{(1-z^2)^m}
\end{equation}
This integral satisfies recursion relations
\begin{align}
K_{n,m}(x)&=K_{n-1,m}(x)-K_{n-1,m-1}(x)\\
\nonumber K_{n,0}(x)&=2\frac{x^{2n+1}}{2n+1}\\
\nonumber K_{0,m}(x)&=\frac{1}{(m-1)}\left[\frac{x}{(1-x^2)^{m-1}}\right. \\ 
\nonumber  &\left.+\frac{2m-3}{2}K_{0,m-1}(x)\right]\\
\nonumber K_{0,1}(x)&=2\textrm{atanh}(x)=\ln \frac{1+x}{1-x}
\end{align}
that eventually yield
\begin{align}
K_{n,1}(x)&=2\left[\textrm{atanh}(x)-\sum^{n-1}_{k=0}\frac{x^{2k+1}}{2k+1}\right]\\
\nonumber K_{n,2}(x)&=\frac{x}{1-x^2}-(2n-1)\textrm{atanh}(x)\\
\nonumber&+\sum^{n-2}_{k=0}\frac{2(n-1-k)}{2k+1}x^{2k+1}
\end{align}

\bibliography{QI_Top_QCD}
\bibliographystyle{quantum}


\end{document}